\documentclass[twocolumn,showpacs,preprintnumbers,amsmath,amssymb]{revtex4-1}
\usepackage{epsfig}

\usepackage[table]{xcolor}
\usepackage{graphicx}
\usepackage{dcolumn}
\usepackage{bm}
\usepackage{esvect}

\usepackage{hyperref}
\hypersetup{colorlinks=true,linkcolor=magenta,anchorcolor=green,citecolor=cyan,filecolor=black,menucolor=black,urlcolor=brown}
\usepackage{slashed}

\newcommand{\be}{\begin{equation}}
\newcommand{\ee}{\end{equation}}
\newcommand{\ba}{\begin{eqnarray}}
\newcommand{\ea}{\end{eqnarray}}

\begin{document}

\title{Non-Relativistic  Formation of Scalar Clumps as a Candidate for Dark Matter}

\author{Philippe Brax}
\affiliation{Institut de Physique Th\'eorique, Universit\'e  Paris-Saclay, CEA, CNRS, F-91191 Gif-sur-Yvette Cedex, France}
\author{Jose A. R. Cembranos}
\affiliation{Departamento de  F\'{\i}sica Te\'orica and IPARCOS,\\
Universidad Complutense de Madrid, E-28040 Madrid, Spain}
\author{Patrick Valageas}
\affiliation{Institut de Physique Th\'eorique, Universit\'e  Paris-Saclay, CEA, CNRS, F-91191 Gif-sur-Yvette Cedex, France}

\begin{abstract}
We propose a new mechanism for the formation of dark matter clumps in the radiation era. We assume that a light scalar field is decoupled from matter and oscillates harmonically around
its vacuum expectation value. We include self-interactions and consider the nonrelativistic regime. The scalar dynamics are described by a fluid approach where the fluid pressure depends on both quantum and self-interaction effects. When the squared speed of sound of the scalar fluid becomes negative, an instability arises and the fluctuations of the scalar energy-density field start growing. They eventually become nonlinear and clumps form. Subsequently, the clumps  aggregate and reach a universal regime. Afterwards, they play the role of cold dark matter.
We apply this mechanism first to a model with a negative quartic term stabilised by a positive self-interaction of order six, and then to axion monodromy, where a subdominant cosine potential corrects a mass term. In the first case, the squared speed of sound  becomes negative when the quartic term dominates, leading to a tachyonic instability. For axion monodromy, the instability starts very slowly after the squared speed of sound  first becomes negative and then oscillates around zero. Initially the density perturbations perform acoustic oscillations due to the quantum pressure. Eventually, they start growing exponentially due to a parametric resonance.
The shape and the scaling laws of the clumps depend on their formation mechanism. When the tachyonic phase takes place, the core density of the clumps is uniquely determined by the energy density at the beginning of the instability. On the other hand, for axion monodromy, the core density scales with the soliton mass and radius.
This difference comes from the crucial role that the quantum pressure plays in both the parametric resonance in the linear regime and in the nonlinear formation regime of static scalar solitons.
 In both scenarios, the scalar-field clumps span a wide range of scales and masses, running from the size of atoms to that of galactic molecular clouds,
and from $10^{-3} \, {\rm gram}$ to thousands of solar masses.
Because of finite-size effects, both from the source and  the lens, these dark matter clumps
are far beyond the reach of microlensing observations. We find that the formation redshift of the scalar clumps can
span a large range in the radiation era; the associated background temperature
can vary from $10 \, {\rm eV}$ to $10^5 \, {\rm GeV}$,
and the  scalar-field mass
from $10^{-26}$ GeV to $10$ GeV.
\end{abstract}

\date{\today}

\maketitle


\section{Introduction}
\label{sec:introduction}

Dark matter is an essential ingredient of both astrophysics and cosmology. On very large scales, its description as a cold and pressureless fluid fits all our observations and is part of the standard model of cosmology \cite{Bergstrom:2012np}. From a more fundamental point of view, particle physicists have tried in the last few decades to find natural models of dark matter involving weakly interacting massive particles (WIMPs). Despite very promising theoretical candidates, such as neutralinos in supersymmetric models \cite{Kowalska:2018toh}, no experimental evidence  of WIMPs has emerged from data so far. This has triggered a recent revival of alternative models, where the origin of dark matter does not follow from the freezing out of particle interactions below a certain temperature. Indeed, ever since the analysis of the strong CP problem and the design of the axion mechanism, the possibility that dark matter could result from the coherent oscillations of a scalar field around the minimum of its potential has been considered \cite{Dine:1981rt,Abbott:1982af,Preskill:1982cy}. Since then, axions \cite{Peccei:1977hh,Wilczek:1977pj,Weinberg:1977ma,Vysotsky:1978dc} and Axion-Like-Particles (ALPs) \cite{Marsh:2015xka} have also been under intense scrutiny, both from the theoretical and the experimental sides. However, they are just one particular example of more general scalar dark-matter models, which can be described from an effective-field-theory point of view as parameterized by their masses and the set of their self-interactions \cite{Urena-Lopez:2019kud}. In the simplest case, called Fuzzy Dark Matter \cite{Hu:2000ke,Hui:2016ltb}, a  massive scalar field oscillating around its vacuum expectation value (vev), and with a sufficiently low mass $m\lesssim 10^{-21}$ eV, could play the role of dark matter. The resulting  properties of these scalar dark-matter models  are similar to the standard cold dark matter (CDM) for the formation of large-scale structures
\cite{Sakharov:1994id,Sakharov:1996xg,Johnson:2008se, Hwang:2009js, Park:2012ru, Hlozek:2014lca, Cembranos:2015oya,Cembranos:2016ugq}, but not for small scales, where distinctive features such as a non-vanishing speed of sound can leave different observational signatures
 \cite{Hlozek:2014lca,Schive:2014dra, Broadhurst:2018fei,Ostriker:2003qj,Cembranos:2005us,Weinberg:2013aya,Pontzen:2014lma,BoylanKolchin:2011de,Moore:1999nt,deBlok:2009sp,Cembranos:2015oya,Cembranos:2018ulm,Armengaud:2017nkf,Brax:2019fzb,Brax:2019npi,Brax:2020tuk}.

In this paper, we focus on one scalar field of mass $m$ whose self-interactions are subdominant compared to the quadratic mass term. We also consider that the oscillating scalar field is mostly time-dependent, with small space-dependent effects. In momentum space, this amounts to considering that momenta are small compared to the mass and that therefore the nonrelativistic approximation of the dynamics is valid. In this case, and after reducing the field to its equivalent quantum-mechanical picture, the Madelung transformation allows one to describe the evolution of the scalar field in terms of a fluid with non-trivial pressure terms. The first one, which is present even in the absence of self-interaction, is the so-called quantum pressure. It has a repulsive effect and allows for the formation of large solitons where the quantum pressure can balance the gravitational attraction \cite{Hui:2016ltb}. These quantum-pressure effects are at the heart of the fuzzy dark matter models, and require a mass term of low value, $m\lesssim 10^{-21}$ eV.   When self-interactions are present and overcome the quantum pressure, which can happen for masses larger than $10^{-21}$ eV, it is known that a repulsive pressure can also result from positive interaction potentials like $\phi^4$. In this case, large solitons can also form by gravitational instability and eventually stabilise when the gravitational attraction becomes balanced by the scalar self-repulsion \cite{Chavanis:2011uv,Brax:2019fzb,Brax:2019npi,Brax:2020tuk}.
In all these cases, dark matter is essentially a smooth fluid with large overdensities in the core of solitonic objects, which have galactic sizes and could play the role of galactic dark matter halos with a smooth inner region. This may alleviate some of the galactic-scale tensions with the data encountered by the standard CDM scenario.

On the other hand, and inspired by the possible representation of dark matter in the form of primordial black holes \cite{Carr:2020xqk} or massive compact halo objects (MACHOs) \cite{2007A&A...469..387T}, it can be envisaged that scalar-field clumps of much smaller sizes could exist and form all (or a large portion) of the dark matter content. In this case, the sign of the quartic self-interactions is crucial \cite{Chavanis:2017loo,Arvanitaki:2019rax}. For axions, this sign is negative leading to an attractive interaction between particles in the nonrelativistic regime. In the relativistic regime, the self-interaction can become dominant and lead to the formation of ``axitons'' as the squared mass of the axion can become negative for large excursions of the field, leading to a potential instability and the formation of clumps \cite{Kolb:1993zz,Schiappacasse:2017ham}. Another type of phenomenon, which  leads to  the creation of ``oscillons'' in some scalar field theories \cite{Amin:2010dc, Amin:2011hj, Olle:2019kbo,Zhang:2020bec}, has been attributed to an instability  where the modes can undergo a parametric resonance \cite{Olle:2019kbo} and the creation of nonlinear overdensities. In this paper, we will deal with similar mechanisms in the nonrelativistic regime. In this case, a tachyonic instability can be triggered when the speed of sound squared coming from the self-interactions becomes negative. This happens for simple models with a negative quartic interaction compensated by a positive one of degree six. This instability is counter-balanced by the quantum pressure on small scales and by the repulsive effects of the order-six term on larger scales,  leading to the creation of clumps mostly determined by the higher-order operator contribution. In another scenario, the model being of the axion monodromy type, with a scalar potential where a large mass term is modulated by small oscillations, the instability felt by the perturbations of the scalar energy density is due to a parametric resonance triggered when the speed of sound squared becomes negative too. In this case, the ensuing parametric-resonance growth of the instability is delayed by a long period of acoustic oscillations sustained by the quantum pressure. In most of these cases, gravity does not play a role and the clumps are formed in the radiation era. Their subsequent evolution first
involves their 2-body collisional aggregation and relaxation towards stable halos, which are next
diluted by the expansion of the Universe. At lower redshifts, these small scalar-field
solitons play the role of dark matter particles, in the same manner as primordial black holes
or MACHOs, and we recover the standard CDM cosmology.
{ Although these scalar clumps are usually much smaller than  galactic cores that can
form in the fuzzy dark matter models, e.g. they can be as small as one angstrom, they can also
reach sub-galactic sizes of the parsec scale, like galactic molecular clouds. Hence these
scenarios lead to a wide range of possible dark matter scales.}

The two types of formation mechanisms that we consider lead to very different properties for the clumps. In the tachyonic case, with a polynomial potential, the density in the core of the clumps is determined by the features of the potential, i.e. the energy density where the self-interactions
change from being attractive to repulsive (which also sets the background energy density at the beginning of the instability). For axion monodromy this is not the case, as the clumps can accommodate a continuous distribution of energy densities in their core. This sharp difference follows from the nature of the energy functional of the clumps as a function of the energy density. In the tachyonic case, the potential energy of the clumps admits a minimum which characterises the density of the clumps, giving a mass-radius
relation $M \sim R^3$.
In the axion monodromy setup, the potential energy is a decreasing function which does not select a unique equilibrium density, resulting in a $M\sim R^5$ mass-radius relation when the self-interaction dominates, and $M \sim R^4$ when gravity becomes the relevant interaction after the nonlinear collapse of the structures triggered by the parametric resonance instability.

As already stated, the dynamics comprise two steps. The first one, which we have just described, results from the type of instability of the fundamental model describing the physics of the scalar field, e.g. a polynomial interaction potential vs  axion monodromy. The second stage happens post-formation and follows a short aggregation phase,  which can influence the final mass and radius of the clumps. We describe in detail how this aggregation process depends on the mass-radius relationship of the clumps and therefore on the initial formation mechanism. Whereas in the polynomial interaction case the aggregation process leads to a significant growth of the size and mass of the clumps, in the axion monodromy case the aggregation is not very efficient and the mass and radii are unaffected.

The results that we present in this paper use two main ingredients. The first one is the leading-order fast harmonic motion of the field, with a frequency set by its mass $m$,
and the second one is the existence of an instability in the growth of the energy density contrast, which is triggered by the negative sign of the speed of sound squared.

The leading-order harmonic motion is guaranteed by the smallness of the perturbations to the scalar potential compared to the leading quadratic term. This  corresponds to models with typically  two scales
associated with two contributions of different origins to the scalar-field potential.
The first contribution, with a large amplitude,  is given by a quadratic term
and gives rise to the leading-order fast harmonic motion. The second contribution, with a small amplitude, is such that its nonlinear orders cannot be neglected.
We will consider two cases, a) when the small-amplitude self-interaction corrections to the quadratic term  are slow varying functions
such as a low-order polynomial, and b) when they show fast oscillations, such as a cosine term. The leading-order harmonic oscillations due to the quadratic term in the scalar potential ensure that the scalar field behaves like dark matter (with a mean density decaying as $1/a^3$ with the expansion of the Universe).
The subleading self-interactions however play a critical role, as they can lead to  instabilities and  the fragmentation of the homogeneous dark-matter distribution.

The two types of instabilities that we exemplify, i.e. the tachyonic and parametric resonance, have been considered in the literature in several contexts. In a recent paper, the case of the ``large-misalignment mechanism'' \cite{Arvanitaki:2019rax} was presented. In this scenario, and taking the cosine axion potential as an example, if the field starts initially close enough to the top of the potential, the instability due to the negative quartic term of the cosine function near the origin is delayed and a parametric resonance instability sets in. This leads to the formation of clumps which can be described as ``solitons'' when the gravitational attraction is balanced by the kinetic pressure and ``oscillons'' when gravity is irrelevant. Their (meta)-stability is entirely due to the scalar self-interactions. This scenario applies to the QCD axion and certain axion monodromy potentials which are flatter than quadratic for large field values.
In our analysis of the axion monodromy models, with large quadratic potentials perturbed by a small cosine interaction, we preserve  the harmonic motion at the leading order throughout our description of the parametric resonance instability. In this dominant-quadratic-term scenario, the nonrelativistic approximation applies throughout. In this case, the parametric resonance instability appears well before the argument of the cosine potential becomes small. Moreover, the speed of sound squared becomes negative well before the parametric resonance starts too. Contrary to the ``large-misalignment mechanism'', where the delay in the growth of perturbations is due to the flatness of the interaction potential initially, in our case the delay is due to the effects of the quantum pressure, which drives initial acoustic oscillations before becoming low enough and allowing the onset of the parametric resonance.
This delay can also be understood as the time it takes for these acoustic oscillations to become tuned to the frequency set by the cosine self-interaction potential
(thanks to their time dependence, due to the expansion of the Universe and the decrease of the
background density), so that a resonance can develop.
Nonrelativistic clumps and their formation have been analyzed numerically in a recent paper \cite{Amin:2019ums}, where a potential with a negative $\phi^4$ interaction term close to the origin was completed by higher order terms, eventually leading to a bounded potential for large field values. In this setting, the tachyonic instability plays a prominent role in the formation of the nonrelativistic clumps. Eventually, nonlinear effects take over and individual clumps form with little scalar interactions between each other. Later in the evolution, this gas of clumps is  affected by the gravitational attraction and they start moving towards each other. In this paper, we also present a similar mechanism for the formation of  clumps through a tachyonic instability and their stabilisation by higher order terms in the scalar potential. Then, we analyze the early aggregation process before the dilution by the expansion of the Universe. We also describe the same process in a thermodynamic way.
We pay particular attention to the parameter space combining  theoretical self-consistency conditions
and standard requirements (the formation of the dark matter clumps should occur before
matter-radiation equality and their size should not exceed the parsec scale).
We also check that their gravitational potential well is too weak to form black holes.
Finally, finite-size effects  imply that they cannot be detected by microlensing observations.
We find that the  scalar field can have a mass $m$ ranging from $10^{-17}$ eV to $10$ GeV,
giving rise to dark matter clumps that range from the size of atoms to that of galactic
molecular clouds.

The paper is arranged as follows. In Sec.~\ref{sec:classical}, we review the classical
field model associated with such a scalar field, and its nonrelativistic regime.
In Sec.~\ref{sec:tachyonic}, we describe our first scenario, associated with the tachyonic instability
where the speed of sound squared becomes negative at low background densities.
We first use a perturbative approach in Sec.~\ref{sec:perturbations},
to follow the growth of the scalar-field density perturbations.
In Sec.~\ref{sec:scalar-field-clumps}, we study the stable isolated scalar-field configurations
that arise in such a model, i.e. the ``solitons'' that correspond to the final dark matter clumps.
We estimate in Sec.~\ref{sec:aggregation} the efficiency of the collisional aggregation of these
scalar clouds, shortly after their formation and before they are diluted by the expansion of the
Universe, and we check in Sec.~\ref{sec:no-BH} that they do not collapse to black holes.
Then, in Sec.~\ref{sec:constraints}, we take into account theoretical constraints
to compute the parameter space of this scenario.
In Sec.~\ref{sec:mass-size-clumps} we compute the scales spanned by the scalar
dark-matter clumps and in Sec.~\ref{sec:micro-lensing} we check that they are far beyond
the reach of microlensing observations.

Next, in Sec.~\ref{sec:AxionM}, we present a different mechanism for clump formation,
associated with a parametric resonance. We take as an example a Lagrangian
inspired from axion monodromy, where a dominant mass term is corrected by a subleading
cosine term. The parametric resonance then arises from the interplay between this
oscillating self-interaction term, the quantum pressure, and the kinetic terms of the scalar field.
We again describe the perturbative growth of the scalar-field density fluctuations
and the stable solitons that can arise. We also compute the parameter space of this second
scenario and the size of the scalar clumps. Again, we check that they do not collapse into
black holes and are much below the observational threshold of microlensing observations.

We present our main conclusions in Sec.~\ref{sec:Conclusion}.
We finally complete our discussion with different appendices on thermodynamical phase transitions, parametric resonance, and soliton profiles.

\section{Classical fields and their nonrelativistic limit}
\label{sec:classical}

\subsection{Classicality}

In the following, we shall be interested in models of scalar dark matter where the dark-matter field can be described classically. This is a reasonable approximation for the quantum field $\phi$, whose nonrelativistic behavior will give rise to dark matter, if the occupation number
$N$ of the associated quantum state is very large. Denoting by $\rho$ the energy density of the field and by $n=\rho/m$ the number density, where $m$ is the mass of the scalar,  the occupation number can be estimated as \cite{Guth2015}
\be
N \simeq  \frac{\rho}{m} \lambda^3_{\rm dB} , \;\;\; \lambda_{\rm dB} = \frac{2\pi}{m v} ,
\ee
where $\lambda_{\rm dB}$ is the de Broglie wavelength of the scalar particles associated to $\phi$. Here $v$ is their typical velocity.
This gives the condition for classicality
\be
N \sim \frac{\rho}{m^4 v^3} \gg 1 .
\label{eq:classical}
\ee

We can envisage two types of situations. In the first one,
the energy density of the scalar field is nearly homogeneously distributed in the Universe and behaves like $\rho\simeq \rho_0/a^3$, where $\rho_0$ is the present dark-matter density in the Universe. Inside  large-scale inhomogeneities such as galaxy halos, the typical velocity of dark-matter particles $v_0$ is small and the classical regime is attained when
\be
m^4 v_0^3  \ll  \rho_0 \sim 10^{-48} \, {\rm GeV}^4 ,
\ee
where we consider low redshifts in the matter era. As we expect $v_0\simeq 10^{-3}$,
this is the case when
\be
\mbox{cosmological inhomogeneities only:} \;\;\;\; m\ll 0.1 \ {\rm eV} .
\ee
In this mass range the field can be treated classically. This also applies at higher redshifts,
as $\rho \propto a^{-3}$ and typically $v \sim a^{-1}$ because of the expansion of the Universe.

Another scenario is the one that we consider in this paper: dark matter is made of scalar-field clumps created in the radiation era and forming a bound state of dark-matter fluid.
Then, in a fashion similar to primordial black holes, these clumps play the role of dark matter particles
and behave at late times as in standard CDM cosmologies.
In this case, the density $\rho$ is large inside the clumps, reflecting the large energy densities at
the time of their formation, and the velocity is negligible as these clumps are equilibrium configurations.
Hence, for such clumps $N$ will be very large and we can treat $\phi$ as a classical field.
In fact, the classicality condition (\ref{eq:classical}) will provide a self-consistency constraint
on the parameter space of the scenarios we study in this paper.

\subsection{Equations of motion}

We focus on scalar-field models characterized by canonical kinetic terms and an interaction
potential $V_{\rm I}(\phi)$. Thus, they are governed by the action
\be
S[\phi] = \int d^4x \; \sqrt{-g} \left[ - \frac{1}{2} g^{\mu\nu} \partial_\mu\phi \partial_\nu\phi
- V(\phi) \right ] ,
\label{eq:S-phi-def}
 \ee
with
\be
V(\phi) = \frac{1}{2} m^2 \phi^2 + V_{\rm I}(\phi) .
\label{eq:V-phi-polynomial-1}
\ee
In this paper, we restrict our study to the nonrelativistic regime, when the self-interactions are
small as compared with the quadratic part,
\be
V_{\rm I} \ll \frac{1}{2} m^2 \phi^2  .
\label{eq:V-I-small}
\ee
At linear order in the gravitational potential $\Phi$ and for $m \gg H$, where $H$ is the Hubble
expansion rate, the equation of motion of the real scalar field $\phi$ in a perturbed
Friedmann-Lema\^itre-Robertson-Walker universe (FLRW) is
\be
\ddot{\phi} + 3 H \dot{\phi} - \frac{1}{a^2} \nabla^2\phi
+ (1+2\Phi) m^2 \phi + \frac{dV_{\rm I}}{d\phi} = 0 ,
\label{eq:phi-pert}
\ee
where $a$ is the scale factor of the Universe, normalised to unity now.
As we are interested in the classical behavior of the field $\phi$ in the nonrelativistic limit,
it is convenient to decompose
\be
\phi=\frac{1}{\sqrt{2m}} (\psi \, e^{-imt} + \psi^\star \, e^{imt}) ,
\label{eq:phi-psi-exp}
\ee
when  the spatial and time variations of $\psi$ are small compared to $m$. This ansatz emphasizes the fact that the scalar field oscillates with a pulsation $m$ as the quadratic terms in the scalar field action
(\ref{eq:S-phi-def}) dominate, following (\ref{eq:V-I-small}). From this we can deduce
the equation of motion of the nonrelativistic complex scalar field $\psi$,
\be
i \left( \dot{\psi} + \frac{3}{2} H \psi \right) = - \frac{\nabla^2 \psi}{2 m a^2} + m \Phi \psi
+ \frac{\partial {\cal V}_{\rm I}}{\partial\psi^\star} ,
\label{eq:psi-NR-motion}
\ee
which is a nonlinear version of the Schr\"odinger equation.
Here we introduced the effective nonrelativistic self-interaction potential
${\cal V}_{\rm I}(\psi,\psi^\star)$, which is obtained from $V_{\rm I}$ by averaging over the leading
oscillations $e^{\pm imt}$ of $\phi$.
For polynomial self-interactions, or analytic potentials that can be defined by
their Taylor expansion, with
\be
V_{\rm I}(\phi) = \Lambda^4 \sum_{p\ge 3} \frac{\lambda_p}{p}
\left( \frac{ \phi}{\Lambda} \right)^p ,
\ee
one obtains \cite{Brax:2019fzb}
\be
{\cal V}_{\rm I}( \psi,\psi^\star )= \Lambda^4 \sum_{p\ge 2}
\frac{\lambda_{2p}}{2p}\frac{(2p)!}{(p!)^2} \left( \frac{\psi \psi^\star}{2m \Lambda^2} \right)^p .
\label{eq:V-I-psi-psi-star}
\ee
It is convenient to introduce the Madel\"ung transform \cite{Madelung_1927}
\be
\psi= \sqrt{\frac{\rho}{m}} e^{iS} .
\ee
This defines the effective density field $\rho$, which coincides with the scalar-field
energy density in this nonrelativistic limit. The phase $S$ defines
an effective curlfree velocity field $\vec v$,
\be
\vec v =\frac{\vec \nabla S}{ma} .
\ee
Then, the equations of motion take a familiar form, i.e. the one of hydrodynamics
\cite{Chavanis:2017loo}.
The real part of the nonlinear Schr\"odinger equation gives the continuity equation
\be
\dot \rho + 3 H \rho + \frac{1}{a} \nabla \cdot (\rho \vec v) = 0 .
\label{eq:continuity-cosmo}
\ee
We can see that the self-interactions due to $V_{\rm I}$  do not modify this continuity equation.
The imaginary part of the nonlinear Schr\"odinger equation becomes the Hamilton-Jacobi relation
\be
\dot S + \frac{(\nabla S)^2}{2ma^2} = - m \Phi - m \frac{d{\cal V}_{\rm I}}{d\rho}
+ \frac{1}{2m a^2} \frac{\nabla^2 \sqrt{\rho}}{\sqrt{\rho}} ,
\label{eq:S-pert}
\ee
where the nonrelativistic self-interaction potential ${\cal V}_{\rm I}(\rho)$ is directly
obtained from ${\cal V}_{\rm I}(\psi,\psi^\star)$ in Eq.(\ref{eq:V-I-psi-psi-star})
with $\psi \psi^\star=\rho/m$,
\be
{\cal V}_{\rm I}( \rho )= \Lambda^4 \sum_{p\ge 2}
\frac{\lambda_{2p}}{2p}\frac{(2p)!}{(p!)^2} \left( \frac{\rho}{2m^2 \Lambda^2} \right)^p .
\label{eq:V-I-rho}
\ee
Then, taking the gradient of Eq.(\ref{eq:S-pert}) gives the hydrodynamical Euler equation,
\be
\dot {\vec v} +  H \vec v + \frac{1}{a} (\vec v \cdot \nabla) \vec v = - \frac{1}{a}
\nabla (\Phi + \Phi_{\rm I} + \Phi_{\rm Q} ) ,
\label{eq:Euler}
\ee
where we used $\nabla(\vec v^{\,2}) = 2 (\vec v \cdot \nabla) \vec v$ as
$\nabla\times\vec v=0$.
The self-interaction potential $\Phi_{\rm I}(\rho)$ is defined by
\be
\Phi_{\rm I}(\rho) = \frac{d{\cal V}_{\rm I}}{d\rho} ,
\label{eq:Phi-I-V-I-def}
\ee
and we have introduced the ``quantum pressure'' term
\be
\Phi_{\rm Q} = - \frac{\nabla^2 \sqrt \rho}{2m^2a^2 \sqrt \rho} .
\label{eq:Phi-Q}
\ee
The continuity equation and the Euler equation will show unstable solutions in the examples
we consider in this article, because of attractive self-interactions $\Phi_{\rm I}$
at low densities.
This description is valid provided the nonlinear terms are small compared to the quadratic terms
in the original action, as in (\ref{eq:V-I-small}). This translates into the conditions
\be
{\cal V}_{\rm I} \ll \rho, \;\;\; \mbox{hence} \;\;\; \Phi_{\rm I} \ll 1 .
\label{eq:NR-small-V-I}
\ee

\subsection{Cosmological background}
\label{sec:background}

\subsubsection{Real scalar field $\phi$}
\label{sec:background-phi}

We now restrict our attention to the cosmological background, where the scalar field
$\bar \phi$ only depends on time. The corresponding equation of motion is
\be
\ddot{\bar\phi} + 3 H \dot{\bar\phi}
+  m^2 \bar\phi + \frac{dV_{\rm I}}{d\phi} = 0 ,
\label{eq:phi-KG}
\ee
whose solution can be written as a slowly varying deformation of the harmonic oscillator,
\be
\bar\phi(t) = \bar\varphi(t) \cos(mt-\bar{S}(t)) .
\label{eq:phi-cos}
\ee
Notice the similarity with the ansatz (\ref{eq:phi-psi-exp}) defining the complex scalar
field $\psi$.
The amplitude of the scalar field evolves in time and decreases with the scale factor
\be
\bar\varphi = \bar\varphi_0 \, a^{-3/2} ,
\label{eq:psi0}
\ee
whilst the phase evolves according to
\be
\bar S(t) = \bar S_0 - \int_{t_0}^t dt \; m \,
\Phi_{\rm I}\left( \frac{m^2 \bar\varphi_0^2}{2 a^3}\right) .
\label{eq:S-S0}
\ee
Hence, at the background level, the scalar field oscillates harmonically at the leading
order, with the high frequency $m$ given by the scalar mass.
The Hubble expansion and the self-interactions give rise to a slow decay of the
amplitude and to a phase shift.
The power-law decay $\bar\varphi \propto a^{-3/2}$ shows that the scalar-field energy density
$\bar\rho_\phi \simeq m^2 \bar\phi^2/2$ decreases like $a^{-3}$ and plays the role of a nonrelativistic
dark-matter component.

\subsubsection{Nonrelativistic limit}
\label{sec:background-psi}

Comparing the solution (\ref{eq:phi-cos}) with the nonrelativistic decomposition
(\ref{eq:phi-psi-exp}),
we can see that, at the background level, the complex scalar field $\bar\psi$
is
\be
\bar\psi(t) = \bar\psi_0 \, a^{-3/2} e^{i\bar{S}} ,
\;\;\; \mbox{with} \;\;\; \bar\psi_0 = \sqrt{\frac{m}{2}} \bar\varphi_0 = \sqrt{\frac{\bar\rho_0}{m}} .
\label{eq:psi-background}
\ee
We can check that the solution defined by $\bar\rho= \bar\rho_0/a^3$ and $\bar S$ given by
Eq.(\ref{eq:S-S0}), which also can be written as
\be
\dot{\bar S} = -  \frac{m\Lambda^4 a^3}{2\bar\rho_0}
\sum_{n=2}^{\infty} \lambda_{2n} \frac{(2n)!}{(n!)^2}
\left( \frac{\bar\rho_0}{2m^2\Lambda^2a^3}\right)^n ,
\label{eq:Sdot-series-rho}
\ee
is indeed the solution of the equations of motion derived from the hydrodynamical action, which read
\ba
&& \dot{\bar S} = - m \frac{d {\cal V}_{\rm I}}{d\rho} ,
\label{eq:dot-S-rho} \\
&& \dot{\bar\rho} + 3 H \bar\rho = 0 .
\ea
Hence, at the background level, the evolution of the scalar field given by the hydrodynamical
equations reproduces the full solution to the scalar-field equation
(\ref{eq:phi-KG}).

\section{Tachyonic instability for smooth self-interactions}
\label{sec:tachyonic}

\begin{figure*}
\begin{center}
\epsfxsize=4.3 cm \epsfysize=6.5 cm {\epsfbox{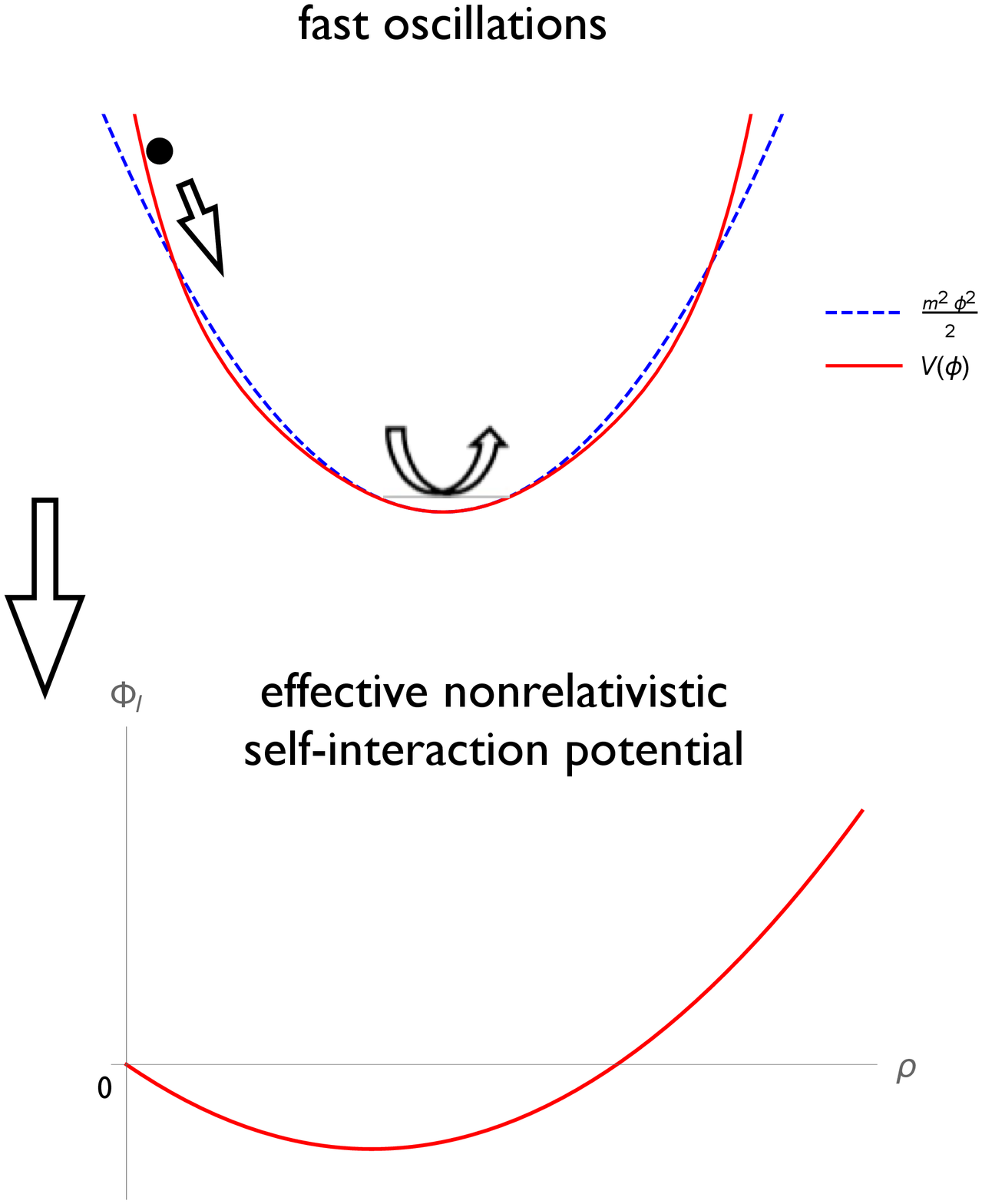}}
\epsfxsize=4.3 cm \epsfysize=7.3 cm {\epsfbox{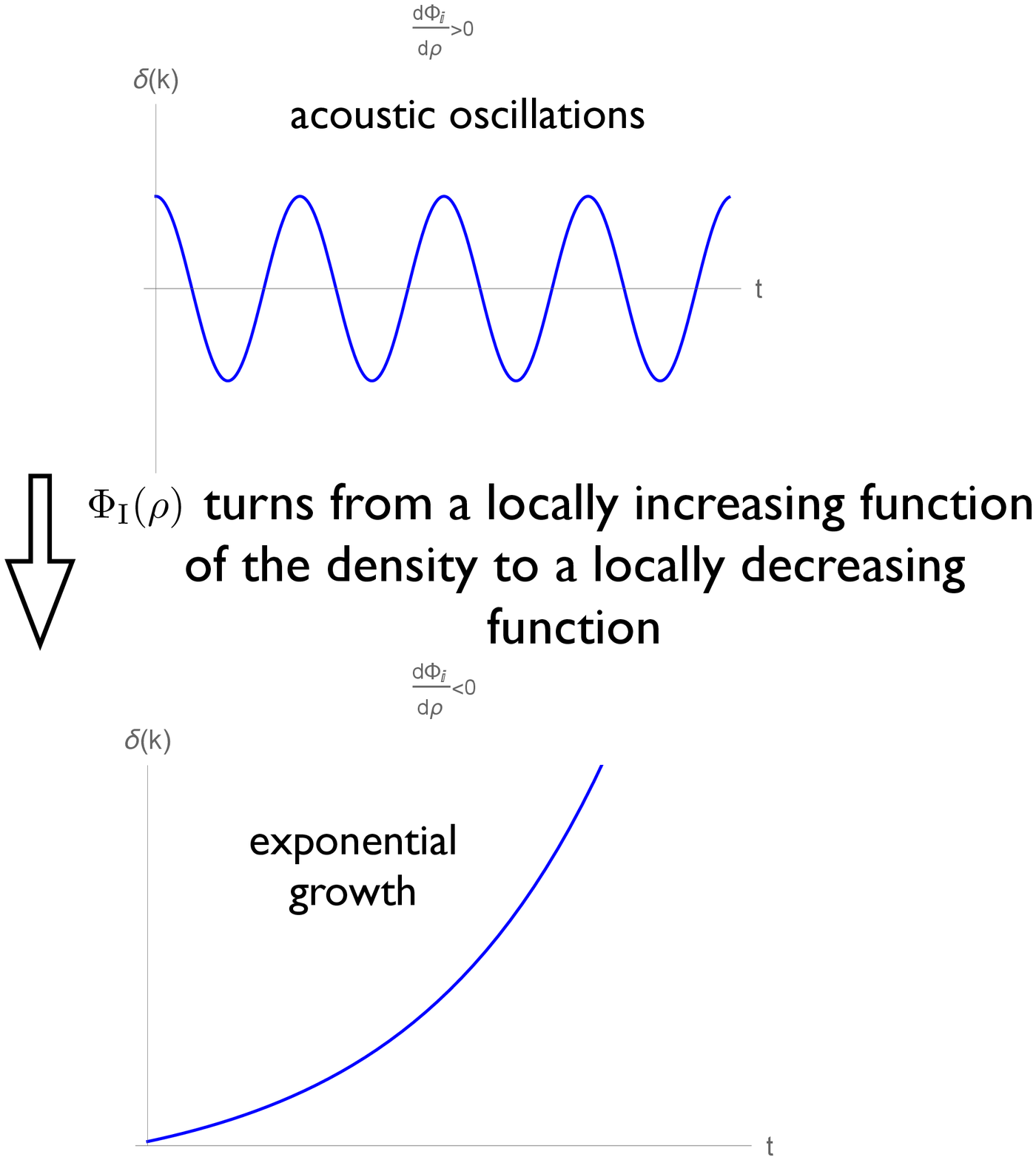}}
\epsfxsize=4.3 cm \epsfysize=6 cm {\epsfbox{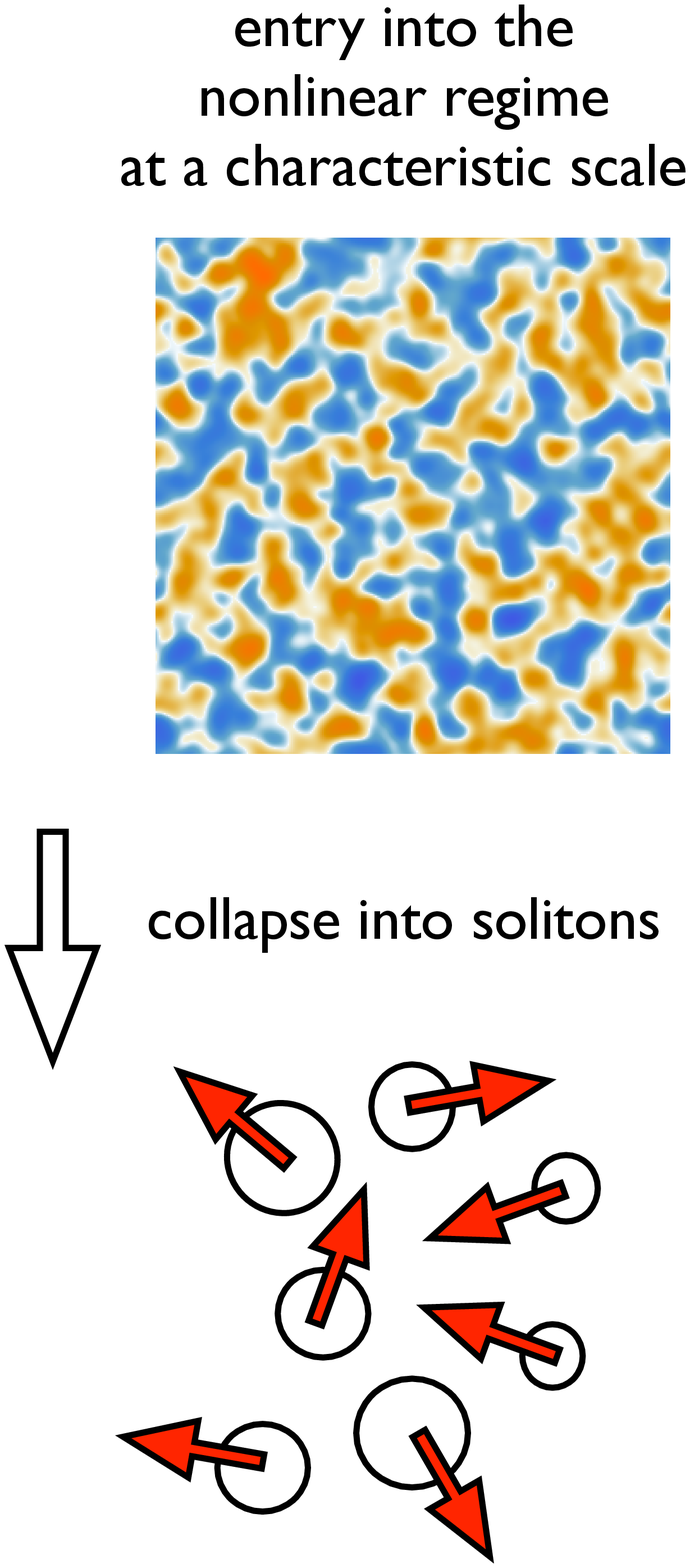}}
\epsfxsize=4.3 cm \epsfysize=6 cm {\epsfbox{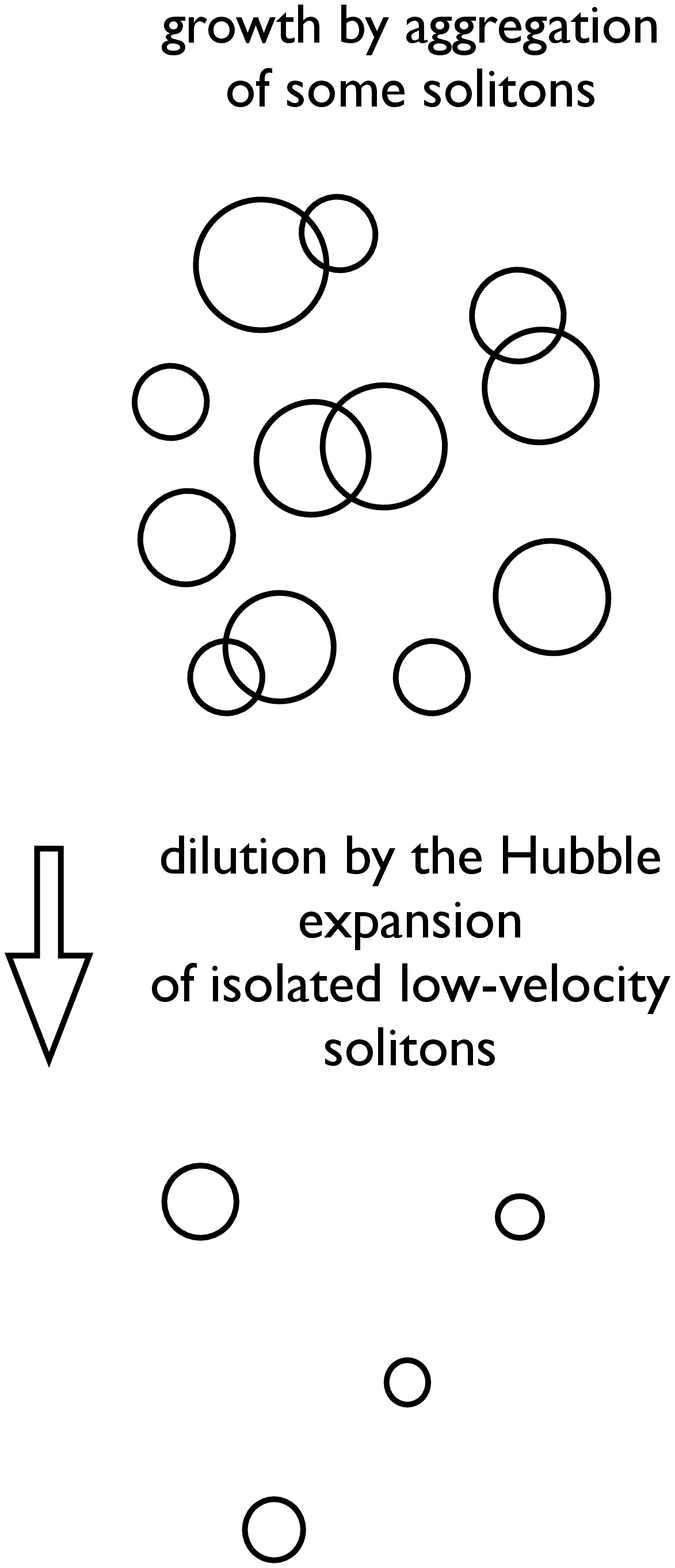}}
\end{center}
\caption{The main stages of the formation of scalar dark-matter clumps for the
tachyonic scenario (\ref{eq:V-I-polynomial}).
Cosmic time grows from the left column to the right colum, and from the upper panel
to the lower panel within each column. See the main text for explanations.}
\label{fig_Plots-poly}
\end{figure*}

\subsection{Polynomial self-interactions}
\label{sec:polynomial}

In the first part of this paper, we consider the scenario illustrated in Fig.~\ref {fig_Plots-poly},
associated with slowly-varying self-interaction potentials.
For { template}, we take a low-order polynomial case where
we directly define the model at the nonrelativistic level,
\be
\Phi_{\rm I} = - c_1 \frac{\rho}{\rho_\Lambda} + c_2 \frac{\rho^2}{\rho_\Lambda^2} , \;\;\;
{\cal V}_{\rm I} = - c_1 \frac{\rho^2}{2\rho_\Lambda} + c_2 \frac{\rho^3}{3\rho_\Lambda^2} ,
\label{eq:Phi-I-polynomial}
\ee
with $c_i>0$. This corresponds to
\be
V_{\rm I}(\phi)= - \frac{c_1 m^4}{3 \rho_\Lambda} \phi^4
+ \frac{2 c_2 m^6}{15 \rho_\Lambda^2}  \phi^6 .
\label{eq:V-I-polynomial}
\ee
We focus on the choice of parameters
\be
c_1 \sim c_2 \ll 1 ,
\ee
meaning that $\rho_\Lambda$ is the characteristic density
that governs the shape of the nonlinear self-interaction potential ${\cal V}_{\rm I}$.
As we will consider redshifts where $\bar\rho \sim \rho_\Lambda$, when
the change of shape of the potential can be felt, we also require $c_i \ll 1$ to fulfil the
nonrelativistic condition (\ref{eq:NR-small-V-I}).
In practice, at densities $\rho \gtrsim \rho_\Lambda$, we can expect higher-order terms
to come into play, if (\ref{eq:Phi-I-polynomial}) is understood as a Taylor expansion in
powers of $\rho$, originating from the Taylor expansion in $\phi$ of $V_{\rm I}(\phi)$.
However, the physics will not change, as long as $\frac{d\Phi_{\rm I}}{d\rho}$ shows
one change of sign, being negative at low densities and positive at high densities.

To facilitate the reading of this section, we already present in Fig.~\ref{fig_Plots-poly}
the formation process of the scalar-field clumps that will play the role of dark-matter particles
at low redshifts. From the first to the fourth column, this goes as follows.

1) The scalar field $\phi$ quickly oscillates in the potential $V(\phi)$, which is dominated
by its quadratic component with a small correction $V_{\rm I}$.
In the nonrelativistic regime, we can integrate over the fast oscillations of $\phi$. The slow dynamics
is then described by the complex scalar field $\psi$, or the hydrodynamics density and velocity
fields $\{\rho,{\vec v}\}$, and the self-interaction potential $\Phi_{\rm I}(\rho)$ defined by
Eq.(\ref{eq:Phi-I-V-I-def}), i.e. Eq.(\ref{eq:Phi-I-polynomial}) in our polynomial example.

2) At early times, when $\frac{d\Phi_{\rm I}}{d\rho}>0$, the scalar-field density perturbations
oscillate as acoustic waves. As the background density $\bar\rho$ decreases with time,
it finally enters the regime where $\frac{d\Phi_{\rm I}}{d\rho}<0$. This quickly leads to
a tachyonic instability ($c_s^2<0$) for some intermediate wave numbers $k$ and an
exponential growth of the density contrast $\delta({\vec k})$.

3) The scalar density field then quickly reaches the nonlinear regime and the overdense regions
collapse to form stable configurations (solitons).

4) Because of their non-negligible velocities, these scalar clouds collide and grow by aggregation,
relaxing towards more massive solitons.
Next, the expansion of the Universe dilutes these scalar clumps, which behave as isolated
CDM clumps.
At lower redshifts, the amplification by gravitational instability of perturbations on much larger
scales will form the cosmic web, galaxies, and clusters, as in the standard $\Lambda$CDM
scenario.

We describe in the following sections these various stages in more detail.

\subsection{Cosmological perturbations}
\label{sec:perturbations}

\subsubsection{Linear theory}
\label{sec:linear-theory}

For small perturbations with respect to the FLRW background, we can linearize the
equations of motion.
As explained in the previous sections, and as illustrated by the first column
in Fig.~\ref{fig_Plots-poly}, in the nonrelativistic regime it is convenient
to work with the fluid approach, where the fast harmonic oscillations of the scalar
field $\phi$ have been integrated out and we are left with the density-dependent
self-interaction potential $\Phi_{\rm I}(\rho)$.
Then, defining the linear density contrast $\delta$ and the divergence $\theta$
of the fluid velocity,
\be
\delta = \frac{\rho-\bar\rho}{\bar\rho} , \;\;\;   \theta = \frac{\nabla\cdot\vec{v}}{a} ,
\label{eq:delta-theta-def}
\ee
the continuity equation  gives the familiar constraint between the density contrast
and the divergence of the velocity field,
\be
\theta = -\dot\delta ,
\label{eq:continuity-linear}
\ee
whilst the Euler equation (\ref{eq:Euler}) implies
\be
\dot\theta + 2 H \theta = - \frac{1}{a^2} \nabla^2 (\Phi + \Phi_{\rm I} + \Phi_{\rm Q} ) .
\ee
Combining these two equations, and upon using the Poisson equation
$\nabla^2 \Phi= 4\pi {\cal G} a^2 \bar \rho \delta$
and the expression (\ref{eq:Phi-Q}) of the quantum potential,
we obtain, in Fourier space, the modified growth equation \cite{Chavanis:2011uv}
\be
\ddot\delta + 2 H \dot\delta + \left( c_s^2 \frac{k^2}{a^2} - 4 \pi {\cal G} \bar\rho \right) \delta
= 0 ,
\label{eq:delta-cosmo}
\ee
where we introduced the speed of sound $c_s$ as
\be
c_s^2 = \frac{k^2}{4 a^2 m^2} + \bar\rho \frac{d\bar\Phi_{\rm I}}{d\bar\rho} .
\label{eq:cs2}
\ee
The first term comes from the quantum potential and only plays a role at short distances. This will be crucial in what follows.

\subsubsection{Exponential instability}
\label{sec:tachyonic-exponential}

As long as $d\bar\Phi_{\rm I}/d\bar\rho > 0$, the only destabilizing force is gravity,
which is negligible in the regime we consider here, i.e. at large density and on short distances,  and is only important at very large
scales. However, when $d\bar\Phi_{\rm I}/d\bar\rho < 0$, the self-interactions lead to an
attractive force that can dominate on intermediate scales, { as also noticed
in \cite{Chavanis:2011zi} }.
Indeed, the quantum pressure always
becomes dominant on very small scales, which are thus stabilized. On very large scales, gravity plays a role too.
In this section, we investigate scenarios where $d\Phi_{\rm I}/d\rho > 0$ at high densities
and $d\Phi_{\rm I}/d\rho < 0$ at low densities. Then, at high redshifts with a large
background density $\bar\rho$, the system is stable, apart from the slow gravitational
instability on large scales, i.e. the Jean's instability, and the scalar field remains homogeneous.
At lower densities, the self-interactions become attractive and destabilize the system,
with a fast growth of perturbations on intermediate scales.
We denote by the subscript $c_s$ the scalar background density and the scale factor when
$d\Phi_{\rm I}/d\rho$ changes sign to become negative,
\be
\frac{d\Phi_{\rm I}}{d\rho}(\rho_{c_s}) = 0 , \;\;\; a_{c_s}
= \left( \frac{3\Omega_{\rm m 0} M_{\rm Pl}^2 H_0^2}{\rho_{c_s}} \right)^{1/3} ,
\label{eq:rho-cs-poly-def}
\ee
where we used $\bar\rho \propto a^{-3}$ from that period until today.
For the simple polynomial case (\ref{eq:Phi-I-polynomial}), this density is given by
\be
\rho_{c_s} = \frac{c_1}{2 c_2} \rho_\Lambda .
\label{eq:rho-cs}
\ee
More generally, this change of slope of $d\Phi_{\rm I}/d\rho$ will occur at a characteristic
density $\rho_\Lambda$ that governs the self-interaction potential ${\cal V}_{\rm I}$.
Typically, as in the polynomial case (\ref{eq:Phi-I-polynomial}), shortly after the time
$t_{c_s}$, e.g. after the Universe has expanded by a factor 2,
$d\Phi_{\rm I}/d\rho$ will be nonzero and of the order of $\Phi_{{\rm I}_{c_s}}/\rho_{c_s}$,
\be
a \sim 2 a_{c_s} : \;\;\; \bar\rho \frac{d\bar\Phi_{\rm I}}{d\bar\rho} \sim
- \left| \Phi_{{\rm I}_{c_s}} \right| \ll 1 ,
\ee
where the last constraint is the nonrelativistic condition (\ref{eq:NR-small-V-I}).
For the polynomial case (\ref{eq:Phi-I-polynomial}), this reads
$\left| \Phi_{{\rm I}_{c_s}} \right| \sim c_1 \ll 1$.
When $a \gtrsim a_{c_s}$, and  considering time scales that are short  compared
to the Hubble time and neglecting gravity, we obtain exponential growing and decaying
modes on intermediate scales,
$\delta_{\pm} \propto e^{\pm \gamma_k t}$, with
\be
q < q_{\rm up} : \;\;\; \gamma_{q} = \frac{q}{2m} \sqrt{q^2_{\rm up}-q^2} ,
\label{eq:gamma-q-qup}
\ee
where we have introduced the upper unstable wave number
\be
\frac{d\bar\Phi_{\rm I}}{d\bar\rho} < 0 : \;\;\;
q_{\rm up} = 2m \sqrt{ - \bar\rho \frac{d\bar\Phi_{\rm I}}{d\bar\rho} } ,
\label{eq:q-up-def}
\ee
and we denote by $q=k/a$ the physical wave number.
The maximum growth rate $\gamma_{\rm max}$ is reached at the wave number
$q_{\rm max}$, with
\be
q_{\rm max} = \frac{q_{\rm up}}{\sqrt{2}} , \;\;\;
\gamma_{\rm max} = \frac{q_{\rm up}^2}{4m}
= m \left| \bar\rho \frac{d\bar\Phi_{\rm I}}{d\bar\rho} \right| .
\label{eq:qmax-def}
\ee
Therefore, wave numbers around $q_{\max}$ become nonlinear first, as long as the
initial power spectrum is not too steep, and structures of physical size
$r\sim 2\pi/q_{\rm max}$ appear.
This perturbative growth of the scalar density perturbations is illustrated by the second
column in Fig.~\ref{fig_Plots-poly}.
Then, shortly after this time $t_{c_s}$, the system fragments into clumps of size
\be
r_{\rm NL} \sim \frac{2\pi}{q_{\rm max}} \sim \frac{1}{m \sqrt{ | \Phi_{{\rm I}_{c_s}} |}}
\gg \frac{1}{m} ,
\label{eq:r-clump}
\ee
and typical density of the order of $\rho_{\rm NL} \sim \rho_{c_s}$, with a mass
\be
M_{\rm NL} \sim \frac{\rho_{c_s}}{m^3 | \Phi_{{\rm I}_{c_s}} |^{3/2}} , \;\;\;
\rho_{\rm NL} \sim \rho_{c_s} \sim \rho_\Lambda .
\label{eq:m-clump}
\ee
Here the subscript ``NL'' refers to the fact that these are the first scalar-field structures
to reach the nonlinear regime, in terms of the density contrast $\delta \sim 1$.

\subsubsection{Constraints from the linear stage}

At the redshift $z_{c_s}$, assuming a standard inflationary scenario with adiabatic
initial conditions, the linear density contrast on subhorizon scales during the radiation era
is \cite{2011itec.book.....G}
\be \delta \sim - 9 \Phi_{i} \ln\left( \frac{k\eta}{\sqrt{3}} \right)
= - 9  \Phi_i \ln\left( \frac{q}{\sqrt{3}H} \right) .
\ee
This holds before the onset of the exponential instability and beyond the quantum pressure scale,
which stops the logarithmic growth. Here, $\eta$ is the conformal time, with $d\eta=dt/a$,
and the initial amplitude is of the order of $\Phi_{i} \sim 10^{-5}$.
Therefore, the exponential instability (\ref{eq:qmax-def}) reaches the nonlinear regime
in less than a Hubble time provided we have
\be
e^{\gamma_{\rm max}/H} > 10^{5} , \;\;\; \mbox{hence} \;\;\;  \gamma_{\rm max} > 12 H_{c_s}  .
\ee
Thus, we obtain the constraint that the growth rate is much greater than the Hubble expansion
rate, $\gamma_{\rm max} \gg H$, which reads
\be
m \left| \bar\rho \frac{d\bar\Phi_{\rm I}}{d\bar\rho} \right| \gg H , \;\;\;
\mbox{hence} \;\;\;  m \left| \Phi_{{\rm I}_{c_s}} \right| \gg H_{\rm c_s} .
\label{eq:exponential-growth}
\ee
This gives a constraint on the parameters $m$ and $\rho_{c_s}$,
\be
\frac{H_0}{m} \Omega_{\gamma_0}^{1/2} \left( \frac{\rho_{c_s}}{M_{\rm Pl}^2 H_0^2} \right)^{2/3}
\ll \left| \Phi_{{\rm I}_{c_s}} \right| \ll 1 ,
\ee
which also reads
\be
10^{-13} \left( \frac{m}{1 \, {\rm GeV}} \right)^{-1} \left( \frac{\rho_{c_s}}{1 \, {\rm GeV}^4} \right)^{2/3} \ll \left| \Phi_{{\rm I}_{c_s}} \right| \ll 1 .
\label{eq:exponential-instability}
\ee

The condition (\ref{eq:exponential-growth}) also ensures that we could neglect the expansion
of the Universe in the analysis above and that the scalar field $\phi$ had already started
fast oscillations in the zeroth-order quadratic potential $m^2\phi^2/2$, as
$m \gg H$ (i.e. the slow-roll regime governed by the Hubble friction is already finished).

We have neglected gravity in this analysis. This is valid provided $\Phi \ll \Phi_{\rm I}$.
The typical gravitational potential associated with these scalar-field clumps is
\be
\Phi \sim \frac{{\cal G} M_{\rm NL}}{r_{\rm NL}}
\sim \frac{\rho_{c_s}}{M_{\rm Pl}^2 m^2 | \Phi_{{\rm I}_{c_s}} |} .
\ee
Therefore, gravity is negligible during the formation of these structures if we have
\be
\Phi \ll \Phi_{\rm I} : \;\;\; \frac{\rho_{c_s}}{M_{\rm Pl}^2 m^2} \ll | \Phi_{{\rm I}_{c_s}} |^2 ,
\label{eq:Phi-small-Phi-I}
\ee
which reads
\be
10^{-37} \left( \frac{m}{1 \, {\rm GeV}} \right)^{-2}
 \frac{\rho_{c_s}}{1 \, {\rm GeV}^4}  \ll | \Phi_{{\rm I}_{c_s}} |^2 \ll 1 .
 \label{eq:weak-gravity-linear}
\ee

\subsection{Scalar-field solitons}
\label{sec:scalar-field-clumps}

Shortly after the entry into the nonlinear regime, the collapse of the first
structures builds scalar-field clumps that can grow through collisions.
We will analyse this aggregation process below in Sec.~\ref{sec:aggregation}.
However, after the scale factor $a(t)$ has increased by a factor two or so,
the expansion of the Universe dilutes these scalar-field clumps. Then, they behave like
isolated compact objects, such as MACHOs, and play the role of CDM particles.

In this section, we describe the way clumps, which are formed by the linear instability
studied previously, eventually settle to equilibrium configurations.
Of course, we cannot describe analytically the full time-dependent evolution of the scalar field,
from the initial instability to the stable configurations that we find below.
This would require numerical simulations that go beyond the present treatment.
However, we check that the scalar-field dynamics admit static configurations,
often called ``solitons'', which are solutions to the equations of motion and are natural
candidates for the end-point of the scalar-field structure-formation process.
In particular, they correspond to minima of the total energy at fixed mass,
which ensures their dynamical stability with respect to small nonlinear perturbations.

Therefore, we expect that the collapse of the first nonlinear structures,
illustrated by the third column in Fig.~\ref{fig_Plots-poly}, will build halos
that are not too far from these solitons. Moreover, as they are later diluted by the
Hubble expansion, these isolated clouds should naturally relax towards these
solitons, possibly radiating a small amount of scalar waves that can be accreted
by those clumps.
This picture is also corroborated by a thermodynamical analysis, which we present
in the appendix~\ref{sec:thermo}.

\subsubsection{Hydrostatic equilibrium as a minimum of the total energy}
\label{sec:Energy-minimum}

Neglecting the expansion of the Universe and using the fact that the velocity field is
curl-free, the continuity and Euler equations (\ref{eq:continuity-cosmo}) and (\ref{eq:Euler}) conserve the total energy
\be
E = E_{\rm kin} + E_{\rm grav} + E_{\rm I} + E_{\rm Q} ,
\label{eq:Etot-def}
\ee
where the kinetic, gravitational, self-interaction and quantum-pressure energies are
given by
\ba
&& E_{\rm kin} = \int d{\vec r} \, \rho \frac{\vec v^{\,2}}{2} , \;\;\;
E_{\rm grav} =  \frac{1}{2} \int d{\vec r} \, \rho \Phi , \nonumber \\
&& E_{\rm I} = \int d{\vec r} \; {\cal V}_{\rm I} , \;\;\;
E_{\rm Q} =  \int d{\vec r} \, \frac{(\nabla\rho)^2}{8 m^2\rho} .
\label{eq:E-rho-v}
\ea
Following \cite{Chavanis:2011zi,Chavanis:2017loo}, we can obtain the properties of
isolated scalar clumps from an energy principle.
Indeed, the conservation of  energy implies that local minima of $E$ are dynamically stable
with respect to small perturbations.
This variational analysis goes beyond linear stability and
infinitesimal perturbations, and we can expect isolated clumps to follow such profiles.
Local minima at fixed mass $M$ are given by the equation $\delta E - \alpha \delta M=0$,
where $\alpha$ is the Lagrangian multiplier associated with the constraint of fixed mass
\cite{Chavanis:2011zi,Chavanis:2017loo}.
For the energy (\ref{eq:E-rho-v}), the first variation with respect to $\rho$ and ${\vec v}$ gives
\be
\delta \rho \frac{{\vec v}^{\,2}}{2} + \rho{\vec v}\cdot\delta{\vec v}
+ \delta\rho ( \Phi + \Phi_{\rm I} + \Phi_{\rm Q} - \alpha ) = 0 .
\label{eq:equilibrium-E}
\ee
This implies that ${\vec v}=0$ and
\be
\Phi + \Phi_{\rm I} + \Phi_{\rm Q} = \alpha .
\label{eq:equilibrium-isolated}
\ee
Thus, we recover the hydrostatic equilibrium of the Euler equation (\ref{eq:Euler}),
$\nabla( \Phi + \Phi_{\rm I} + \Phi_{\rm Q} ) = 0$.
In the following we analyse the solutions to this equation.

\subsubsection{Gaussian ansatz for the radial profile}
\label{sec:Gaussian-profile}

It is not possible to obtain an explicit solution of Eq.(\ref{eq:equilibrium-isolated}), but
we can understand the main features of the equilibrium by minimizing the energy
over a class of trial functions.
Thus, { as in \cite{Chavanis:2011zi,Chavanis:2017loo} },
let us consider static Gaussian spherical density profiles at constant mass $M$,
\be
\rho(r) = \rho_c e^{-(r/R)^2} ,  \;\; \mbox{with} \;\; \rho_c = \frac{M}{\pi^{3/2} R^3} .
\label{eq:Gaussian-density-trial}
\ee
For the polynomial case (\ref{eq:Phi-I-polynomial}), their energies are
\ba
&& E_{\rm grav} = - \frac{\cal G}{\sqrt{2}} M^{5/3} \rho_c^{1/3} , \;\;\;
E_{\rm Q} = \frac{3\pi M^{1/3} \rho_c^{2/3}}{4 m^2} , \nonumber \\
&& E_{\rm I} = M \left[ - \frac{c_1}{2^{5/2}} \frac{\rho_c}{\rho_\Lambda}
+ \frac{c_2}{3^{5/2}} \frac{\rho_c^2}{\rho_\Lambda^2} \right] .
\label{eq:E-grav-Q-I}
\ea
Let us neglect the gravitational energy, in agreement with (\ref{eq:Phi-small-Phi-I}).
If we only had the quadratic term in $E_{\rm I}$, both $E_{\rm Q}$ and
$E_{\rm I}$ would be increasing functions of $\rho$. Then, the minimum of the energy
would be at $\rho_c=0$. Indeed, both the quantum pressure and the self-interactions
would be repulsive, so that there would be no stable state and the scalar cloud would
keep expanding and diluting (until gravity comes into play).
Therefore, for small stable clumps to exist, the linear attractive term in $E_{\rm I}$
must balance the quantum pressure before it is dominated by the quadratic repulsive term.
For $c_1 \sim c_2$ the transition between the attractive and repulsive regimes occurs
at $\rho \sim \rho_\Lambda \sim \rho_{c_s}$, as in Sec.~\ref{sec:linear-theory}.
Therefore, stable solitons exist provided $E_{\rm I} \gtrsim E_{\rm Q}$ at
$\rho \sim \rho_\Lambda$.
This gives
\be
\mbox{solitons exist if} \;\; M > M_{\rm min} , \;\;\;
M_{\rm min} \sim \frac{\rho_{\Lambda}}{c_1^{3/2} m^3} .
\label{eq:M-stable}
\ee
With $|\Phi_{{\rm I}_{c_s}}| \sim c_1$, we find that the initial clumps (\ref{eq:m-clump})
formed by the linear instability are actually of the order of the smallest stable mass $M_{\rm min}$.
We have seen in the derivation of (\ref{eq:M-stable}) that this threshold also corresponds
to a core density $\rho_c \sim \rho_\Lambda$. At higher masses, the quantum pressure
becomes negligible and the density is set by the minimum of $E_{\rm I}$. This gives
again $\rho_c \sim \rho_{\Lambda}$. Therefore, for all masses above the threshold
$M_{\rm min}$ we have
\be
M > M_{\rm min} : \;\;\; \rho_c \sim \rho_{\Lambda} , \;\;\;
R \sim \left( \frac{M}{\rho_\Lambda} \right)^{1/3} .
\label{eq:clumps-scaling}
\ee
This also means that the total energy $E$ of these solitons is dominated by their
self-interaction energy and it scales linearly with their mass,
\be
M \gg M_{\rm min} : \;\;\;   E_Q \ll E_{\rm I} , \;\;\; E \simeq E_{\rm I} \sim c_1 \, M .
\label{eq:clumps-scaling-E}
\ee
The solitons (\ref{eq:clumps-scaling}) correspond to the regime III-a
in Fig.~5 of the dense axion stars studied in \cite{Chavanis:2017loo},
where the results (\ref{eq:M-stable})-(\ref{eq:clumps-scaling-E}) were also derived.

\subsubsection{Numerical computation of the radial profile}
\label{sec:numerical-profile}

\begin{figure}
\begin{center}
\epsfxsize=8.8 cm \epsfysize=6 cm {\epsfbox{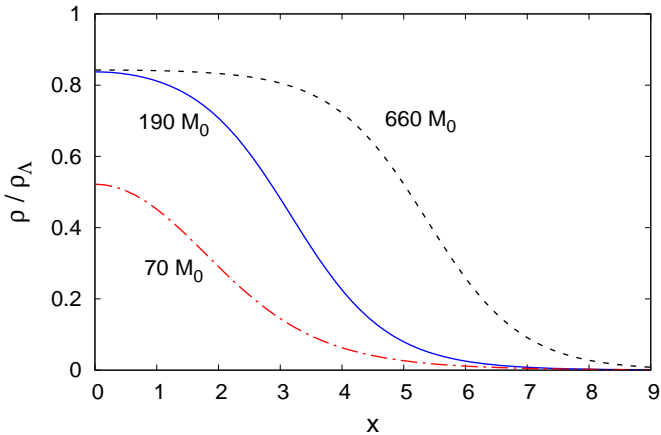}}
\end{center}
\caption{Radial density profile for the equilibrium (\ref{eq:equilibrium-isolated}).
We show the masses $M=70, 190$ and $660 \; M_0$, from left to right.}
\label{fig_rho_x_poly}
\end{figure}

A numerical computation of the soliton profiles confirms the analysis of the previous section.
Neglecting the gravitational energy, the equation of equilibrium (\ref{eq:equilibrium-isolated})
that describes minima of the total energy at fixed mass reads
\be
\frac{d^2y}{dx^2} + \frac{2}{x} \frac{dy}{dx} = 2 y \left[ - y^2 + \frac{c_2}{c_1} y^4 + \tilde\alpha \right] ,
\label{eq:y-x-profile}
\ee
where we introduced the dimensionless variables
\be
y = \sqrt{\frac{\rho}{\rho_\Lambda}} , \;\;\; x = \sqrt{c_1} m r , \;\;\; \tilde\alpha = - \frac{\alpha}{c_1} .
\ee
Then, the soliton mass reads
\be
\frac{M}{M_0} = 4\pi \int_0^\infty dx \, x^2 y^2 , \;\;\;
M_0= \frac{\rho_{\Lambda}}{c_1^{3/2} m^3} .
\label{eq:M0-def}
\ee
As expected, $M_0$ also sets the order of magnitude of the lower mass threshold
$M_{\rm min}$ of Eq.(\ref{eq:M-stable}).

We solve the boundary-value problem (\ref{eq:y-x-profile}) with a double-shooting method
(integrating from both boundaries and matching at an intermediate point)
for given values of $\tilde\alpha$.
Next, integrating the density over the radius gives the total mass $M$ as a function of $\tilde\alpha$.
We show in Fig.~\ref{fig_rho_x_poly} the density profiles that we obtain for the masses
$M=70, 190$ and $660 \; M_0$, when we take $c_1=c_2$.
We find that at large masses the core density stabilizes at values of the order of $\rho_\Lambda$
while the mass grows as $R^3$ with the characteristic radius $R$.
At large radii, $r \gg R$, the density shows an exponential tail, which is governed by
the quantum pressure.
We also find a lower value for the mass $M_{\rm min}$ of these equilibrium solutions,
with $M_{\rm min} \simeq 67 M_0$, in agreement with the scalings of Eq.(\ref{eq:M-stable}).
Thus, the numerical computation confirms the analytical predictions (\ref{eq:M-stable})
and (\ref{eq:clumps-scaling}). Because the self-interaction potential selects the unique density
scale $\rho_\Lambda$, which sets the scale of both the minima of ${\cal V}_{\rm I}$
and $\Phi_{\rm I}$, the equilibrium profiles have very simple properties.
They show a flat core at a density of the order of $\rho_\Lambda$ and higher masses
are obtained by increasing the radius, with $M \propto R^3$.

We discuss in more details in the appendix~\ref{app:trajectory-polynomial} the properties
of these solitonic profiles, interpreting the differential equation (\ref{eq:y-x-profile}) as the
damped motion of a particle $y(x)$ with time $x$ in a potential $U(y)$. This provides
another simple explanation of the behaviors found in Fig.~\ref{fig_rho_x_poly}.

As already advocated, within the nonrelativistic approximation, these solitons are stable configurations minimising the energy functional (\ref{eq:Etot-def}) for a given
value of the mass $M$. This is a feature of the nonrelativistic approximation, where the oscillation pulsation of the background field is $m$. For models  where the field probes higher harmonics of the scalar potential, for instance in axionic cases where the term in $-\phi^4$ becomes of the same
order as the quadratic term for large field values, the pulsation can vary at high enough density, leading to an instability of the solitons for large densities. This instability implies that the solitons can have a lifetime which can be much shorter than the age of the Universe \cite{Visinelli:2017ooc}. This is not the case here, as the scalar field always follows harmonic oscillations at the leading order.  Numerical simulations of a related case to the one presented in this paper confirm this observation \cite{Amin:2019ums}.

\subsection{Aggregation}
\label{sec:aggregation}

The perturbative analysis of Sec.~\ref{sec:perturbations} shows that the fragmentation
process starts at the redshift $z_{c_s}$ where the squared sound speed of
Eq.(\ref{eq:cs2}) changes sign to become negative.
Moreover, the typical size $r_{\rm NL}$ of these nonlinear structures is initially of the order of
$2\pi/q_{\rm max}$ as given by Eq.(\ref{eq:r-clump}), their density of the order of
$\rho_\Lambda$ and their mass $M_{\rm NL}$ given by Eq.(\ref{eq:m-clump}).
The comparison with Eqs.(\ref{eq:M-stable})-(\ref{eq:clumps-scaling}) shows that
this also corresponds to the lowest mass associated with stable solitons, as well as
with their core density,
\be
M_{\rm NL} \sim M_{\rm min} , \;\;\; \rho_{\rm NL} \sim \rho_{\Lambda} .
\label{eq:M-NL-M-min}
\ee
Therefore, we could expect these structures to relax towards stable solitons
of mass of the order of $M_{\rm min}$.
However, after formation and before gravity comes into play, these halos can grow
(or be destroyed) through direct collisions.
The typical peculiar velocity ${\vec v}_{\rm NL}$ at the formation time $t_{c_s}$
can be estimated from the linear theory, at its limit of validity when $\delta \sim 1$.
From the linearized continuity equation (\ref{eq:continuity-linear}) and the growth rate
$\delta({\vec k},t) \propto e^{\gamma_k t}$, we obtain
\be
v \sim \frac{\gamma_k\delta}{q} .
\ee
At the onset of the formation of the clumps, when the modes of physical wave number
$q_{\rm max}$ reach the nonlinear regime, we obtain from
Eqs.(\ref{eq:q-up-def})-(\ref{eq:qmax-def})
\be
v^2_{\rm NL} \sim | \Phi_{{\rm I}_{c_s}} |  .
\label{eq:v2-Phi-I}
\ee
If we assume that the halos aggregate after each collision, and relax to a more massive
equilibrium soliton with the scalings (\ref{eq:clumps-scaling}), their number
density decreases with time as
\be
\frac{d n}{dt} + 3 H n = - n^2 \sigma v ,
\label{eq:collisions-1}
\ee
with a cross section $\sigma \sim 4 \pi R^2$ and a typical relative velocity $v$.
This relies on the hydrodynamical picture, where scalar-field solitons behave as spheres of
a barotropic fluid with a large pressure. In the regime where quantum pressure dominates,
the wave-like nature of the system as described by the Schr\"odinger equation could
lead to true solitonic behaviors, where the solitons cross each other (as in the one-dimensional
cubic Schr\"odinger equation). However, in this paper we focus on a different regime
where the self-interactions dominate over the quantum pressure.
Thus, the bulk of the solitons and the scalings (\ref{eq:clumps-scaling}) are only determined
by the shape of the self-interactions, while the quantum pressure only governs the low-density
tail of the solitons.
Then, we can expect the system to behave like a fluid rather than a set of waves.
Thanks to the linear scaling with mass of the total energy (\ref{eq:clumps-scaling-E}),
this aggregation model conserves the total energy and can proceed without radiating
significant scalar-field waves.

The Hubble expansion rate decreases as $H(t) \propto a^{-2}$ in the radiation era
while the velocity dispersion decays as $v \propto 1/a$ with the expansion of the Universe,
if we neglect for simplicity the velocity changes due to collisions.
Assuming the mass distribution of the solitons remains peaked around a characteristic
mass $M(t)$, we have $M(t) \propto 1/(a^3 n)$ by conservation of the effective scalar-field
density $\rho$ in a comoving volume. This expresses the growth of the halos
as they merge and the falloff of their comoving number density.
Then, the cross section grows as $\sigma^2 \propto (a^3 n)^{-2/3}$.
This gives for the solution of Eq.(\ref{eq:collisions-1})
\be
n(t) = n_i \left( \frac{a}{a_i} \right)^{-3} \left[ 1 + \frac{n_i \sigma_i v_i}{6 H_i}
\left( 1 - \left( \frac{a_i}{a} \right)^2 \right) \right]^{-3} ,
\label{eq:nt-aggregation}
\ee
where the subscript $i$ stands for the initial condition at the formation time, $t_{c_s}$.
The first factor corresponds to the dilution by the expansion of the Universe and the
second factor to the mergings of the clumps.
At late times the comoving number density $n_c$ goes to a finite value,
\be
a \gg a_i : \;\;\; n_c = n_{c i} \left( 1 + \frac{n_i \sigma_i v_i}{6 H_i}
\right)^{-3} ,
\ee
which corresponds to a typical size and mass of the final solitons of the order of
\be
R_\infty = R_i \left( 1 + \frac{n_i \sigma_i v_i}{6 H_i} \right) , \;\;\;
M_\infty = M_i \left( 1 + \frac{n_i \sigma_i v_i}{6 H_i} \right)^3 .
\label{eq:R-Ri-growth}
\ee
At the initial time, of the order of $t_{c_s}$, we have from Eqs.(\ref{eq:r-clump}) and
(\ref{eq:v2-Phi-I}), in agreement with the analysis of Sec.~\ref{sec:scalar-field-clumps}
and with the relationship (\ref{eq:M-NL-M-min}),
\be
R_i \sim \frac{1}{m \sqrt{ | \Phi_{{\rm I}_{c_s}} |}} , \;\; v_i \sim \sqrt{ | \Phi_{{\rm I}_{c_s}} |} , \;\;
\sigma_i \sim R_i^2 , \;\; n_i \sim \frac{1}{R_i^3} .
\label{eq:Ri-vi-sigmai-ni}
\ee
This gives
\be
\frac{n_i \sigma_i v_i}{H_i} \sim \frac{m | \Phi_{{\rm I}_{c_s}} |}{H_{c_s}} \gg 1 ,
\ee
where we used the constraint (\ref{eq:exponential-growth}) associated with the exponential
growth of small perturbations at $z_{c_s}$.
Thus, we have a significant merging and growth of the scalar clouds.
Then, from Eq.(\ref{eq:R-Ri-growth}) the typical size and mass of the scalar clumps
formed at the end of the aggregation process is
\be
R_{\rm clump} \sim \frac{v_i}{H_i} \sim \frac{\sqrt{|\Phi_{{\rm I}_{c_s}}|}}{H_{c_s}} , \;\;\;
M_{\rm clump} \sim \frac{\bar\rho_{c_s}}{H_{c_s}^3} |\Phi_{{\rm I}_{c_s}}|^{3/2} .
\label{eq:R-ballistic}
\ee
This size corresponds to the distance that can be travelled by an initial soliton during a Hubble time,
sweeping material along the way, before the expansion of the Universe dilutes the scalar clouds
and halts collisions.
This ballistic approximation follows from the fact that we did not include the change of velocity
after collisions (but we included the growth of the cross section with the rise of the halo mass).
This is clearly an upper bound and we can expect a broad distribution
of halo sizes, $R_i \leq R \leq R_\infty$, with a typical size at a lower value associated
with Brownian-like trajectories.
Because of this significant aggregation process, the scalar-cloud  masses grow much beyond
the threshold (\ref{eq:M-stable}). This implies that the quantum pressure is negligible and
the radial profile of the solitonic solutions (\ref{eq:equilibrium-isolated}) is close to a top-hat,
as in (\ref{eq:clumps-scaling}).
We will check in Sec.~\ref{sec:constraints} below that gravity remains negligible
despite this growth of the soliton mass.
This aggregation process and the final dilution by the Hubble expansion towards
a collection of isolated dark-matter solitons are illustrated by the fourth column
in Fig.~\ref{fig_Plots-poly}.

In this section we have discussed the merging of the initial solitons by aggregation using an effective description based on the master equation
(\ref{eq:collisions-1}). This  provides a phenomenological understanding of the complex processes which occur when scalar-field configurations collide. A more precise characterisation of the dynamics of multi-soliton states and their collisions would require numerical simulations and a quantitative comparison with our effective results based on (\ref{eq:collisions-1}).
 Numerical studies of soliton collision have been performed in the recent past, for instance with Fuzzy Dark Matter in mind \cite{Guzman:2018evm,Schwabe:2016rze}. In the self-interacting case of interest here, semi-analytic methods combined with numerical studies have been used in the case of quartic interactions \cite{Cotner:2016aaq} and also for bounded potentials with an attractive quartic behaviour for small field values \cite{Amin:2019ums}. The latter case is the closest to the one presented in this section. Numerically, various types of phenomena have been observed. Merging of solitons occurs as well as orbiting solitons in a binary system and even bouncing. When merging happens, a certain amount of scalar energy has been seen to be  radiated away. This phenomenon was also observed in  the case of negative quartic interactions \cite{Hertzberg:2020dbk}, where it has been obtained that up to thirty percent of the initial soliton mass can be radiated away. This quantitative result has been obtained in a different part  of the soliton phase diagram, i.e. where gravity is responsible for the existence and stability of the solitons whereas in our case gravity is negligible. In our case we hope that the compact nature of the solitons could lead to a reduced rate of scalar wave emission.
 All in all, a better quantitative understanding of the effect of this possible radiation loss on the asymptotic number of clumps is certainly important. One analytic possibility would be to include loss terms in (\ref{eq:collisions-1}) which could be fitted with numerical results. Such an analysis requires new numerical simulations and a comparison with modified master equations with loss terms. We plan to come back to this topic in forthcoming works.

{

\subsection{No collapse into black holes}
\label{sec:no-BH}

We now check that the scalar-field clumps formed in this process do not collapse eventually
into black holes. This is avoided if the gravitational potential $\Phi$ at the surface of the
stable solitons obtained above is weak and far in the Newtonian regime, $| \Phi | \ll 1$.
From Eq.(\ref{eq:R-ballistic}) we have
\be
| \Phi | \sim \frac{{\cal G} M_{\rm clump}}{R_{\rm clump}} \sim
\frac{\bar\rho_{c_s} | \Phi_{{\rm I}_{c_s}} |}{M_{\rm Pl}^2 H_{c_s}^2}
\ll | \Phi_{{\rm I}_{c_s}} | \ll 1 .
\label{eq:Phi-small-poly}
\ee
Here we used the fact that the scalar-field energy density is subdominant in the radiation era,
so that $\bar\rho_{c_s} \ll M_{\rm Pl}^2 H_{c_s}^2$ from the Friedmann equation.
Therefore, the clumps are far in the weak-gravity regime and do not form black holes.
This is consistent with the fact that gravity is always subdominant with respect to the
scalar-field self-interactions.

}

{

\subsection{Parameter space}
\label{sec:constraints}

\begin{figure}
\begin{center}
\epsfxsize=8.8 cm \epsfysize=6 cm {\epsfbox{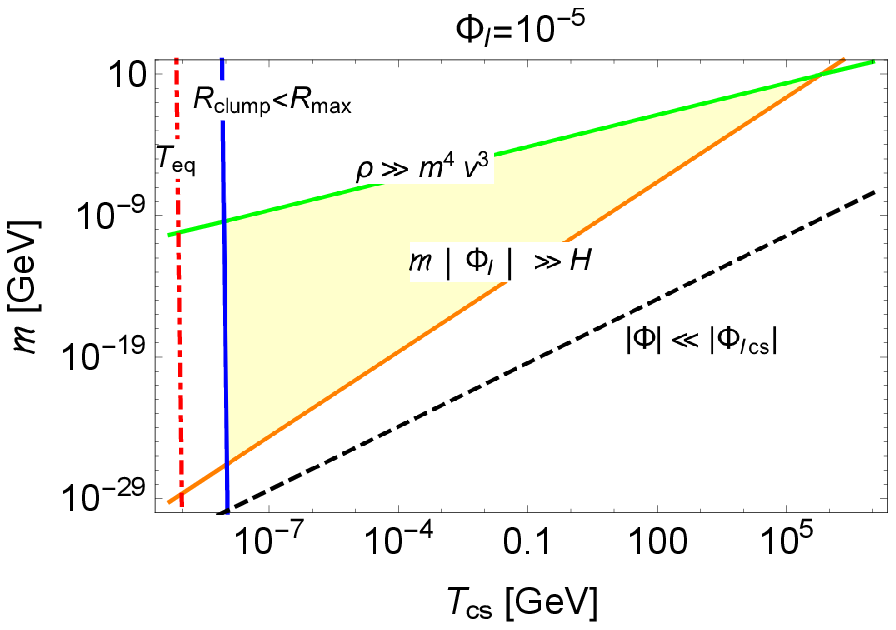}}\\
\epsfxsize=8.8 cm \epsfysize=6 cm {\epsfbox{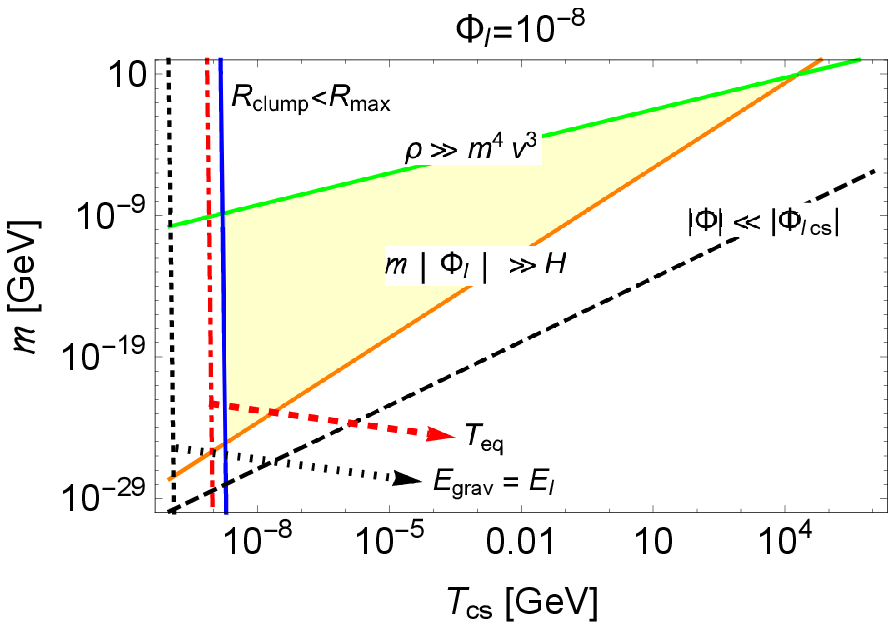}}
\end{center}
\caption{The yellow shaded area is the domain of validity, in the plane $(T_{c_s},m)$,
of the scenario described in this paper associated with potentials of the form
(\ref{eq:V-phi-polynomial-1}).
The upper panel shows the case $| \Phi_{\rm I_{c_s}} |=10^{-5}$ and the lower panel the case
$| \Phi_{\rm I_{c_s}} |=10^{-8}$.
From the left and turning clockwise, the constraints that delimit the allowed domain
are associated with the maximum size of the clumps, the classicality condition, and the condition
that the instability rate is much greater than the Hubble rate.
The left red dot-dashed line is the temperature $T_{\rm eq}$ at matter-radiation equality.
The lower black dashed line is the condition for gravity to be negligible during the formation process
(the linear stage of the tachyonic instability) while the left black dotted line in the lower panel
is the condition for gravity to be negligible in the final nonlinear solitons (it does not appear
in the upper panel as it is slightly to the left of this panel boundary).
Within the region delimited by the previous conditions, we automatically have
$T_{c_s} > T_{\rm eq}$ and negligible gravity.}
\label{fig_T-m-poly}
\end{figure}

The scenario described in the previous sections leads to the formation of scalar clouds,
or solitons, at times shortly after $t_{c_s}$. This is due to an exponential instability,
which leads to a fragmentation of the homogeneous background and the formation
of clumps of initial size (\ref{eq:r-clump}). This is followed within a Hubble time by a strong
aggregation process, where these scalar clouds merge to reach sizes up to (\ref{eq:R-ballistic}).
The profiles of these halos should relax to the solitonic solutions (\ref{eq:equilibrium-isolated}),
which for the large final masses, $M_{\rm clump} \gg M_{\rm min}$, are approximately top-hats
at the density $\rho_\Lambda$, from (\ref{eq:clumps-scaling}).
These scalar clouds form the dark matter ``particles'' that play the role of the WIMPs
or primordial black holes of other dark matter models.
In this section, we derive the parameter space of the model allowed by theoretical constraints.
This is shown in Fig.~\ref{fig_T-m-poly} in the $(T_{c_s},m)$ plane,
for the choices $| \Phi_{\rm I_{c_s}} |=10^{-5}$
and $| \Phi_{\rm I_{c_s}} |=10^{-8}$ (upper and lower panels).

First, we require the size of the scalar clumps to be below $R_{\max} = 1 \; {\rm pc}$,
so that they remain much below the size of small galaxies and can build
realistic dark matter profiles in galactic halos.
The typical size of the clumps formed at the end of the aggregation phase
was obtained in Eq.(\ref{eq:R-ballistic}), which also reads
\be
R_{\rm clump} = \frac{ | \Phi_{{\rm I}_{c_s}} |^{1/2} 3^{1/2} M_{\rm Pl} } {T_{c_s}^2} .
\label{eq:R-clump-T}
\ee
This gives the constraint
\be
R_{\rm clump} < R_{\max} : \;\;\;
T_{c_s} > \frac{  | \Phi_{{\rm I}_{c_s}} |^{1/4} 3^{1/4} M_{\rm Pl}^{1/2} } {R_{\max}^{1/2} } ,
\label{eq:Rclump-Rmax}
\ee
which is shown by the blue solid line labeled ``$R_{\rm clump} < R_{\max}$''
on the left in Fig.~\ref{fig_T-m-poly}, with the choice $R_{\max} = 1 \; {\rm pc}$.

Second, we require the formation of the scalar clumps to occur before the time
of matter-radiation equality.
This ensures that we recover the standard CDM scenario at lower redshifts.
Therefore, we impose the lower bound
\be
T_{c_s} > T_{\rm eq} , \;\;\; \mbox{with} \;\;\; T_{\rm eq} \simeq 1 \, {\rm eV} ,
\label{eq:T-eq-poly}
\ee
which is shown by the red dot-dashed line labeled ``$T_{\rm eq}$''  on the left
in Fig.~\ref{fig_T-m-poly}. We can see that for $| \Phi_{{\rm I}_{c_s}} | \gtrsim 10^{-9}$
this constraint is automatically satisfied once we verify the first constraint (\ref{eq:Rclump-Rmax}),
$R_{\rm clump} < R_{\max}$.

Next, we also have three theoretical self-consistency conditions.
First, the condition (\ref{eq:exponential-growth}) for an exponential instability
gives a lower bound on the scalar-field mass $m$,
\be
m |\Phi_{\rm I_{c_s}}| \gg H_{c_s} : \;\;\;
m \gg \frac{T_{c_s}^2}{\sqrt{3} | \Phi_{\rm I_{c_s}} | M_{\rm Pl} } .
\label{eq:m-cs-lower-bound-exponential}
\ee
This corresponds to the orange solid line labeled ``$m |\Phi_{\rm I}| \gg H$'' in Fig.~\ref{fig_T-m-poly}.
Here, we take a factor $10^3$ to ensure the left and right hand sides
are separated by at least three orders of magnitude.

Second, the classicality condition (\ref{eq:classical}) provides an upper bound on the scalar mass
$m$,
\be
m \ll \rho_{c_s}^{1/4} | \Phi_{\rm I_{c_s}} |^{-3/8} ,
\ee
where we used Eq.(\ref{eq:Ri-vi-sigmai-ni}) for $v_i$.
This can be written in terms of the temperature $T_{c_s}$ as
\be
\frac{\rho}{m^4 v^3} \gg 1 : \;\;\;
m \ll  \frac{M_{\rm Pl}^{1/8} H_0^{1/8} T_{c_s}^{3/4}}
{(3 \Omega_{\gamma 0})^{3/16} | \Phi_{\rm I_{c_s}} |^{3/8}} .
\label{eq:m-upper-bound-classical}
\ee
This is shown by the green solid line labeled ``$\rho \gg m^4 v^3$'' in Fig.~\ref{fig_T-m-poly}.
Here, we again take a factor $10^3$ to ensure the left and right hand sides
are separated by at least three orders of magnitude.

Third, we assumed that the gravitational force is negligible during the formation process.
This is given by the constraint (\ref{eq:Phi-small-Phi-I}), which also reads
\be
| \Phi | \ll | \Phi_{\rm I_{c_s}} | : \;\;\;
m \gg \frac{H_0^{1/4} T_{c_s}^{3/2}}
{(3\Omega_{\gamma 0})^{3/8} | \Phi_{\rm I_{c_s}} | M_{\rm Pl}^{3/4} } .
\label{eq:small-gravity-poly}
\ee
This corresponds to the black dashed line labeled ``$| \Phi | \ll | \Phi_{\rm I_{c_s}} |$''
in Fig.~\ref{fig_T-m-poly}.
We can see that it is automatically verified when the previous conditions are satisfied.

We can check that gravity remains small in the final solitons that are built after the nonlinear collapse
and the aggregation stage. This is satisfied provided we have
$| E_{\rm grav} | \ll | E_{\rm I} |$, where $E_{\rm grav}$ and $E_{\rm I}$ are the gravitational
and self-interaction energies of the final solitons.
From Eq.(\ref{eq:R-ballistic}) and with $E_{\rm I} \sim M c_1 \sim M | \Phi_{I_{c_s}} |$,
this gives the condition
\be
| E_{\rm grav} | < | E_{\rm I} | : \;\;\; T_{c_s} >
\frac{3^{1/4} M_{\rm Pl}^{1/2} H_0^{1/2}}{8\pi \Omega_{\gamma 0}^{3/4}} .
\label{eq:small-grav-sol-poly}
\ee
This is shown by the vertical black dotted line labeled ``$E_{\rm grav} = E_{\rm I}$''
on the left in the lower panel in Fig.~\ref{fig_T-m-poly}.
This line does not depend on the choice of $| \Phi_{I_{c_s}} |$ and it does not appear
in the upper panel because it is located slightly to the left of this panel boundary.
We can see that in both cases it is located to the left of the maximum-radius boundary
(\ref{eq:Rclump-Rmax}). Therefore, the condition (\ref{eq:small-grav-sol-poly}) is automatically
satisfied and the solitons always remain governed by the self-interactions.

Thus, as shown in Fig.~\ref{fig_T-m-poly}, the parameter space of the model takes the form
of a triangle in the $(T_{c_s},m)$ plane. It is delimited by the maximum clump size
(\ref{eq:Rclump-Rmax}), the exponential-instability condition (\ref{eq:m-cs-lower-bound-exponential}),
and the classicality condition (\ref{eq:m-upper-bound-classical}).
The requirements that the formation occurs before the matter-radiation equality,
(\ref{eq:T-eq-poly}), and that gravity remains small, (\ref{eq:small-gravity-poly}),
are automatically satisfied.
Gravity also automatically remains small in the final scalar clumps, (\ref{eq:small-grav-sol-poly}).
Thus, we can see that the scalar-field mass spans the range
\be
10^{-26} \, {\rm GeV} \lesssim m \lesssim 10 \; {\rm GeV} ,
\ee
while the background temperature at the redshift $z_{c_s}$ covers the range
\be
10 \, {\rm eV} \lesssim T_{c_s} \lesssim 10^5 \, {\rm GeV} .
\ee
This gives a wide range of temperatures and masses in the allowed parameter space.

}

{

\subsection{Mass and size of the scalar clumps}
\label{sec:mass-size-clumps}

\begin{figure}
\begin{center}
\epsfxsize=8.8 cm \epsfysize=6. cm {\epsfbox{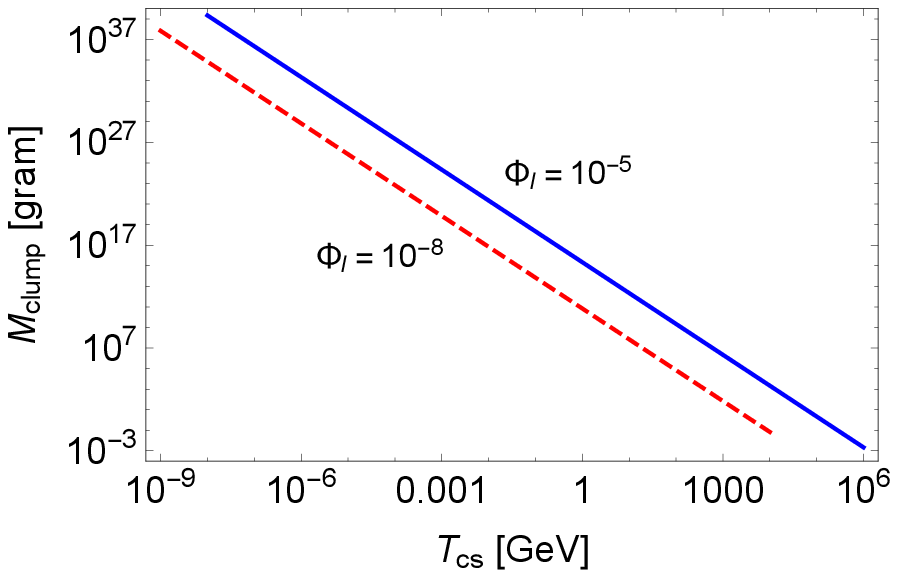}}\\
\epsfxsize=8.8 cm \epsfysize=6. cm {\epsfbox{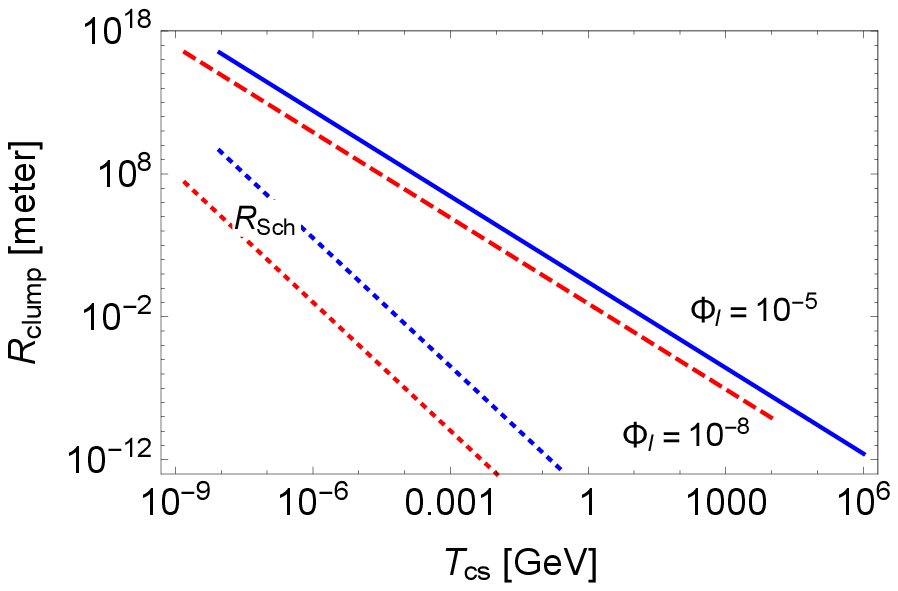}}
\end{center}
\caption{
{\it Upper panel:} mass of the clumps as a function of the background temperature $T_{c_s}$
at the onset of the tachyonic instability, for $| \Phi_{\rm I_{c_s}} |=10^{-5}$ (upper blue solid line)
and $| \Phi_{\rm I_{c_s}} |=10^{-8}$ (lower red dashed line).
{\it Lower panel:} radius of the clumps. The lower dotted lines show the Schwarzschild radius
$R_{\rm Sch}$.
}
\label{fig_M-Tcs-poly}
\end{figure}

The typical size and mass of the clumps formed at the end of the aggregation phase
were obtained in Eq.(\ref{eq:R-ballistic}). This gave Eq.(\ref{eq:R-clump-T})
for the radius, and for the mass:
\be
M_{\rm clump} = \frac{ | \Phi_{{\rm I}_{c_s}} |^{3/2} 3^{3/4} M_{\rm Pl}^{7/2} H_0^{1/2} }
{ \Omega_{\gamma 0}^{3/4} T_{c_s}^3 } .
\label{eq:M-clump-T}
\ee
The clump mass and radius are independent of the scalar-field mass $m$ and only depend on the
redshift $z_{c_s}$ when the tachyonic instability appears.
We show in Fig.~\ref{fig_M-Tcs-poly} the clump mass and radius as a function of $T_{c_s}$.
We also display the Schwarzschild radius of the clumps,
\be
R_{\rm Sch} = 2 {\cal G} M .
\label{eq:R-Sch-def}
\ee
It is much smaller than the radius of the clumps, in agreement with the result (\ref{eq:Phi-small-poly})
that the clumps are in the weak-gravity regime and do not form black holes.

We can see that the clumps cover a huge range of masses and radii, from microscopic
to sub-galactic scales.
Thus, their mass goes from $10^{-3} \, {\rm gram}$ up to
$10^{37} \, {\rm gram} \sim 10^4 \, M_\odot$, and their radius from $0.01 \, {\rm angstrom}$
to $1 \, {\rm parsec}$.
At low mass, their core density is of the order of $\rho \sim 10^{27} \, {\rm gram/cm^3}$,
much above that of neutron stars, while at large mass it is of the order of
$\rho \sim 10^{-13} \, {\rm gram/cm^3} \sim 10^{17} \bar\rho_0$, which remains much greater
than the current mean density $\bar\rho_0$ of the Universe.
At the large-mass end, these clumps are thousand times more massive than the Sun,
like the most massive stars, but have much greater radii, up to the parsec.
Thus, they are similar to galactic molecular clouds and do not correspond to the
standard stellar-mass MACHOs (massive compact halo objects), which are strongly
constrained by microlensing observations.

}

{

\subsection{Evading microlensing constraints}
\label{sec:micro-lensing}

\begin{figure}
\begin{center}
\epsfxsize=8.8 cm \epsfysize=6. cm {\epsfbox{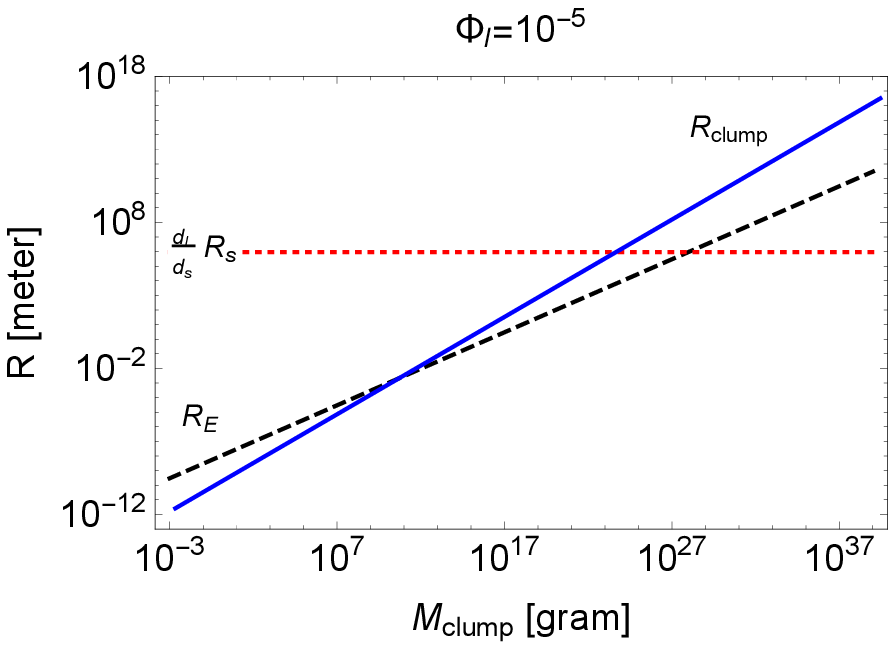}}\\
\epsfxsize=8.8 cm \epsfysize=6. cm {\epsfbox{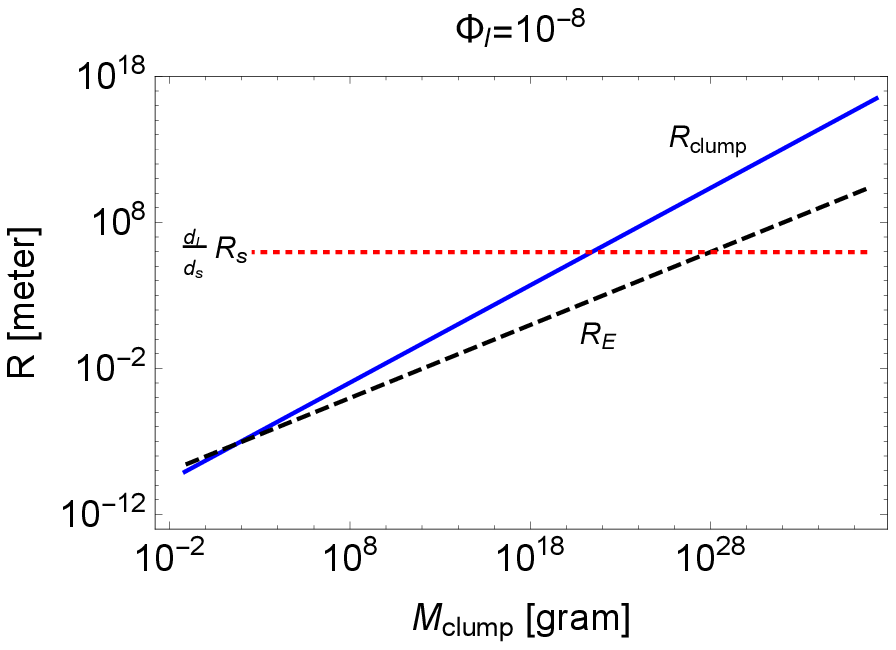}}
\end{center}
\caption{Characteristic radii in the lens plane. We show the clump radius $R_{\rm clump}$
(blue solid line), the Einstein radius $R_E$ (black dashed line), and the outer
impact parameter $\frac{d_L}{d_s} R_s$ of a source of one solar radius aligned with the lens
(red dotted line). We take $d_L = 1 \, {\rm kpc}$ and $d_s = d_{\rm M31} \simeq 770 \, {\rm kpc}$.}
\label{fig_M-R-poly}
\end{figure}

Massive compact halo objects, such as primordial black holes, can be constrained by
microlensing observations.
Indeed, such MACHOs located in the Milky Way halo would cause a time-varying amplification
of background stars when then cross their line of sight.
Monitoring the Andromeda galaxy (M31) with the Subaru Hyper Suprime-Cam (HSC), the number
of observed microlensing events has provided strong upper bounds on the abundance
of primordial BH in the mass range $10^{-11} < M_{\rm BH} < 10^{-6} M_\odot$
\cite{Niikura:2017zjd}.
At low BH mass, the microlensing sensitivity is strongly decreased by finite-source-size
and wave-optics effects
\cite{Niikura:2017zjd,Sugiyama:2019dgt,Smyth:2019whb}.
In this section, we show that the scalar-field solitons produced in our scenario only produce
very small magnifications of distant stars, much below observational thresholds.

For pointlike lenses, the relevant scale in the lens plane is the Einstein radius $R_E$,
\be
R_E = \sqrt{4 {\cal G} M d_L (1-d_L/d_s)} ,
\label{eq:R_E-def}
\ee
where $d_L$ and $d_s$ are the distances from the observer to the lens and to the source
\cite{Schneider_1992,Bartelmann:2010fz}.
For $d_L = 1 \, {\rm kpc}$ and $d_s = d_{\rm M31} \simeq 770 \, {\rm kpc}$, this gives
\be
R_E \simeq 10^{-8} \sqrt{ \frac{M_{\rm clump}}{1 \, {\rm g}}} \; {\rm meter} ,
\ee
which is shown by the black dashed line in Fig.~\ref{fig_M-R-poly}.
We can see that at large masses $R_E$ becomes smaller than the radius $R_{\rm clump}$
of the clumps. Therefore, in contrast with the case of primordial black holes, we must take into
account the effects associated with the finite size of these lenses and this will
make high-mass solitons evade detection by microlensing.
The flat red dotted line in Fig.~\ref{fig_M-R-poly} shows the impact parameter
$\frac{d_L}{d_s} R_s$ of a circular source of one solar radius, $R_s=R_\odot$,
which is aligned with the lens and the observer.
At low clump mass, $\frac{d_L}{d_s} R_s$ is much greater than the clump size and the
Einstein radius, which means that the finite size of the source plays a significant role.
For the case of primordial black holes, this finite-source effect significantly decreases
the lensing magnification. This implies that microlensing observations cannot constrain
small black holes below $10^{22} \, {\rm gram}$
\cite{Niikura:2017zjd,Sugiyama:2019dgt,Smyth:2019whb}.
This will also prevent the detection of low-mass solitons in our case.
For small lenses, wave-optics effects also decrease the magnification as compared with the
geometrical-optics prediction that neglects finite-lens effects. However, these
wave-optics effects are subdominant and smeared out by the finite-size effects of the lens
\cite{Sugiyama:2019dgt}. Therefore, in this paper we do not consider the subdominant
wave-optics effects and focus on the dominant finite-size effects, which already
reduce the microlensing magnification to a very small level.

To simplify the computation, we approximate the lens by a disk of constant surface density
$\Sigma$. This should be a good approximation as the solitons have a flat core and a shallow
envelope that shows a fast exponential decrease, see Fig.~\ref{fig_rho_x_poly}.
Then, with the optical axis centered on the lens disk, we define the normalized radius $x_0$
of the lens, in the lens plane, as
\be
x_0 = \frac{R_{\rm clump}}{R_E} ,
\ee
and the normalized impact parameter $y$ of a source at radius $r$ in the source plane, as
\be
y = \frac{d_L r}{d_s R_E} .
\ee
In particular, the outer normalized impact parameter $y_s$, for a circular source of radius $R_s$
in the source plane that is aligned with the lens and the observer, is
\be
y_s = \frac{d_L R_s}{d_s R_E} .
\ee
For such axially symmetric lenses, the lens equation is \cite{Schneider_1992,Bartelmann:2010fz}.
\be
y = x - \frac{m(x)}{x} ,
\ee
where the dimensionless lens mass within radius $x$ is
\be
m(x) = 2 \int_0^x dx' \; x' \kappa(x') ,
\ee
with $\kappa$ the lens convergence. For a constant surface density disk, we have
$\kappa = \frac{R_E^2}{R_{\rm clump}^2} = 1/x_0^2$ inside the disk, and $\kappa=0$
outside of the disk. This gives
\be
x < x_0 : \;\;\; m(x) = \frac{x^2}{x_0^2} , \;\;\; x > x_0 : \;\;\;m(x) = 1 ,
\ee
and the lens mapping becomes \cite{Schneider_1992}
\ba
&& |x| < x_0 : \;\;\; y = x \left(1 - \frac{1}{x_0^2} \right) , \nonumber \\
&& |x| > x_0 : \;\;\; y = x - \frac{1}{x} .
\label{eq:lens-mapping}
\ea
It is useful to define the quantity $y_0$ by
\be
y_0 = \left| x_0 - \frac{1}{x_0} \right| .
\ee
We show in Fig.~\ref{fig_lens-poly} the normalized radii $x_0$, $y_0$, and $y_s$ in the lens plane,
as a function of the clump mass.
The inversion of the lens mapping (\ref{eq:lens-mapping}) provides the position $x(y)$
of the image as a function of the position $y$ of the source. By axial symmetry, we can take
$y \geq 0$. If there are several solutions $x_i(y)$, the lensing of the distant source gives rise
to several images on the sky.

\begin{figure}
\begin{center}
\epsfxsize=8.8 cm \epsfysize=6. cm {\epsfbox{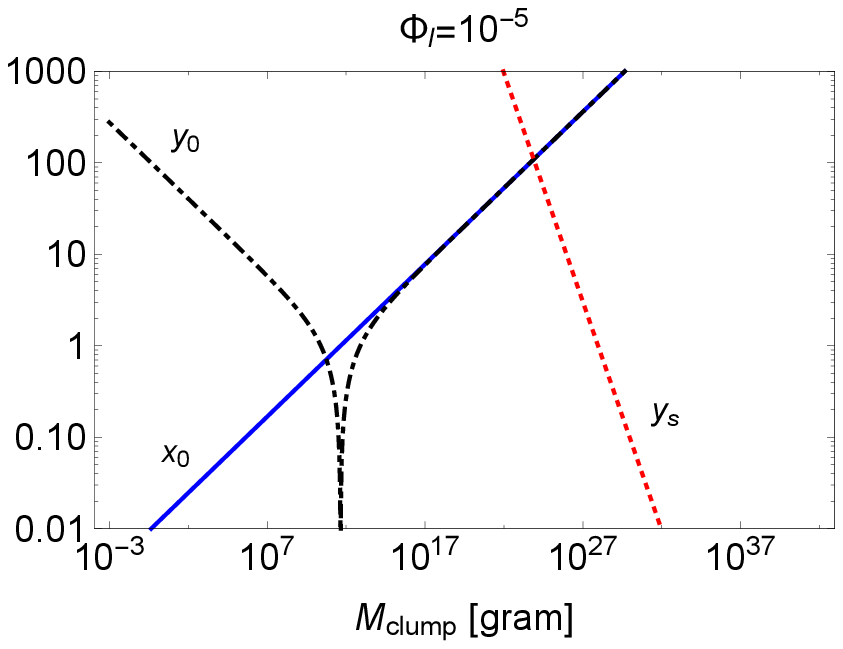}}\\
\epsfxsize=8.8 cm \epsfysize=6. cm {\epsfbox{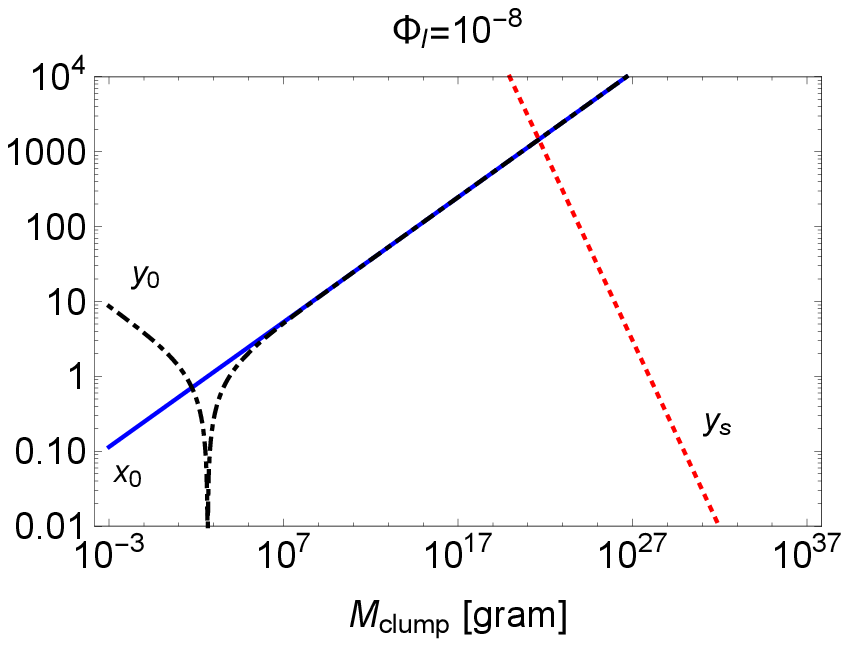}}
\end{center}
\caption{
Lensing radii normalized to the Einstein radius, in the lens plane.
We show the normalized scalar-clump radius $x_0$ (blue solid line),
the source radius $y_s$ (red dotted line) for a star of one solar radius,
and the characteristic quantity $y_0=| x_0-1/x_0|$.
We take $d_L = 1 \, {\rm kpc}$ and $d_s = d_{\rm M31} \simeq 770 \, {\rm kpc}$.}
\label{fig_lens-poly}
\end{figure}

For $x_0 < 1$, the size of the lens is small and there can be strong lensing effects
for small impact parameter. Thus, there are three images at small impact parameter
\cite{Schneider_1992},
\be
x_0 < 1 , \;\; y < y_0 : \;\;\; x_\pm = \frac{y\pm\sqrt{y^2+4}}{2} , \;\;\;
x_c = \frac{x_0^2}{x_0^2-1} y .
\label{eq:x0-1m-y-y0m}
\ee
The images $x_\pm$ are outside of the lens disk, the image $x_c$ is inside the disk.
As light can propagate through the scalar cloud, the central image $x_c$ is a true solution.
The magnifications associated with these images are
\ba
&& \mu_\pm = \pm \frac{1}{4} \left[ \frac{y}{\sqrt{y^2+4}} + \frac{\sqrt{y^2+4}}{y} \pm 2 \right] ,
\nonumber \\
&& \mu_c = \left( 1 - \frac{1}{x_0^2} \right)^{-2} ,
\label{eq:mu-pm-muc}
\ea
and the total magnification is their sum
\be
\mu = | \mu_+ | + | \mu_- | + | \mu_c | =  \frac{y^2+2}{y \sqrt{y^2+4}}
+ \left( 1 - \frac{1}{x_0^2} \right)^{-2} .
\ee
For large impact parameter, only the image $x_+$ exists
\be
x_0 < 1 , \;\; y > y_0 : \;\;\; x_+ =  \frac{y + \sqrt{y^2+4}}{2} .
\ee
For $y \to \infty$ we recover $x_+ \simeq y$, as the deflection angle decreases at large
impact parameter.
The magnification is
\be
\mu = \mu_+ = \frac{1}{4} \left[ \frac{y}{\sqrt{y^2+4}} + \frac{\sqrt{y^2+4}}{y} + 2 \right] ,
\ee
which goes to unity (no magnification) at large distance $y$.

For $x_0 > 1$, the size of the lens is large and there is always only one image.
At small impact parameter it is inside the disk,
\be
x_0 > 1 , \;\; y < y_0 : \;\;\; x_c = \frac{x_0^2}{x_0^2-1} y , \ee
while at large impact parameter it is outside of the disk,
\be
x_0 > 1 , \;\; y > y_0 : \;\;\; x_+ =  \frac{y + \sqrt{y^2+4}}{2} .
\ee
The total magnification is then either $\mu = \mu_c$ or $\mu = \mu_+$,
with these quantities already given in (\ref{eq:mu-pm-muc}).

As seen in Fig.~\ref{fig_lens-poly}, at small scalar-clump mass the size of the source is large,
$y_s \gg y_0$, which means that finite-source effects must be taken into account.
Therefore, we integrate the magnification over the surface of the source.
Approximating the source as a disk of radius $y_s$ of constant surface brightness,
the maximum total magnification is obtained when the source is centered on the optical axis,
that is, it is aligned with the lens and the observer.
This gives for the maximum total magnification
\be
\bar\mu_0 = \frac{1}{\pi y_s^2} \int_0^{y_s} d^2 {\vec y} \; \mu(y)
= \frac{2}{y_s^2} \int_0^{y_s} dy \; y \sum_i | \mu_i (y) | .
\label{eq:mu-bar0-1}
\ee
In the last expression we used the axial symmetry and we explicitly wrote the sum
over the images $i$ of the source, to include the case when there are several images.
On the other hand, the magnification $\mu$ is also obtained from the Jacobian matrix
$A_{ij} = \frac{\partial y_i}{\partial x_j}$ as $\mu = \frac{1}{\det(A)} = \frac{x}{y} \frac{dx}{dy}$,
where we used the axial symmetry in the last expression.
Therefore, the total magnification (\ref{eq:mu-bar0-1}) also reads
\be
\bar\mu_0 = \frac{2}{y_s^2} \sum_i \left| \int dx_i \, x_i \right| .
\label{eq:mu-bar0-2}
\ee
As is well known, because gravitational light deflection does not involve emission, absorption,
or frequency shift, the specific intensity and the surface brightness are not modified.
Then, the magnification is the ratio of the solid angles subtended by the image and the source
in the absence of lensing \cite{Schneider_1992}, $d^2x/d^2y$.
We recover this ratio in Eq.(\ref{eq:mu-bar0-1}), which we sum over the number of images.

From the above analysis, we have four cases associated with $x_0 \gtrless 1$ and
$y_s \gtrless y_0$. We obtain
\ba
&& x_0 < 1 , \;\; y_s < y_0 : \;\;\; \bar\mu_0 = \frac{x_{+s}^2 - x_{-s}^2 + x_{cs}^2}{y_s^2} ,
\nonumber \\
&& x_0 < 1 , \;\; y_s > y_0 : \;\;\; \bar\mu_0 = \frac{x_{+s}^2}{y_s^2} ,
\nonumber \\
&& x_0 > 1 , \;\; y_s < y_0 : \;\;\; \bar\mu_0 = \frac{x_{cs}^2}{y_s^2} ,
\nonumber \\
&& x_0 > 1 , \;\; y_s > y_0 : \;\;\; \bar\mu_0 = \frac{x_{+s}^2}{y_s^2} ,
\ea
where $x_{+s}$, $x_{-s}$, and $x_{cs}$ are the positions of the images associated with
a source at position $y_s$.
We can see in Fig.~\ref{fig_lens-poly} that for low clump mass, where $x_0 < 1$, we have
$y_s \gg y_0$. Thus, we obtain
\ba
x_0(M_{\rm clump}) < 1 : \;\;\; \bar\mu_0 & = & \frac{x_{+s}^2}{y_s^2}
= \left( \frac{1+\sqrt{1+4/y_s^2}}{2} \right)^2 \nonumber \\
& \simeq & 1 + \frac{2}{y_s^2} \simeq 1 ,
\label{eq:x0-m1-ys-large}
\ea
which is very close to unity as $y_s \gg 1$.
For intermediate clump mass, where $x_0>1$ and $y_s > y_0$, we obtain again
\ba
&& x_0(M_{\rm clump}) > 1 \;\; \mbox{and} \;\; y_s > y_0  : \nonumber \\
&& \bar\mu_0 = \frac{x_{+s}^2}{y_s^2} \simeq 1 + \frac{2}{y_s^2} \simeq 1 ,
\label{eq:x0-p1-ys-large}
\ea
which is again very close to unity as we still have $y_s \gg 1$.
Finally, for large clump mass, where $x_0>1$ and $y_s < y_0$, we obtain
\ba
&& x_0(M_{\rm clump}) > 1 \;\; \mbox{and} \;\; y_s < y_0  : \nonumber \\
&& \bar\mu_0 = \frac{x_{cs}^2}{y_s^2} = \left( \frac{x_0^2}{x_0^2-1} \right)^2
\simeq 1 + \frac{2}{x_0^2} \simeq 1 ,
\label{eq:x0-p1-ys-small}
\ea
which is very close to unity as $x_0 \gg 1$.
Numerically, we find that $\bar\mu_0 -1 < 10^{-3}$ over all clump masses.
This is much below the observational threshold $\mu_T=1.34$ \cite{Niikura:2017zjd}.
Therefore, microlensing observations do not constrain the models studied in this paper.
At low clump masses, this is because the finite-source effects decrease the lensing magnification.
The same effect prevents the detection of small primordial black holes.
At large masses, the microlensing inefficiency is due to the finite-lens effect, because the
scalar-clump radius is much greater than the Einstein radius, $x_0 \gg 1$.
This is different from primordial black hole scenarios, where large masses can be constrained
by microlensing because the Schwarzschild radius is much smaller than the Einstein radius,
as $R_E = \sqrt{ 2 R_{\rm Sch} d_L (1-d_L/d_s)} \gg R_{\rm Sch}$.
Instead, our massive scalar clumps have a very large radius and are similar to galactic
molecular clouds, rather than compact objects. This leads to small gravitational potential
wells, hence to very small deflection angles and lensing magnifications.
As noticed in Sec.~\ref{sec:aggregation}, in a more realistic computation the clumps
are expected to have a finite range of masses and radii below the peak values
(\ref{eq:R-ballistic}). However, the very small magnification $\bar\mu_0 -1 < 10^{-3}$
ensures that our result should not change once we take into account the finite
width of the clump mass function.
}

\section{Axion Monodromy}
\label{sec:AxionM}

\begin{figure*}
\begin{center}
\epsfxsize=4.4 cm \epsfysize=6.5 cm {\epsfbox{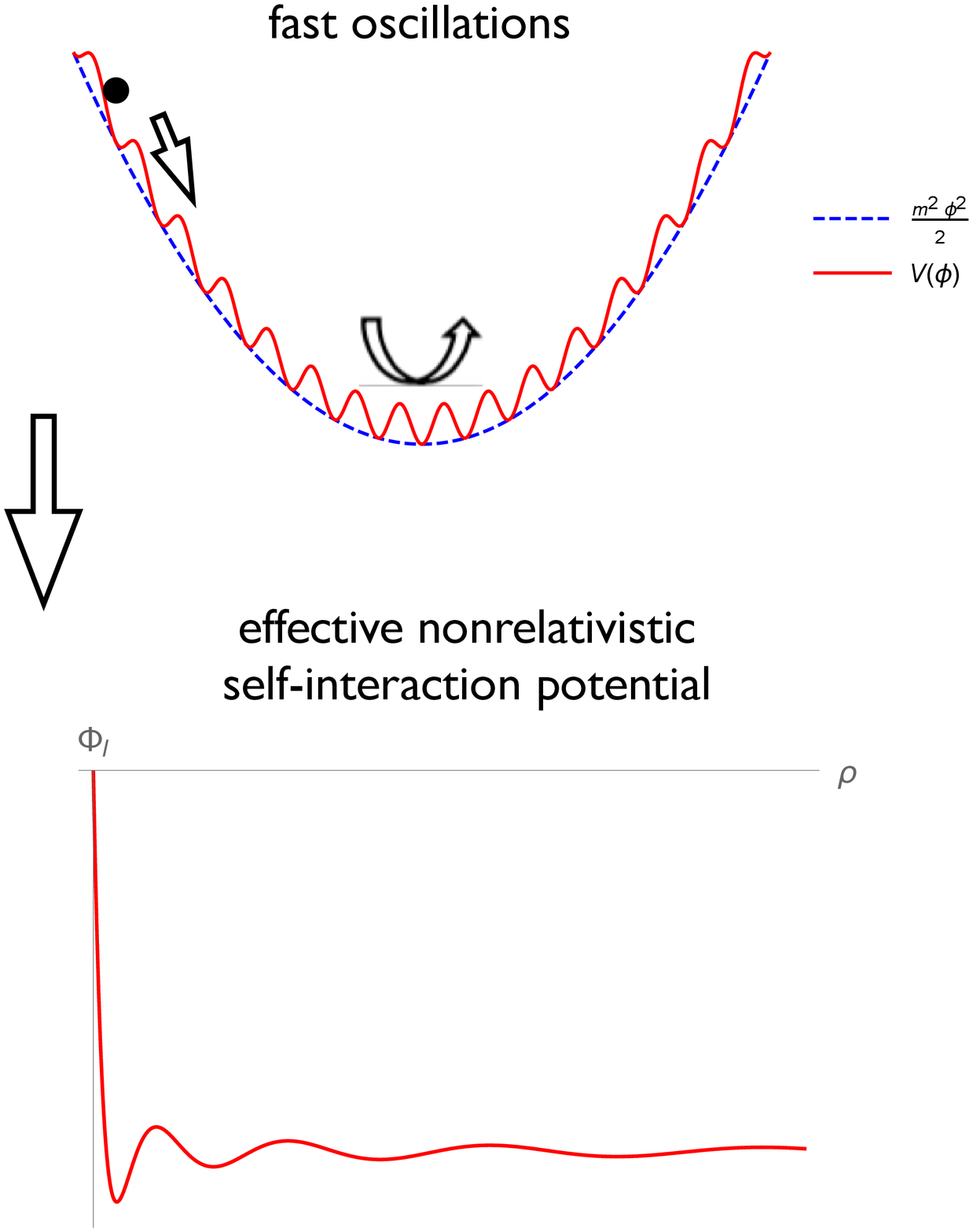}}
\epsfxsize=4.4 cm \epsfysize=7.3 cm {\epsfbox{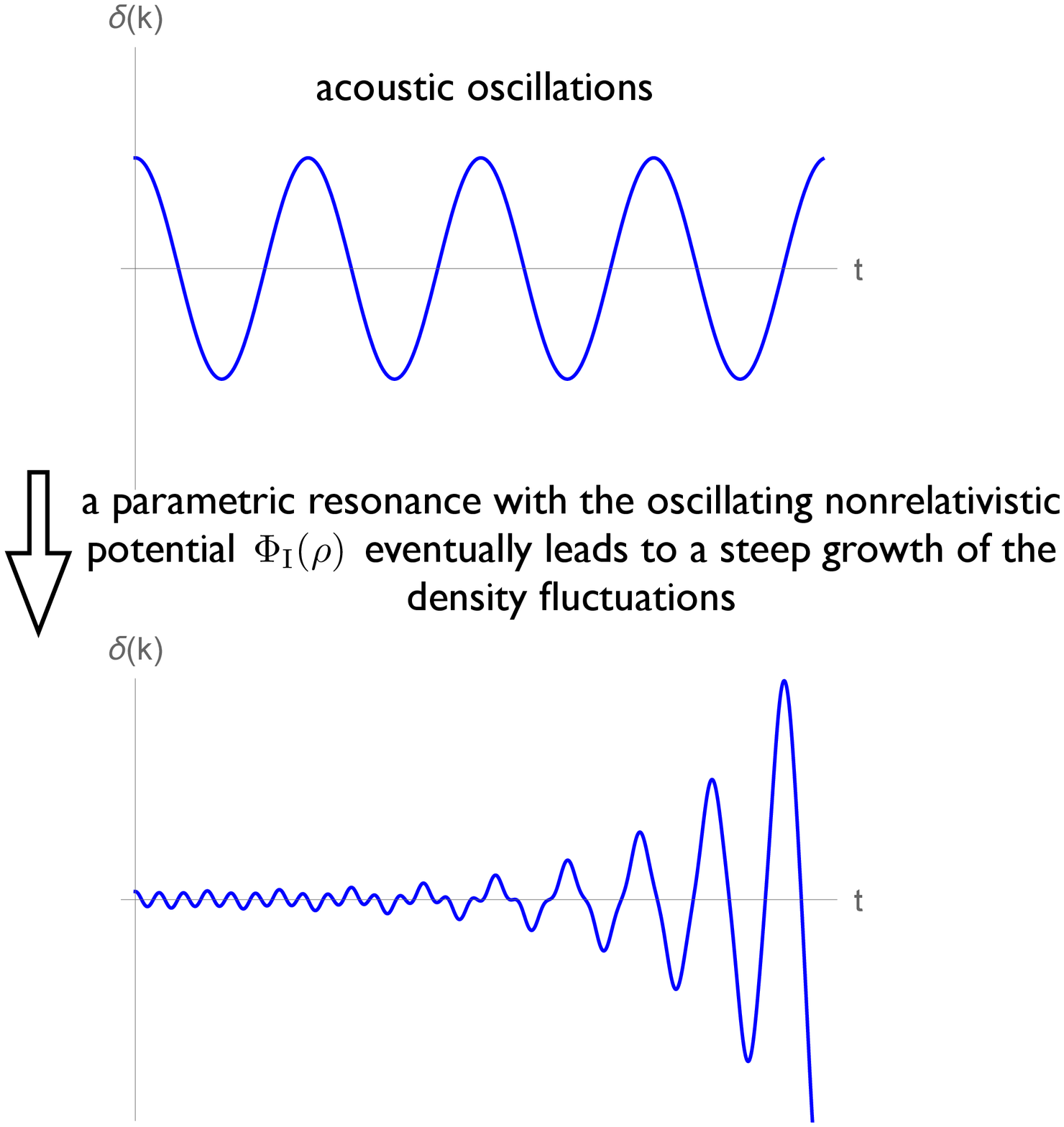}}
\epsfxsize=4.4 cm \epsfysize=6 cm {\epsfbox{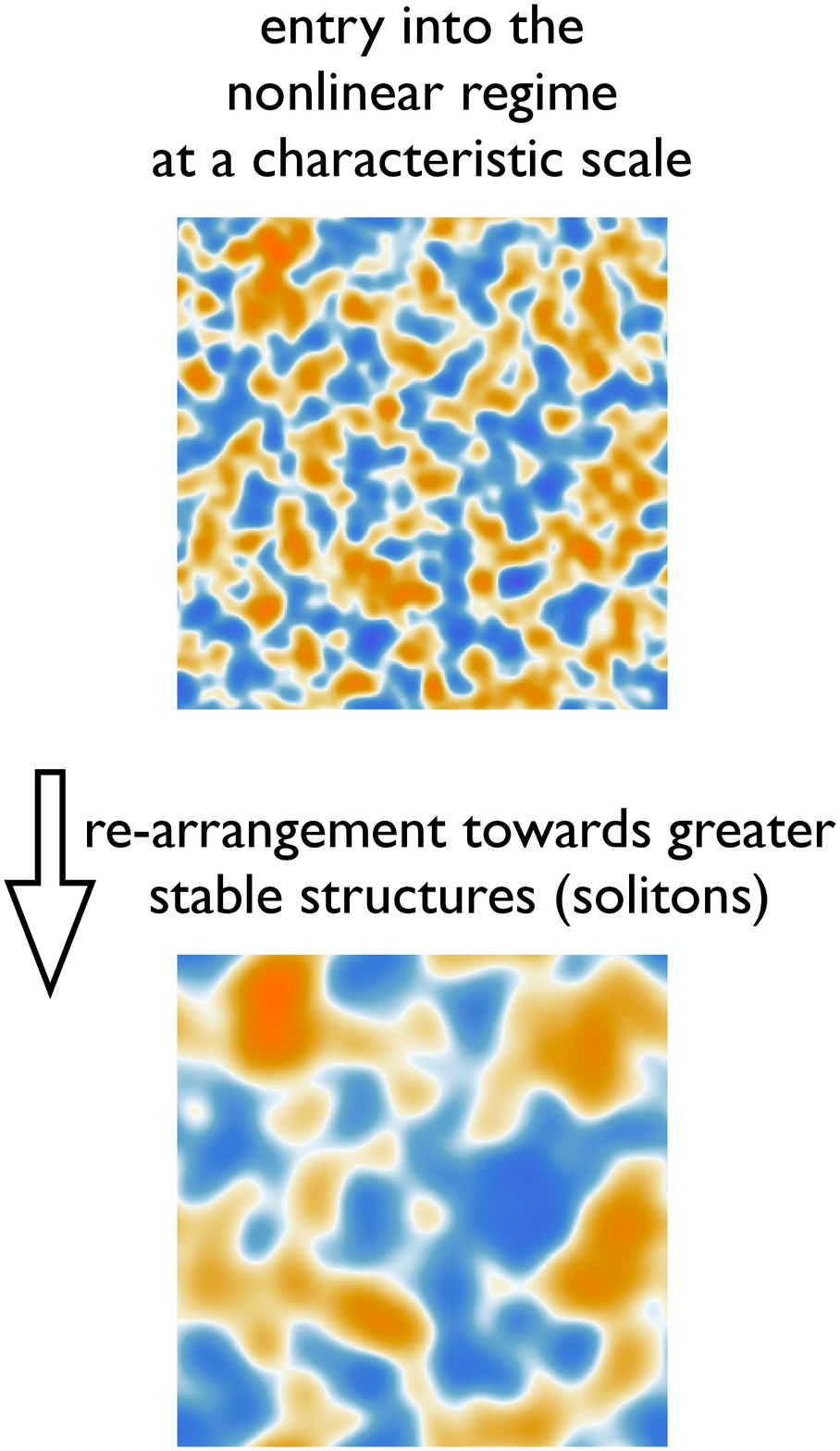}}
\epsfxsize=4.4 cm \epsfysize=5.7 cm {\epsfbox{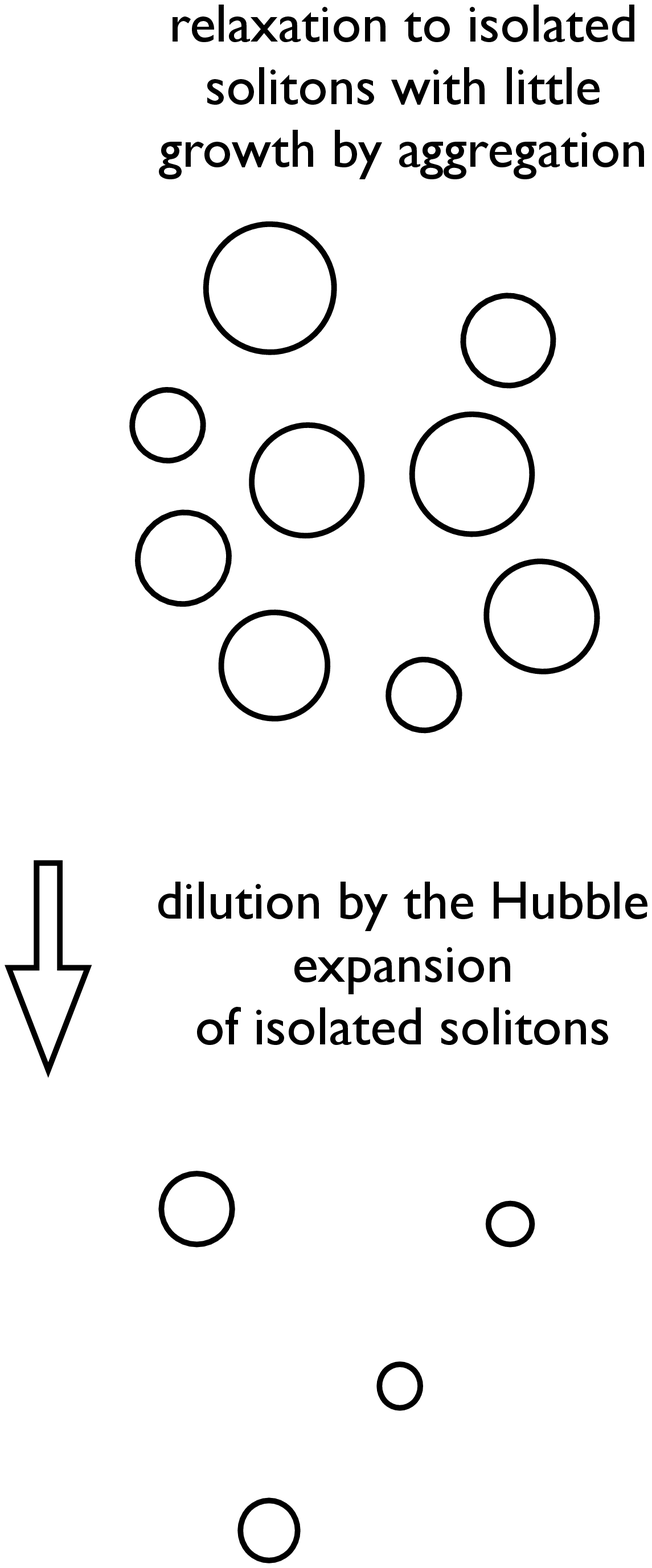}}
\end{center}
\caption{
The main stages of the formation of scalar dark-matter clumps for the
parametric-resonance scenario (\ref{eq:V-phi-cosine}).}
\label{fig_Plots-cos}
\end{figure*}

In this section we present another mechanism for the formation of clumps. In this case parametric resonance plays the main role. We consider this effect in the context of axion monodromy potentials as it is illustrated schematically in Fig.~\ref{fig_Plots-cos}.

\subsection{Cosine potential}

Axions have been a long standing candidate for dark matter. In the case of the QCD axion, which arises after the breaking of the Peccei-Quinn symmetry, the potential term
arises non-perturbatively and is periodic. This is a generic feature of axions or axion-like-particles, where the axion field can be seen as a Goldstone mode of a globally broken symmetry. The potential terms for these scalar (or pseudo-scalar) fields possesses the remaining discrete symmetry $\phi\to \phi+f$ which is associated to the original $U(1)$ symmetry and arises either from non-perturbative effects or from soft breaking terms prior to the symmetry breaking phenomenon. Generically, this gives rise to cosine potentials. In the misalignement mechanism, the axion field is constant until the Hubble rate drops below the axion mass. Then oscillations start and the axion becomes a good dark matter candidate. In certain string scenarios, where the axions come from the integrated forms of string theory along closed cycles of the compactification manifold, the coupling of these fields to internal fluxes can give rise to additional polynomial interactions in the axion field \cite{McAllister:2014mpa}. This has been called axion monodromy as the potential does not remain periodic due to these fluxes but is shifted when $\phi\to \phi+f$. In the following we will focus on the potential \cite{Berges:2019dgr}
\be
V(\phi) = \frac{m_0^2}{2} \phi^2 + M_I^4 \left[ 1- \cos(\phi/f) \right] ,
\;\;\; \frac{M_{\rm I}^4}{f^2} \ll m_0^ 2 .
\label{eq:V-phi-cosine}
\ee
We can absorb the quadratic part of the cosine into the mass term
and write $V(\phi) = \frac{m^2}{2} \phi^2 + V_{\rm I}(\phi)$, with
\ba
&& m^2 = m_0^2 + \frac{M_{\rm I}^4}{f^2} \simeq m_0^2 ,
\label{eq:m-m0-correction}  \\
&& V_{\rm I}(\phi) = M_{\rm I}^4 \left[ 1 - \cos(\phi/f) - \frac{\phi^2}{2 f^2} \right]  .
\ea
For $\phi \ll f$ we recover an attractive quartic potential, with
$\lambda_4 = - M_{\rm I}^4/(6 f^4) < 0$.
In the following, we consider that the mass term dominates over the cosine interaction. Notice that this is different from \cite{Berges:2019dgr}, where the two terms have the same order of magnitude.
In the nonrelativistic regime, where we average over the fast oscillations of the scalar field,
the effective interaction potential becomes \cite{Brax:2019fzb}
\be
\Phi_{\rm I}(\rho) = \frac{8\rho_b}{\rho_a} \left[ \frac{2J_1(\sqrt{\rho/\rho_b})}{\sqrt{\rho/\rho_b}}
- 1\right] ,
\label{eq:Phi-I-J1}
\ee
with
\be
\rho_a = \frac{8 m^4 f^4}{M_{\rm I}^4} , \;\;\; \rho_b = \frac{m^2 f^2}{2} ,
\;\;\; \rho_b \ll \rho_a .
\label{eq:rhoa-rhob}
\ee
This corresponds to the integrated potential
\be
{\cal V}_{\rm I}(\rho) = \frac{8\rho_b}{\rho_a} \left[ - \rho + 4 \rho_b - 4 \rho_b
J_0(\sqrt{\rho/\rho_b}) \right]  ,
\label{eq:V-I-J0}
\ee
obtained by averaging the potential $V_{\rm I}(\phi)$ over the period of the fast
leading-order oscillations.
This gives for the squared-sound speed associated with the quantum pressure and the
self-interactions
\be
c_s^2 = \frac{k^2}{4 a^2 m^2}  - \frac{8 \rho_b}{\rho_a} J_2( \sqrt{\bar\rho/\rho_b}) .
\label{eq:cs2-J2}
\ee
Thus, at large background densities the self-interaction contribution to the squared
sound speed oscillates around zero, with increasing amplitude as time goes on.
From the asymptotic behavior of the second-order Bessel function, we obtain at large
background densities for this contribution
\be
\bar\rho \gg \rho_b : \;\;\; \left. c_s^2 \right |_{\rm I}
\simeq \frac{8 \rho_b}{\rho_a} \sqrt{\frac{2}{\pi}} \left( \frac{\bar\rho}{\rho_b} \right)^{-1/4}
\cos\left( \sqrt{\frac{\bar\rho}{\rho_b}} - \frac{\pi}{4} \right) .
\label{eq:cs2-I-cosine-cos}
\ee
As the speed of sound squared becomes negative quasi-periodically, there will be instabilities that we will spell out in the following section.

As for the tachyonic case, to facilitate the reading of the next sections, we already present
in Fig.~\ref{fig_Plots-cos} the main stages of the formation of the scalar dark-matter clumps.

1) The scalar field $\phi$ again quickly oscillates in the potential $V(\phi)$, dominated
by its quadratic component with a small correction $V_{\rm I}$.
This self-interaction contribution now shows fast oscillations, such as the cosine
in Eq.(\ref{eq:V-phi-cosine}). Integrating out the fast leading-order oscillations
of $\phi$, the nonrelativistic self-interaction potential $\Phi_{\rm I}(\rho)$ defined by
Eq.(\ref{eq:Phi-I-V-I-def}), i.e. Eq.(\ref{eq:Phi-I-J1}) in our example,
now shows oscillations with a decaying amplitude at large densities.

2) At early times, the scalar-field density perturbations again oscillate as acoustic waves.
As the background density $\bar\rho$ decreases with time,
the amplitude of the self-interactions grows and they finally become relevant.
As in the scenario presented in the first part of this paper and illustrated in
Fig.~\ref{fig_Plots-poly}, a tachyonic instability would develop at late times,
associated with the first region connected to the origin where $\frac{d\Phi_{\rm I}}{d\rho} < 0$
(the first significant drop of $\Phi_{\rm I}$ seen in the lower left panel in Fig.~\ref{fig_Plots-cos}).
However, at much earlier times, still in the region where $\Phi_{\rm I}(\rho)$
shows many oscillations, a parametric resonance triggered by these oscillatory features
develops and amplifies the scalar density perturbations.

3) The scalar density field then quickly reaches the nonlinear regime.
Because of the intricate properties and time-dependent nature of the parametric resonance,
the length scales and densities that first become nonlinear do not correspond to those
associated with stable isolated structures.
This suggests that the system will undergo a significant redistribution towards greater structures
that can form stable solitons.

4) The relative velocities are now rather modest and we do not expect significant
collisional aggregation. Finally, the expansion of the Universe again dilutes the scalar clumps,
which then behave as isolated CDM particles.
At much lower redshifts, gravitational instability will again build the cosmic web and galaxies
as in the standard $\Lambda$CDM scenario.

We describe in the following sections these various stages in more detail.

\subsection{Dynamics of the scalar density field}
\label{sec:linear-dynamics-cosine}

\subsubsection{Acoustic oscillations of the density contrast}

From Eqs.(\ref{eq:cs2-J2}) and (\ref{eq:cs2-I-cosine-cos}),
the squared sound-speed $c_s^2$ becomes negative on subhorizon scales
for the first time at the redshift $z_{c_s}$, when $\bar\rho = \rho_{c_s}$ with
\be
\frac{H_{c_s}^2}{4 m^2} = \frac{8 \rho_b}{\rho_a} \sqrt{\frac{2}{\pi}}
\left( \frac{\rho_{c_s}}{\rho_b} \right)^{-1/4}  ,
\label{eq:Hi-def}
\ee
where we assumed that we are in the large-density regime (\ref{eq:cs2-I-cosine-cos}),
\be
\rho_{c_s} \gg \rho_b \; .
\ee
This gives the useful relationship
\be
\frac{\rho_{c_s}}{\rho_b} \sim \left( \frac{\rho_b}{\rho_a} \right)^4 \left( \frac{m}{H_{c_s}} \right)^8 .
\label{eq:rho-cs-rhob}
\ee
Since we have $\rho_b/\rho_a \ll 1$, Eq.(\ref{eq:Hi-def}) also implies
\be
z \leq z_{c_s} : \;\;\; \frac{H}{m} \ll 1.
\label{eq:end-slow-roll-cos}
\ee
Thus, the slow-roll stage of the evolution of the scalar field $\phi$, when it was governed
by the Hubble friction, finished long before $z_{c_s}$ and the scalar field shows fast oscillations
in its mainly quadratic potential $m^2\phi^2/2$. This justifies the effective description in
terms of the hydrodynamical variables $\{\rho,{\vec v}\}$ and of the self-interaction
potential $\Phi_{\rm I}(\rho)$, as illustrated in the first column in
Fig.~\ref{fig_Plots-cos}.
Equation (\ref{eq:Hi-def}) also reads
\be
H_{c_s} \sim 10^{-2} \left( \frac{m}{1 \, {\rm GeV}} \right)^{16/19}
\left( \frac{\rho_b}{\rho_a} \right)^{8/19} \left( \frac{\rho_b}{1 \, {\rm GeV}^4}\right)^{2/19} \,
{\rm GeV} .
\ee
Using the approximation (\ref{eq:cs2-I-cosine-cos}), the evolution equation
(\ref{eq:delta-cosmo}) of the linear density contrast reads
\ba
&& \ddot\delta + \frac{1}{t} \dot\delta + \frac{H_{c_s}^4}{4 m^2} \frac{k^2}{k_{c_s}^2}
\Biggl \lbrace \frac{k^2}{k_{c_s}^2} \left( \frac{t}{t_{c_s}} \right)^{-2} +
\left( \frac{t}{t_{c_s}} \right)^{-5/8} \nonumber \\
&& \times \cos\left[ \sqrt{\frac{\rho_{c_s}}{\rho_b}}
\left( \frac{t}{t_{c_s}} \right)^{-3/4} \right] \Biggl \rbrace \delta = 0 ,
\label{eq:delta-t-cos}
\ea
where we introduced $k_{c_s} = a_{c_s} H_{c_s}$.
Here we used $H=1/(2t)$ in the radiation era and in the cosine term we discarded the
constant phase $-\pi/4$, which can be absorbed in a small change of $t_{c_s}$ or of the
origin of time.
Making the change of time coordinate
\be
\eta = - \ln\left[ \frac{1}{2} \sqrt{\frac{\rho_{c_s}}{\rho_b}}
\left( \frac{t}{t_{c_s}} \right)^{-3/4} \right] \ll - 1 ,
\ee
we obtain
\ba
&& \frac{d^2\delta}{d\eta^2} + \frac{H_{c_s}^2}{9 m^2} \frac{k^2}{k_{c_s}^2}
\left[ \frac{k^2}{k_{c_s}^2} + e^{11(\eta-\eta_{c_s})/6}
\cos\left( 2 e^{-\eta} \right) \right] \delta = 0 . \nonumber \\
&&
\label{eq:delta-eta-cos}
\ea
The time coordinate $\eta$ grows with cosmic time but it is restricted to large negative values,
as the asymptotic form (\ref{eq:cs2-I-cosine-cos}) of the Bessel function only applies
when the argument of the cosine is large.
As shown in Appendix~\ref{sec:Mathieu-1}, at early times the density contrast shows
acoustic oscillations with a constant amplitude, driven by the quantum pressure term.
For moderate wave numbers, we obtain
\ba
&&  \frac{k^2}{k_{c_s}^2} \ll \frac{m}{H_{c_s}} e^{-\eta} , \;\;\;
\frac{H_{c_s} k}{m k_{c_s}} e^{\eta+11(\eta-\eta_{c_s})/12} \ll  1 :\nonumber \\
&& \delta(\eta) \simeq \delta_i \cos\left( \frac{H_{c_s} k^2}{3 m k_{c_s}^2} (\eta-\eta_i) \right)
\Biggl [ 1 + \frac{H_{c_s}^2 k^2}{36 m^2 k_{c_s}^2} \nonumber \\
&& \times \, e^{2\eta+11(\eta-\eta_{c_s})/6} \cos\left( 2 e^{-\eta} \right) \Biggl ] ,
\label{eq:delta-delta0-Hcs-1}
\ea
whereas for large wave numbers we have
\ba
&&  \frac{k^2}{k_{c_s}^2} \gg \frac{m}{H_{c_s}} e^{-\eta} , \;\;\;
\frac{H_{c_s}}{m} e^{\eta+11(\eta-\eta_{c_s})/6} \ll  1 :\nonumber \\
&& \delta(\eta) \simeq \delta_i \cos\left( \frac{H_{c_s} k^2}{3 m k_{c_s}^2} (\eta-\eta_i) \right)
+  \delta_i \sin\left( \frac{H_{c_s} k^2}{3 m k_{c_s}^2} (\eta-\eta_i) \right) \nonumber \\
&& \times \sin\left( 2 e^{-\eta} \right) \frac{H_{c_s}}{12 m} e^{\eta+11(\eta-\eta_{c_s})/6} .
\label{eq:delta-delta0-Hcs-2}
\ea
Because $H_{c_s} \ll m$ and $e^{\eta} \ll 1$, the density contrast only
starts growing beyond its initial value $\delta_i \sim 10^{-5}$ long after the redshift $z_{c_s}$,
at the time $t_g(k)$ with
\begin{eqnarray}
k< k_g :& & \; t_g(k) = t_{g\infty} \left( \frac{k}{k_g} \right)^{-16/23} , \;\;
\nonumber\\
k> k_g :& & \; t_g(k) = t_{g\infty}\, ,
\label{eq:tg-k}
\end{eqnarray}
where we define
\be
k_g = k_{c_s}  \left( \frac{\rho_{c_s}}{\rho_b} \right)^{11/68}
\left( \frac{m}{H_{c_s}} \right)^{11/34} \gg k_{c_s} ,
\ee
and
\be
t_{g\infty} = t_{c_s} \left( \frac{\rho_{c_s}}{\rho_b} \right)^{4/17}
\left( \frac{m}{H_{c_s}} \right)^{8/17}  \gg t_{c_s} .
\label{eq:tg-infty-def}
\ee
Thus, the time $t_g(k)$ decreases at higher wave numbers, up to $k_g$.
At greater wave numbers, $t_g(k)=t_{g\infty}$ is constant and fluctuations on these very small
scales start growing simultaneously at $t_{g\infty}$.
At that time, the argument of the cosine is of the order of
\ba
| \tau_{g\infty} | & \sim & \sqrt{\frac{\rho_{c_s}}{\rho_b}}  \left( \frac{t_{g\infty}}{t_{c_s}} \right)^{-3/4}
\sim \left( \frac{\rho_{c_s}}{\rho_b} \right)^{11/34}  \left( \frac{m}{H_{c_s}} \right)^{-6/17}
\nonumber \\
& \sim & \left( \frac{\rho_b}{\rho_a} \right)^{22/17} \left( \frac{m}{H_{c_s}} \right)^{38/17}  ,
\label{eq:tau-g}
\ea
where in the last expression we used the relation (\ref{eq:rho-cs-rhob}).
This is still a large value if $\rho_b/\rho_a$ is not too small. Then, the squared-sound speed
(\ref{eq:cs2-I-cosine-cos}) can still show many oscillations as the background density decreases.
If $|\tau_{g\infty}| \lesssim 1$, we have $\rho/\rho_b \lesssim 1$ and we are in the low-density
regime of the self-interactions, where we approximate the Bessel functions by their low-order
Taylor expansion. Then, we recover the polynomial case
(\ref{eq:Phi-I-polynomial})-(\ref{eq:V-I-polynomial}) with a tachyonic instability, associated
with the negative value $\lambda_4<0$ of the quartic term of the potential $V_{\rm I}(\phi)$
for $\phi \ll f$.
Thus, we are back to the physics analyzed in the previous
sections~\ref{sec:perturbations}-\ref{sec:constraints}.
Therefore, in the following we consider in more details the case $| \tau_{g\infty} | \gg 1$.
We will see that in this scenario a parametric resonance takes place before the
tachyonic instability can set in,
\be
\mbox{parametric resonance for } \;\; | \tau_{g\infty} | \gg 1 .
\label{eq:tau-large}
\ee

\subsubsection{Mathieu-equation resonances}

Changing time coordinate to
\be
\tau = - e^{-\eta} = - \frac{1}{2} \sqrt{\frac{\rho_{c_s}}{\rho_b}}
\left( \frac{t}{t_{c_s}} \right)^{-3/4}  \ll -1 ,
\label{eq:tau-def-t}
\ee
and writing $\delta(\tau)$ as
\be
\delta(\tau) = (-\tau)^{-1/2} y(\tau) ,
\ee
the evolution equation (\ref{eq:delta-eta-cos}) becomes
\be
\frac{d^2 y}{d\tau^2} + [ A(\tau) - 2 q(\tau) \cos(2\tau) ] y = 0 ,
\label{eq:Mathieu}
\ee
where we shifted the argument of the cosine by a phase $\pi$ (corresponding to
a negligible shift of $\tau$) to recover the standard sign of the Mathieu equation, and
\ba
A(\tau) & = & \left( \frac{1}{4} + \frac{H_{c_s}^2 k^4}{9 m^2 k_{c_s}^4} \right) \frac{1}{\tau^2}
\nonumber \\
& = & \frac{1}{4 \tau^2} + \frac{4}{9} \left( \frac{\tau}{\tau_{g\infty}} \right)^{-2}
\left( \frac{k}{k_g} \right)^4  ,
\label{eq:A-k-tau-def}
\ea
\ba
&& q(\tau) =  \frac{H_{c_s}^2 k^2}{18 m^2 k_{c_s}^2} \frac{(-\tau_{c_s})^{11/6}}{(-\tau)^{23/6}}
= \frac{2}{9} \left( \frac{\tau}{\tau_{g\infty}} \right)^{-23/6} \left( \frac{k}{k_g} \right)^2 .
\nonumber \\
&&
\label{eq:q-k-tau-def}
\ea
Here $\tau_{g\infty}$ is the value of $\tau$ at the time $t_{g\infty}$ introduced in
(\ref{eq:tg-infty-def}), and it is of the order of (\ref{eq:tau-g}).
For wave numbers smaller than $k_g$, we can also write in terms of $\tau_g(k)$,
associated with the time $t_g(k)$ of (\ref{eq:tg-k}),
\ba
k < k_g : && A(k,\tau) = \frac{1}{4 \tau^2}
+ \frac{4}{9} \left( \frac{\tau}{\tau_g(k)} \right)^{-2} \left( \frac{k}{k_g} \right)^{68/23} , \nonumber \\
&& q(k,\tau) = \frac{2}{9} \left( \frac{\tau}{\tau_g(k)} \right)^{-23/6} .
\label{eq:A-q-low-k}
\ea
Equation (\ref{eq:Mathieu}) has the form of a Mathieu equation with slowly-varying coefficients.
The coefficients $A(t)$ and $q(t)$ grow with cosmic time as $|\tau|$ decreases.
For constant coefficients $A$ and $q$, Floquet theory shows that the Mathieu equation
has solutions of the form $e^{\pm i\nu\tau} P(\pm\tau)$, where $P(\tau)$ is periodic
of period $\pi$ and $\nu$ is the characteristic Mathieu exponent
\cite{Abramowitz,McLachlan}.
When $\nu$ has a nonzero imaginary part, $\mu=|{\rm Im}(\nu)| \neq 0$,
there is a growing and a decaying solution, $y_\pm \sim e^{\pm \mu \tau}$,
up to an oscillating prefactor.
This gives instability bands in the plane $(q,A)$ of the parameters, see \cite{McLachlan}.
These instability bands touch the $A$-axis, at $q=0$, at the discrete values $A_n=n^2$, where
$n=1,2,3, \dots$. Their width $\Delta A$ grows with $q$ for $q>0$.

\begin{figure*}
\begin{center}
\epsfxsize=8.8 cm \epsfysize=6 cm {\epsfbox{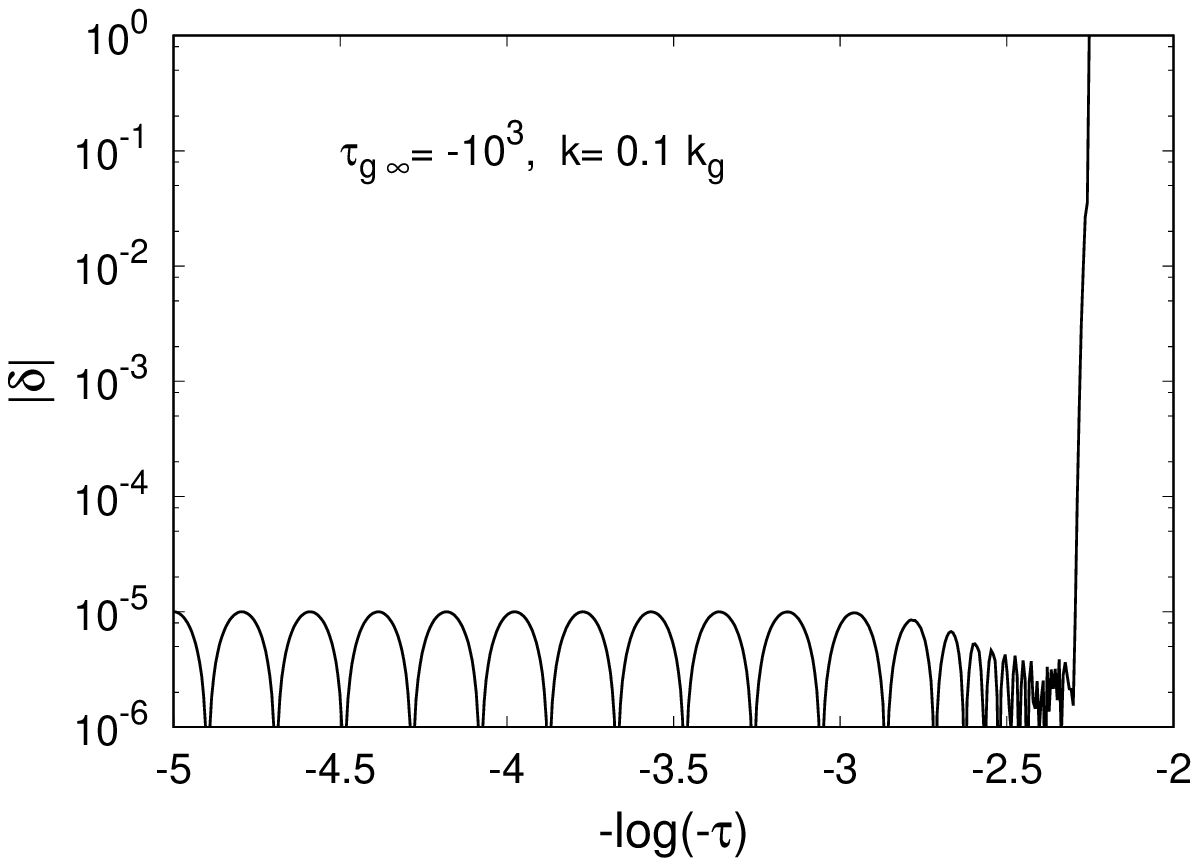}}
\epsfxsize=8.8 cm \epsfysize=6 cm {\epsfbox{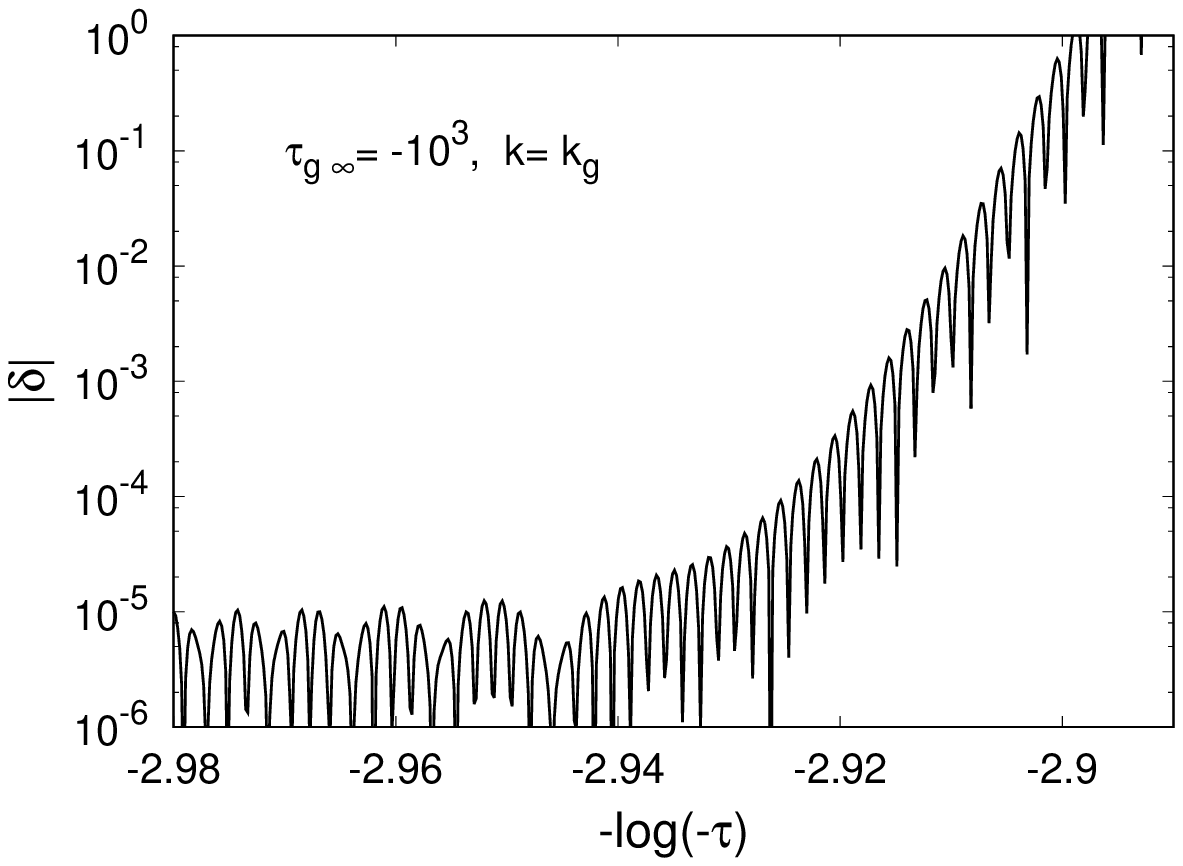}}\\
\epsfxsize=8.8 cm \epsfysize=6 cm {\epsfbox{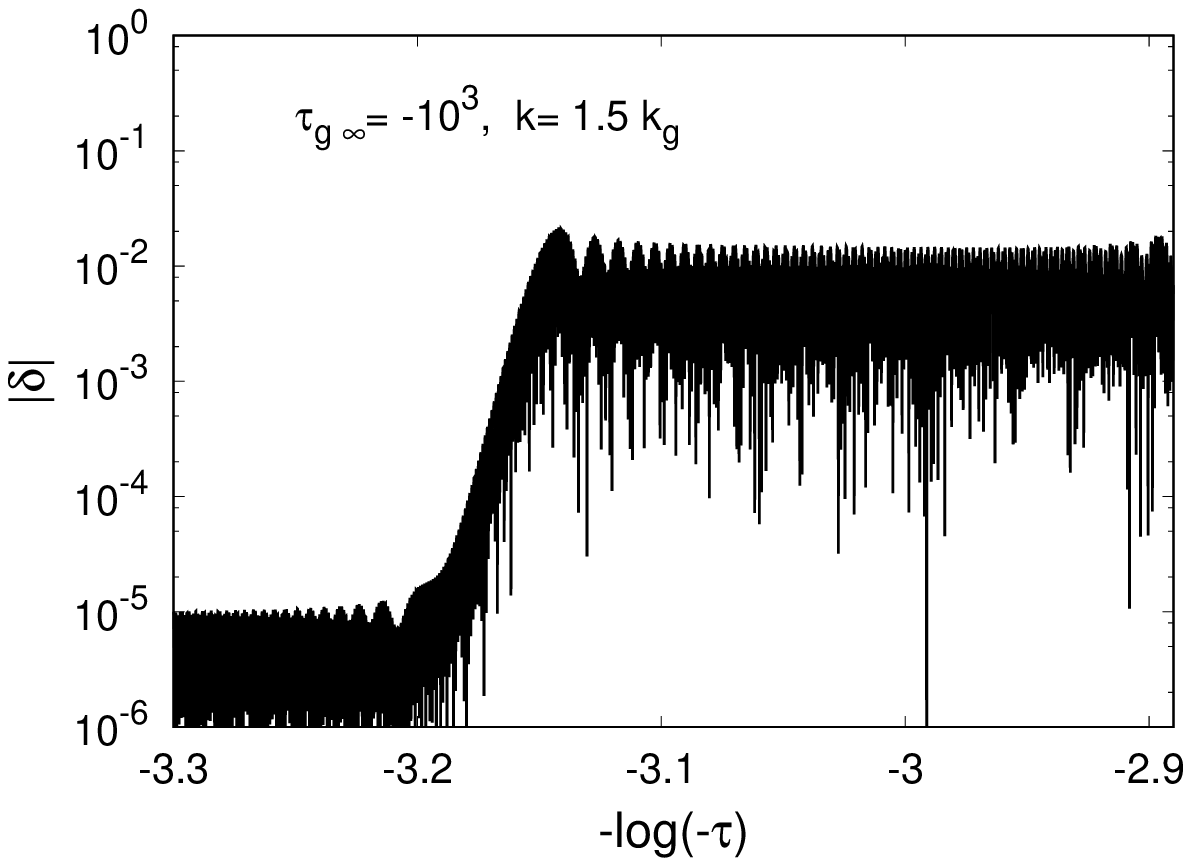}}
\epsfxsize=8.8 cm \epsfysize=6 cm {\epsfbox{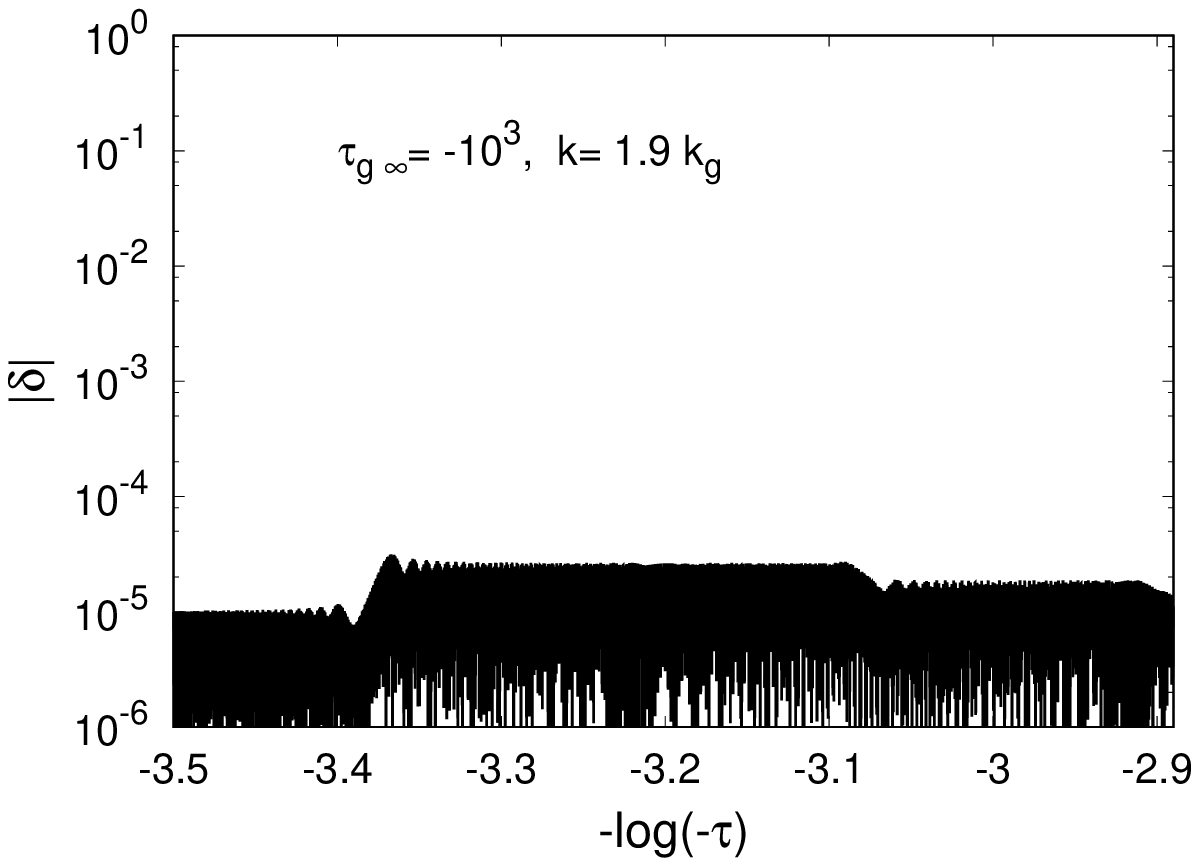}}
\end{center}
\caption{Evolution of the modulus of the density contrast $\delta({\vec k},\tau)$,
for different wave numbers $k$, from Eq.(\ref{eq:Mathieu}).
For illustration, the characteristic time $\tau_{g\infty}$ is set to $-10^3$ and the initial condition to
$|\delta_i|=10^{-5}$.
For $k=1.5 k_g$ and $k=1.9 k_g$ (lower panels), the oscillations of $\delta(\tau)$ are so fast
that they cannot be distinguished in the figure and the curve fills all the area under its upper
envelope.}
\label{fig_Mathieu}
\end{figure*}

\paragraph{Intermediate wave numbers}

Let us first consider wave numbers of the order of $k_g$. For $t \lesssim t_{g\infty}$ we have
$q\lesssim 1$ and $A\lesssim 1$. Therefore, we are in the first stability region of the
Mathieu-equation stability chart, see \cite{McLachlan}.
This agrees with the perturbative analysis of the Appendix~\ref{sec:Mathieu-1}
and the results (\ref{eq:delta-delta0-Hcs-1})-(\ref{eq:delta-delta0-Hcs-2}).
After that time, $A(t)$ and $q(t)$ grow beyond unity along a line
$A \sim q^{12/23}$ and enter a first unstable region at $q \simeq 0.4$.
Then, $y(\tau)$ grows exponentially as $e^{\mu\tau}$,
where $\mu=|{\rm Im}(\nu)|$ is the imaginary part of the characteristic Mathieu exponent
\cite{McLachlan,Abramowitz}.
Indeed, $A(\tau)$ and $q(\tau)$ evolve on a time scale given by $|\tau|$ whereas
the $\cos(2\tau)$ term oscillates on the much shorter time scale $\pi \ll |\tau|$.
Then, it takes a time $\Delta\tau$ for the density contrast $\delta$ to grow from
$\delta_i \sim 10^{-5}$ to unity, with
\be
\Delta \tau \sim \frac{5 \ln 10}{\mu} .
\ee
This is much shorter than $|\tau|$ as soon as $\mu$ is of the order of unity. Therefore,
the density contrast reaches the nonlinear regime soon after the time $t_{g\infty}$,
on comoving scales $x_g \sim 1/k_g$.

This case is shown by the upper right panel in Fig.~\ref{fig_Mathieu}. It clearly
shows the oscillations with almost constant amplitude until the time $t_{g\infty}$
and the exponential rise shortly after $t_{g\infty}$. The density contrast quickly grows
by a factor of $10^5$.

\paragraph{Low wave numbers}

Let us now consider low wave numbers, $k \ll k_g$.
From Eq.(\ref{eq:A-q-low-k}), we can see that they remain
in the stability region $\{ q \lesssim 1, A \lesssim 1\}$ until the time $t_g(k)>t_{g\infty}$,
in agreement with the perturbative result (\ref{eq:delta-delta0-Hcs-1}).
They enter the first unstable region with $q \simeq 0.9$ and $A \simeq 0$, along
the $q$-axis. Since dark-matter clumps have already formed at the latest at the time
$t_{g\infty}$, on the scale $x_g$, this is no longer relevant.

This case is shown by the upper left panel in Fig.~\ref{fig_Mathieu}.
It clearly shows the oscillations with constant amplitude until the time $t_{g}(k)>t_{g\infty}$
and the exponential rise at $t_{g}(k)$.
In agreement with the analysis above, the oscillation frequency is lower than for
the case $k=k_g$ and the instability appears later.

\paragraph{High wave numbers}

Higher wave numbers have larger values of $q$ and $A$, along a line $A \sim q^2$.
This mostly goes through the stable regions of the Mathieu equation,
except for very narrow instability bands that are missed in the first-order perturbative result
(\ref{eq:delta-delta0-Hcs-2}), apart from the first one.
Indeed, the first signs of these higher instability bands appear as secular terms in higher
orders of perturbation theory \cite{Bender-Orszag}.
Nevertheless, we can check that the instability rate does not diverge at high wave numbers.

First, we can obtain a conservative lower bound on the time $t_g^{\rm min}$ when the
density contrast becomes of order unity.
Indeed, Eq.(\ref{eq:bound-delta}) is valid at all orders (\ref{eq:delta-n-upper})
of the perturbative expansion (\ref{eq:delta-eta-perturbative}) and provides an upper bound
on secular terms.
This gives $|\delta| < 3 |\delta_i|$ until the time $t_{g \rm min}$, with
\be
t_{g \rm min} = t_{c_s} \left( \frac{m}{H_{c_s}} \right)^{8/11} \gg t_{c_s} .
\label{eq:t-min-def}
\ee
This time does not depend on the wave number. This implies that there is no ultraviolet
divergence; the time when the density contrast becomes of the order of unity does not go
to zero at high $k$ and remains above the finite value (\ref{eq:t-min-def}).

We can check that this agrees with estimates obtained from the stability chart of the
constant-coefficients Mathieu equation.
First, let us consider the behavior of large wave numbers,
$k \gg k_g$, when they cross high-order instability bands.
Let us recall that for large $n$, not too far from the $A$-axis, the $n^{\rm th}$ instability band
occurs at $A_n$ with an exponentially small width $\Delta A_n$
\cite{AVRON198176,ANAHTARCI2012243},
\be
A_n \simeq n^2 , \;\;\; \Delta A_n \sim \frac{8 (q/4)^n}{[(n-1)!]^2}
\left[ 1 - \frac{q^2}{4n^3} + \dots \right] .
\label{eq:An-Delta-An}
\ee
At time $t$, we have for wave numbers greater than $k_g$,
$A \simeq \frac{4 \tau_{g\infty}^2 k^4}{9 \tau^2 k_g^4}$
and $q = \frac{2 (-\tau_{g\infty})^{23/6}  k^2}{9 (-\tau)^{23/6} k_g^2}$.
This gives $n \simeq \frac{2 \tau_{g\infty} k^2}{3 \tau k_g^2}$ and we are inside an instability
band when $n$ is very close to an integer.
We can check that the corrective term in the bracket in (\ref{eq:An-Delta-An})
is negligible for $|\tau| \geq |\tau_{g\infty}|$ and $k \gg k_g$.
Then, we obtain for the width $\Delta A$ the asymptotic upper bound
\be
n \gg 1 : \;\;\; \frac{\Delta A}{A} \lesssim
\exp\left[ - \frac{4 \tau_{g\infty} k^2}{3 \tau k_g^2} \ln\frac{k}{k_g} \right] .
\ee
The time spent inside the instability region is
$\frac{\Delta\tau}{|\tau|} = \frac{1}{2} \frac{\Delta A}{A}$.
Therefore, with a growth exponent $\mu_n$, the density contrast grows during the
time spent in the $n^{\rm th}$ instability band by a factor
\be
n \gg 1 : \;\;\; e^{\mu_n \Delta\tau_n} \lesssim \exp \left[ \mu_n |\tau_n|
e^{-\frac{4 \tau_{g\infty} k^2}{3 \tau_n k_g^2} \ln(k/k_g) } \right] .
\label{eq:mu-high-k-high-n}
\ee
As $\mu_n$ decreases at high $n$, we can see that the growth becomes negligible
at high $k$.
We can resum the cumulative growth due to the crossing of successive instability bands
by a given wave number $k$. From $n \simeq \frac{2 \tau_{g\infty} k^2}{3 \tau k_g^2}$,
we obtain the crossing time $\tau_n$ of the $n^{\rm th}$ band,
$\tau_n \simeq \frac{2\tau_{g\infty} k^2}{3 n k_g^2}$. As $|\tau_n| \gg |\tau_{g\infty}|$
for all $n$, we can apply Eq.(\ref{eq:mu-high-k-high-n}) for all $n \gg 1$.
Neglecting the decrease of $\mu_n$ with $n$, we obtain the conservative estimate
of the cumulative growth factor $G$ by the time $t_{g\infty}$,
\be
G_{n_0,N} = \prod_{n=n_0}^{N} e^{\mu_n \Delta\tau_n} = e^{S_{n_0,N}} ,
\ee
where $N=\frac{2 k^2}{3 k_g^2}$ is the final band reached at the time $t_{g\infty}$,
$n_0 \gg 1$ is the lowest value where we can use Eq.(\ref {eq:mu-high-k-high-n}),
and
\ba
&& n_0 \gg 1 : \;\;\; S_{n_0,N} \lesssim \sum_{n=n_0}^N \mu  |\tau_n|
e^{-\frac{4 \tau_{g\infty} k^2}{3 \tau_n k_g^2} \ln(k/k_g) } \nonumber \\
&& \sim \mu  | \tau_{g\infty} | \left( \frac{k}{k_g} \right)^{2-2n_0}
\sum_{\ell=0}^{N-n_0} \frac{1}{\ell+n_0} \left( \frac{k}{k_g} \right)^{-2\ell} .\;\;\;
\hspace{0.3cm}
\ea
The sum over $\ell$ converges and the limit $N\to\infty$ provides an upper bound.
This also shows that the cumulative growth is dominated by the lower bands,
$n \sim n_0$. This gives
\be
n_0 \gg 1 : \;\;\;  G_{n_0,N} \lesssim \exp\left[ | \tau_{g\infty} |
\left( \frac{k}{k_g} \right)^{2-2n_0} \right] ,
\label{eq:G-high-k-n0-N}
\ee
where we take the upper bound $\mu \lesssim 1$. Thus,
the cumulative growth due to the crossing of high-order bands,
for instance $n\geq 10$, decreases at high wave numbers. Therefore, there is no ultraviolet divergence due to the crossing of high-order instability bands by high wave numbers.

To estimate the growth associated with the crossing of the first few instability bands,
we evaluate the growth obtained for the first band $n=1$, which should be the largest one.
From Eqs.(\ref{eq:A-k-tau-def})-(\ref{eq:q-k-tau-def}), we can see
that high wave numbers, $k \gg k_g$, cross the first instability band, $n=1$,
at time $t_1 \sim t_{g\infty} (k/k_g)^{-8/3}$ with $A\simeq 1$ and $q \sim (k/k_g)^{-17/3} \ll 1$.
At low $q$, the width of the first instability band is $\Delta A_1 \sim q$, with a growth rate
$\mu_1 \sim q$ \cite{Fukunaga:2019unq}. This gives a growth factor
\be
n= 1 , \;\; k \gg k_g : \;\;\; e^{\mu_1 \Delta\tau_1} \sim \exp\left[ | \tau_{g\infty} |
\left( \frac{k}{k_g} \right)^{-28/3} \right] ,
\label{eq:mu-high-k-n-1}
\ee
which again goes to unity at large $k$.
Moreover, we can infer that Eq.(\ref{eq:mu-high-k-n-1}) provides the extension
down to $n_0=1$ of Eq.(\ref{eq:G-high-k-n0-N}), which was only valid for large $n_0$
and neglected the decrease of $\mu$ at low $q$.

This high-wave number case is shown by the two lower panels in Fig.~\ref{fig_Mathieu}.
In agreement with the analysis above, the instability appears earlier for higher $k$
but the amount of growth decreases as the instability bands are narrower with lower
growth rates. Moreover, higher-order instability bands crossed at later times do not
significantly change the amplitude of the density contrast. In the case $k=1.9 k_g$,
shown in the lower right panel, the instability band crossed at $-\log(-\tau) \simeq -3.1$
actually leads to a small decrease of $|\delta|$. This can happen depending on the
phase of the density contrast at the entry of the narrow instability band, if it starts with a greater
weight on the decaying mode.
At higher $k$, there is no significant change from the initial amplitude $|\delta_i|=10^{-5}$ of
the oscillations.

This analysis shows that the growth factor decreases at high wave numbers.
Therefore, only a finite range of wave numbers above $k_g$ has been able to
show a significant growth of the density contrast by the time $t_{g\infty}$.
This agrees with the finiteness of the bound (\ref{eq:t-min-def}), which provides
a lower bound for the earliest instability time of the fastest-growing mode $k$.

This linear growth of the scalar density perturbations by a parametric resonance
is illustrated by the second column in Fig.~\ref{fig_Plots-cos}.

\subsubsection{Initial nonlinear scalar structures}
\label{sec:nonlinear-cosine}

Thus, we can conclude that the density contrast becomes of the order of unity at a time
$t_{\rm NL}$ with
\be
t_{g \rm min} \leq t_{\rm NL} \leq t_{g \infty} ,
\ee
for wave numbers $k_{\rm NL}$ somewhat greater than $k_g$.
We can obtain an upper bound for the highest unstable wave number from
Eq.(\ref{eq:mu-high-k-n-1}), which as we explained above is not modified by the
crossing of higher-order instability bands.
This gives for the wave numbers where the Mathieu-equation instability bands
can have some significant effect,
\be
k \leq k_g \left( \frac{\rho_{c_s}}{\rho_b} \right)^{33/952}
\left( \frac{m}{H_{c_s}} \right)^{-9/238} .
\label{eq:k-max-2}
\ee
This provides an upper bound for the wave numbers where the density contrast first
becomes of the order of unity.
The small exponents show that this upper bound is not many orders of magnitude greater
than $k_g$.

Thus, we can consider that the density contrast reaches the nonlinear regime at times
of the order of $t_{g\infty}$, on comoving scales $x_g \sim 1/k_g$.
This gives a typical size for the first nonlinear structures in physical coordinates
$r_{\rm NL} \sim a_{g \infty}/k_g$,
which yields
\be
r_{\rm NL} \sim \frac{1}{H_{c_s}}  \left( \frac{\rho_{c_s}}{\rho_b} \right)^{-3/68}
\left( \frac{m}{H_{c_s}} \right)^{-3/34} < \frac{1}{H_{c_s}} ,
\label{eq:r-clump-tg}
\ee
and a typical mass
\be
M_{\rm NL} \sim \frac{\rho_{c_s}}{H_{c_s}^3}  \left( \frac{\rho_{c_s}}{\rho_b} \right)^{-33/68}
\left( \frac{m}{H_{c_s}} \right)^{-33/34} .
\label{eq:M-clump-init-cosine}
\ee
Thus, we obtain a typical size that is somewhat smaller than $1/H_{c_s}$, but not
by a great factor as the exponents in Eq.(\ref{eq:r-clump-tg}) are rather small.
Using the relation (\ref{eq:rho-cs-rhob}) we can also write $r_{\rm NL}$ and $M_{\rm NL}$ as
\ba
&& r_{\rm NL} \sim \frac{1}{m} \left( \frac{\rho_b}{\rho_a} \right)^{-3/17}
\left( \frac{m}{H_{c_s}} \right)^{19/34} , \nonumber \\
&& M_{\rm NL} \sim \frac{\rho_b}{m^3} \left( \frac{\rho_b}{\rho_a} \right)^{35/17}
\left( \frac{m}{H_{c_s}} \right)^{209/34} ,
\label{eq:M-NL-cosine}
\ea
and the typical density as
\be
\rho_{\rm NL} \sim \bar\rho_{g\infty} \sim \rho_b \left( \frac{\rho_b}{\rho_a} \right)^{44/17}
\left( \frac{m}{H_{c_s}} \right)^{76/17} .
\label{eq:rho-NL-cosine}
\ee
Comparing with Eq.(\ref{eq:tau-g}) we find
\be
\rho_{\rm NL} \sim \rho_b \, | \tau_{g\infty} |^2 \gg \rho_b,
\label{eq:rho-NL-tau}
\ee
as we assumed $ | \tau_{g\infty} | \gg 1$ following (\ref{eq:tau-large}).

In contrast with the polynomial case (\ref{eq:m-clump}), the typical density $\rho_{\rm NL}$
at the entry into the nonlinear density contrast regime is not only set by a characteristic
density scale of the self-interaction potential, such as $\rho_b$ or $\rho_a$ that would
play the role of $\rho_\Lambda$ in Eq.(\ref{eq:m-clump}).
This is clearly shown by the new factor $m/H_{c_s}$ that involves both the scalar-field mass $m$
and the Hubble expansion rate, which could be seen as an external parameter.
This is due to the importance of the quantum pressure.

In the polynomial case, the instability was triggered by the change of sign of the self-interactions
contribution to the squared speed of sound, see Eq.(\ref{eq:rho-cs-poly-def}).
The quantum pressure then only determined the lower bound for the scales
where the instability can develop, see Eqs.(\ref{eq:gamma-q-qup})-(\ref{eq:qmax-def}),
introducing in this manner the length scale $1/m$.
Thus, the quantum pressure only played a secondary role. This is also seen in the solitons
found in the polynomial case, see Eqs.(\ref{eq:M-stable})-(\ref{eq:clumps-scaling}).
The quantum pressure only sets the minimal mass $M_{\rm min}$ and radius $R_{\rm min}$
of the solitons, but their typical core density $\rho_\Lambda$ and their scaling law
(\ref{eq:clumps-scaling}) do not involve the quantum pressure, which plays a negligible role
at high masses (and only governs the low-density tails of these scalar clouds).

In contrast, in the case of the Bessel-type self-interaction potential (\ref{eq:V-I-J0}),
which decays at large densities in an oscillatory manner, the instability is
triggered by a parametric resonance between the oscillations of the potential and
the harmonic oscillator built by the combination of the scalar-field kinetic term and
its quantum pressure. This harmonic oscillator corresponds to the term
$\ddot\delta+ c_s^2 \frac{k^2}{a^2} \delta$ in the equation of motion (\ref{eq:delta-cosmo}),
where we only include the quantum pressure
contribution to $c_s^2$, or to the term $y''+Ay$ in the generalized Mathieu equation
(\ref{eq:Mathieu}). Thus, instead of a tachyonic instability we have a parametric resonance.
It clearly involves the interplay between the scalar-field kinetic terms, its quantum pressure,
and its self-interactions. This explains the appearance of the new factor $m/H_{c_s}$
in Eq.(\ref{eq:rho-NL-cosine}), as compared with Eq.(\ref{eq:m-clump}).
As could be expected, this mulitplicative factor can be expressed in terms of
$|\tau_{g\infty}|$ in (\ref{eq:rho-NL-tau}), which measures the possibility for the parametric
resonance to take place, and its advance before the tachyonic instability that would be found
at low densities, see the discussion above (\ref{eq:tau-large}).

\subsection{Scalar-field solitons}
\label{sec:scalar-field-clumps-cosine}

\subsubsection{Equilibrium profiles}

\begin{figure}
\begin{center}
\epsfxsize=8.8 cm \epsfysize=6 cm {\epsfbox{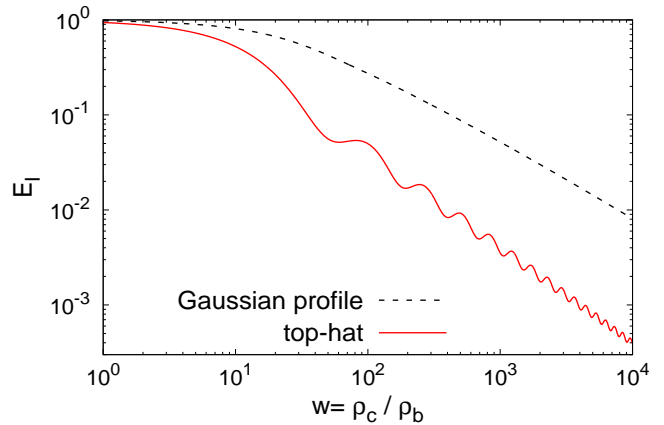}}
\end{center}
\caption{Self-interaction energy ${\cal E}_{\rm I}(\rho_c/\rho_b)$ for a Gaussian
profile (upper dashed line) and a top-hat profile (lower solid line). We can see the negligible influence of the oscillations compared to the overall decrease. The absence of minimum as seen for the Gaussian profile entails that a continuous distribution of densities can be present for clumps in axion monodromy models.}
\label{fig_E_I}
\end{figure}

As in Sec.~\ref{sec:scalar-field-clumps}, we now look for the equilibrium profiles of isolated
scalar-field halos.
Local minima of the energy $E$ at fixed mass $M$ are still given by
Eqs.(\ref{eq:equilibrium-E})-(\ref{eq:equilibrium-isolated}), which coincide with the hydrostatic
equilibrium of the Euler equation (\ref{eq:Euler}).
For the trial Gaussian density profile (\ref{eq:Gaussian-density-trial}), the
gravitational and quantum-pressure energies $E_{\rm grav}$ and $E_{\rm Q}$
are still given by Eq.(\ref{eq:E-grav-Q-I}), while the self-interaction energy $E_{\rm I}$ reads
\be
E_{\rm I}^{\rm G} = - \frac{8\rho_b}{\rho_a} M + \frac{8\rho_b}{\rho_a} M \,
{\cal E}_{\rm I}^{\rm G}( \rho_c / \rho_b ) ,
\label{EI-cosine}
\ee
where $\rho_c$ is the density at the center of the halo and
we introduced the function
\be
{\cal E}_{\rm I}^{\rm G}(w)  = \frac{16}{\sqrt{\pi} \, w} \int_0^{\infty} du \, u^2 \left[ 1
- J_0\left( \sqrt{w} \, e^{-u^2/2} \right) \right]  .
\label{eq:EI-Gaussian-cosine}
\ee
On the other hand, for a top-hat profile we obtain the same form (\ref{EI-cosine})
but with a scaling function ${\cal E}_{\rm I}^{\rm T.H.}(w)$ given by
\be
{\cal E}_{\rm I}^{\rm T.H.}(w) = \frac{4}{w} \left[ 1 - J_0(\sqrt{w}) \right] .
\label{eq:EI-top-hat-cosine}
\ee
We display the functions ${\cal E}_{\rm I}^{\rm G}(w)$ and ${\cal E}_{\rm I}^{\rm T.H.}(w)$
in Fig.~\ref{fig_E_I}.
The top-hat profile shows the decaying oscillations arising from the self-interaction
potential (\ref{eq:V-I-J0}). The regular Gaussian profile erases these small oscillations,
through the smooth radial integration, and only shows a smooth decay.
At large densities, $w \gg 1$, we have ${\cal E}_{\rm I}^{\rm G}(w) \sim (\ln w)^{3/2}/w$
and ${\cal E}_{\rm I}^{\rm T.H.}(w) \sim 1/w$, while at low densities we have
${\cal E}_{\rm I}(0)=1$.

The first term proportional to $M$ in Eq.(\ref{EI-cosine}) plays no role, as we consider
minima of the total energy at constant mass. This had to be the case, because it
originates from the linear term in $\rho$ in the self-interaction potential
(\ref{eq:V-I-J0}), which could be absorbed as a small correction to the quadratic
term $\phi^2$ of the potential $V(\phi)$, see also Eq.(\ref{eq:m-m0-correction}).
This corresponds to a small change of the scalar-field mass and should not alter the physics.

Neglecting logarithmic corrections, we write
\be
\left. E_{\rm I} \right|_M \sim M \, \frac{\rho_b^2}{\rho_a (\rho_b+ \rho_c)} ,
\label{EI-cosine-1}
\ee
which gives the correct asymptotes at both low and high core densities, except for numerical
prefactors.
Here, the subscript $|_M$ means that we have removed the irrelevant constant contribution
$- \frac{8\rho_b}{\rho_a} M$.
Then, looking for a minimum with respect to $\rho_c$ of the sum of the self-interaction
and quantum-pressure energies, $\left. E_{\rm I} \right|_M+E_{\rm Q}$,
we obtain from Eqs.(\ref{eq:E-grav-Q-I}) and (\ref{EI-cosine-1}) that the minimum $\rho_c$
is nonzero for masses above a lower threshold $M_{\rm min}$, with
\be
M_{\rm min} \sim \left( \frac{\rho_b}{\rho_a}\right)^{-3/2} \frac{\rho_b}{m^3} ,
\label{eq:M-stable-cosine}
\ee
and for higher masses it scales as
\be
M \gg M_{\rm min} : \;\;\; \rho_c \sim \rho_b \left( \frac{M}{M_{\rm min}} \right)^{2/5} .
\label{eq:M-rhoc-cosine}
\ee
Below the mass $M_{\rm min}$ the self-interactions are not strong enough to resist the
quantum pressure and the halo keeps on extending with a density that goes to zero.
The existence of a critical mass $M_{\rm min}$ is thus common to the cosine potential
(\ref{eq:V-phi-cosine}) studied in this section and to potentials such as the polynomial case
(\ref{eq:Phi-I-polynomial}).
This is because small halo masses require small radii for the density to be large enough
for the self-interactions to become important, but small radii further increase the impact
of the quantum pressure, as it involves gradients of the density.

On the other hand, the cosine potential (\ref{eq:V-phi-cosine}) does not select
a unique density $\rho_\Lambda$, up to factors of unity.
This could be seen from the analysis of linear perturbations in
Sec.~\ref{sec:linear-dynamics-cosine}, where we obtained instabilities for a range of
densities, which peak at a density $\rho_g$ that can be many orders of magnitude above
the potential scale $\rho_b$.
In the context of static isolated solitons, this is also seen from the self-interaction energy
$E_{\rm I}$ shown in Fig.~\ref{fig_E_I}, which does not display a unique minimum but
keeps decreasing at large densities, possibly showing an infinite series of local minima
along the way.
Then, Eq.(\ref{eq:M-rhoc-cosine}) shows that the interplay between the self-interactions and
the quantum pressure select a mass-dependent typical density $\rho_c$ for the
equilibrium profile. This core density grows with $M$ as $\rho_c \propto M^{2/5}$.
Thus, while in the polynomial case the quantum pressure played no
significant role in the soliton profiles (apart from setting their minimum mass) and their
scaling law (\ref{eq:clumps-scaling}), becoming negligible at high masses,
for the Bessel potential (\ref {eq:V-I-J0}) the quantum pressure plays a key role at all soliton
masses. There, the soliton profile is always set by the balance between the self-interactions
and the quantum pressure.

The self-interactions and quantum-pressure energies of these solitons scale as
\ba
&& E_{\rm I} \sim - \frac{\rho_b}{\rho_a} M + \frac{\rho_b}{\rho_a} M_{\rm min}
\left( \frac{M}{M_{\rm min}} \right)^{3/5} , \nonumber \\
&& E_{\rm Q} \sim \frac{\rho_b}{\rho_a} M_{\rm min}
\left( \frac{M}{M_{\rm min}} \right)^{3/5} .
\label{eq:clumps-scaling-E-cosine}
\ea
As in the polynomial scenario (\ref{eq:clumps-scaling-E}), the total energy $E \simeq E_{\rm I}$
is dominated by the self-interactions energy.
However, this leading term $- \frac{\rho_b}{\rho_a} M$ does not play any role in the
determination of the equilibrium profile, which is set by the balance between the
self-interactions and quantum-pressure contributions associated with the subleading terms
$\propto M^{3/5}$ (but note that both contributions are positive in
Eq.(\ref{eq:clumps-scaling-E-cosine})).

Thus, the importance of the quantum pressure term for scenarios with Bessel-type
potentials like (\ref{eq:V-I-J0}), in contrast with the polynomial scenario (\ref{eq:Phi-I-polynomial})
mostly governed by its self-interactions, appears both for the parametric resonance
studied in Sec.~\ref{sec:linear-dynamics-cosine} and for the isolated soliton profiles
studied in this section.

\subsubsection{Numerical computation of the radial profile}
\label{sec:numerical-profile-cosine}

\begin{figure}
\begin{center}
\epsfxsize=8.8 cm \epsfysize=6 cm {\epsfbox{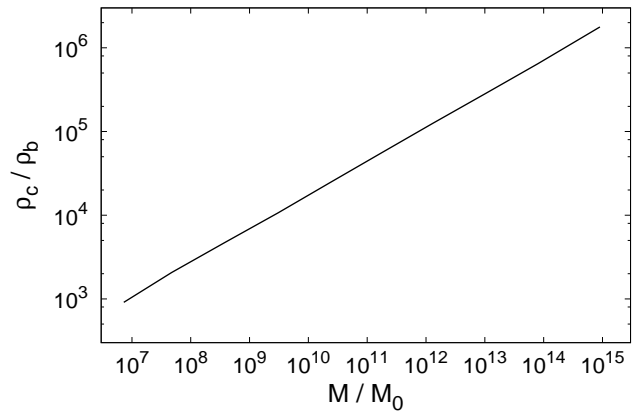}}
\end{center}
\caption{Mass - core density relation for the spherical-equilibrium soliton profiles obtained
from Eq.(\ref{eq:y-x-profile-cosine}).}
\label{fig_rho_M_cosine}
\end{figure}

\begin{figure*}
\begin{center}
\epsfxsize=5.9 cm \epsfysize=5 cm {\epsfbox{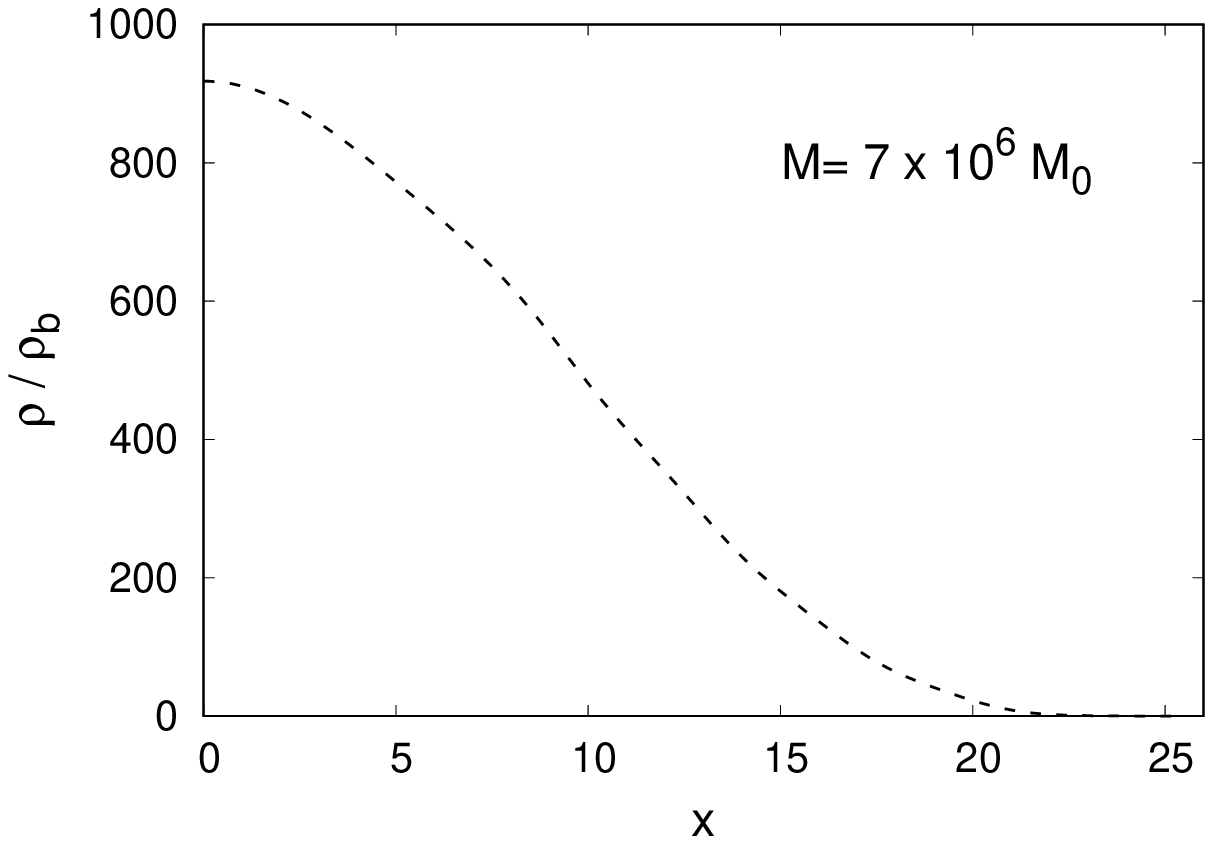}}
\epsfxsize=5.9 cm \epsfysize=5 cm {\epsfbox{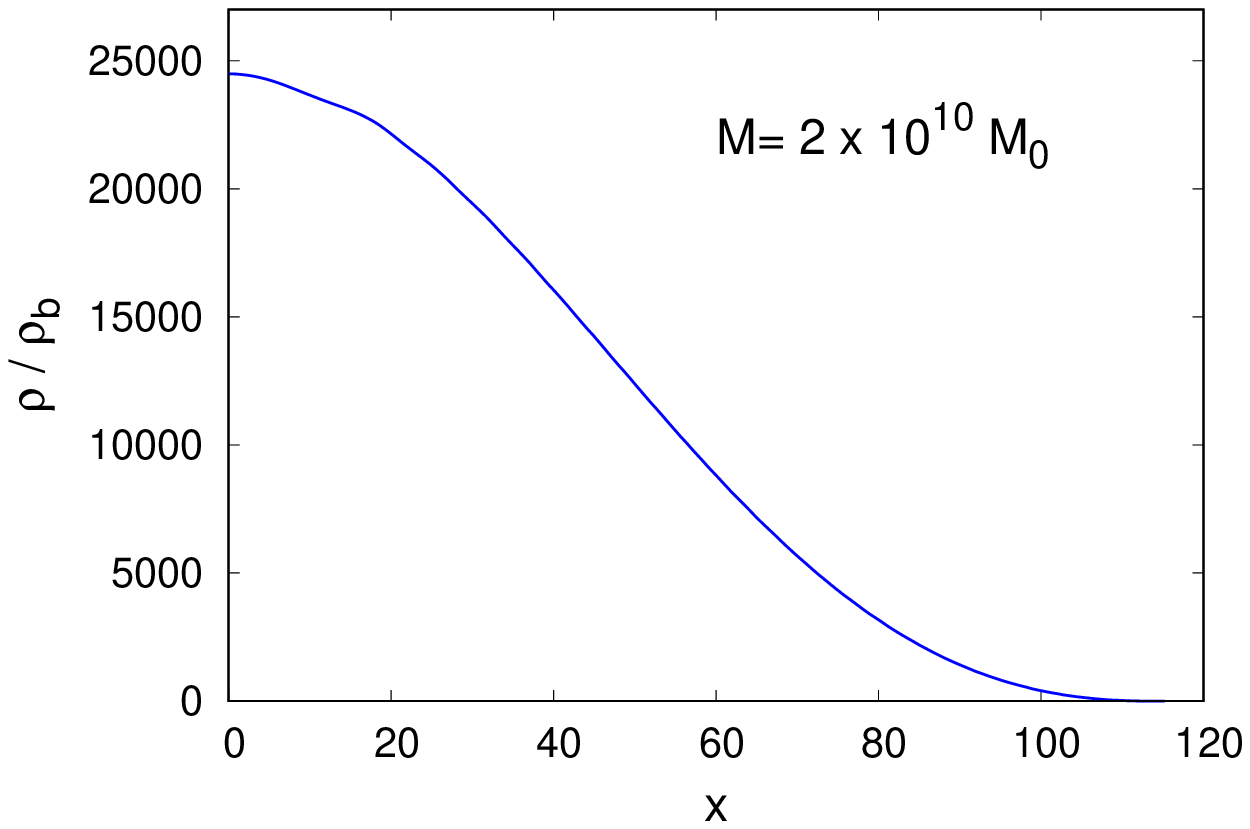}}
\epsfxsize=5.9 cm \epsfysize=5 cm {\epsfbox{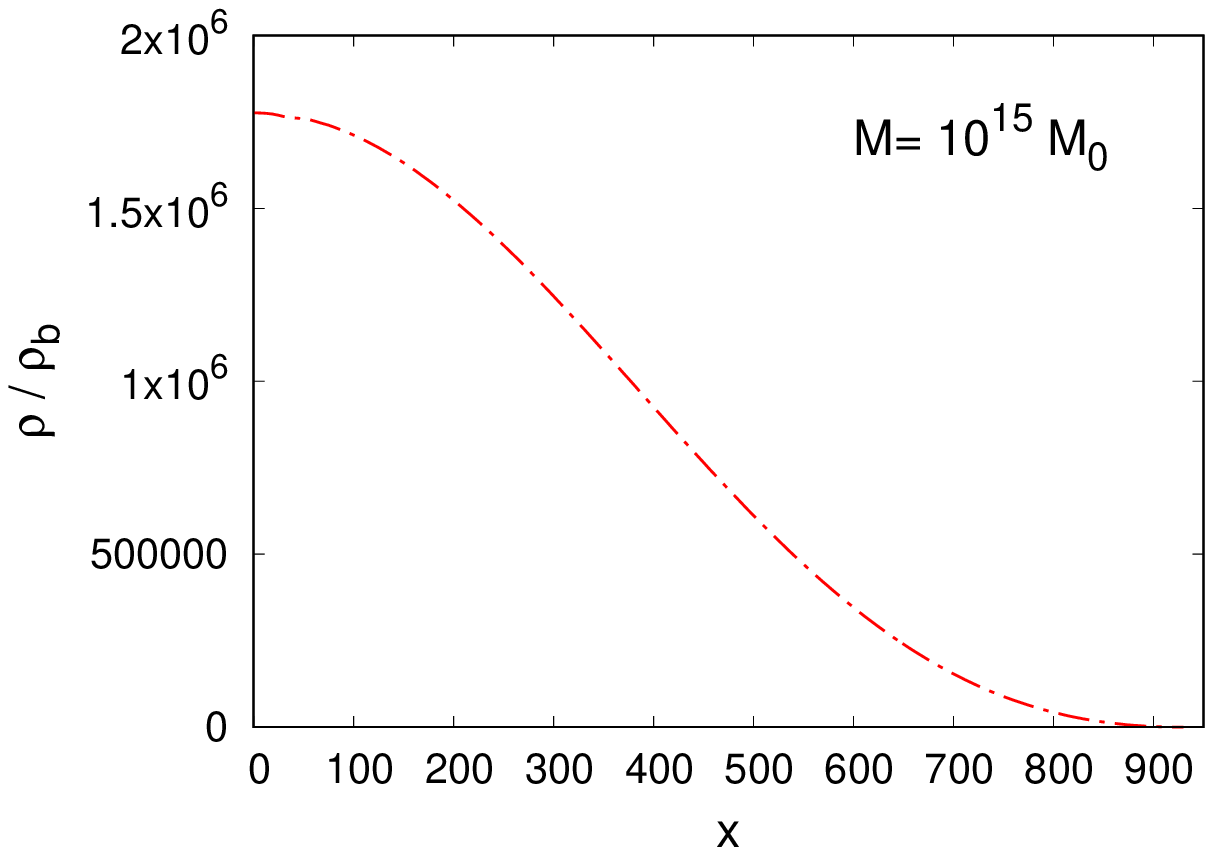}}
\end{center}
\caption{Density profiles of the spherical-equilibrium solitons obtained
from Eq.(\ref{eq:y-x-profile-cosine}). We show the cases of soliton mass
$M=7 \times 10^6 M_0$, $2 \times 10^{10} M_0$ and $10^{15} M_0$
from left to right.}
\label{fig_rho_x_cosine}
\end{figure*}

As for the polynomial case studied in Sec.~\ref{sec:numerical-profile}, we confirm
the analytical results with a numerical computation of the soliton profiles.
Neglecting the gravitational energy, the equation of equilibrium (\ref{eq:equilibrium-isolated})
that describes minima of the total energy at fixed mass now reads
\be
\frac{d^2y}{dx^2} + \frac{2}{x} \frac{dy}{dx} = J_1(y) + \tilde\alpha y ,
\label{eq:y-x-profile-cosine}
\ee
where we introduced the dimensionless variables
\be
y = \sqrt{\frac{\rho}{\rho_b}} , \;\;\; x = \sqrt{\frac{32 \rho_b}{\rho_a}} m r ,
\;\;\; \tilde\alpha = - \frac{1}{2} - \alpha \frac{\rho_a}{16\rho_b} .
\ee
The soliton mass is also given by the integral (\ref{eq:M0-def}), but $M_0$ is now given by
\be
M_0 =  \left( \frac{32 \rho_b}{\rho_a}\right)^{-3/2} \frac{\rho_b}{m^3} ,
\ee
which also sets the order of magnitude of the lower mass threshold
$M_{\rm min}$ of Eq.(\ref{eq:M-stable-cosine}).

We again solve the boundary-value problem (\ref{eq:y-x-profile-cosine}) with
a double-shooting method. We first show in Fig.~\ref{fig_rho_M_cosine}
the mass - density relation of the equilibrium profiles that we obtain in this fashion.
As expected we recover a mass-dependent core-density, with a slope that agrees
with the analytical prediction (\ref{eq:M-rhoc-cosine}).
We show in Fig.~\ref{fig_rho_x_cosine} the soliton profiles obtained for three masses $M$.
Even though the Bessel function $J_1(y)$ in Eq.(\ref{eq:y-x-profile-cosine}) is not scale-free,
its cosine-like oscillations are mostly erased by the smooth density profiles,
as was the case for the self-interaction energy $E_{\rm I}^{\rm G}$ shown in Fig.~\ref{fig_E_I}
for the Gaussian profile ansatz.
Then, the profiles obtained at these vastly different masses and densities are quite similar
and regular, without significant oscillations.
As for the polynomial case studied in Fig.~\ref{fig_rho_x_poly}, they show an exponential
tail at large radii.
These smooth behaviors explain why we recover the simple analytic prediction
(\ref{eq:M-rhoc-cosine}).

We discuss in more detail in the appendix~\ref{app:trajectory-cosine} the properties
of these solitonic profiles, interpreting again the differential equation (\ref{eq:y-x-profile-cosine})
as the damped motion of a particle $y(x)$ with time $x$ in a potential $U(y)$. This provides
another simple explanation for the behaviors found in Fig.~\ref{fig_rho_x_cosine}.
In particular, it clearly explains why the solitonic profiles obtained for the polynomial case
in Fig.~\ref{fig_rho_x_poly} and those obtained for the cosine model in Fig.~\ref{fig_rho_x_cosine}
show different behaviors.

\subsection{Mergings in the nonlinear regime}
\label{sec:merging-cosine}

\subsubsection{Initial relaxation onto the soliton scaling law}
\label{sec:relaxation-cosine}

We have seen in Sec.~\ref{sec:nonlinear-cosine} that, at the entry of the density contrast into the nonlinear regime, the first structures have a mass $M_{\rm NL}$ and a density
$\rho_{\rm NL}$ given by Eqs.(\ref{eq:M-NL-cosine})-(\ref{eq:rho-NL-cosine}).
Comparing with the minimum mass $M_{\rm min}$ and the minimum density
$\rho_{\rm min} \sim \rho_b$ of the solitons found in
Eqs.(\ref{eq:M-stable-cosine})-(\ref{eq:M-rhoc-cosine}), we obtain
\be
\frac{M_{\rm NL}}{M_{\rm min}} \sim | \tau_{g\infty} |^{11/4} \gg 1 , \;\;\;
\frac{\rho_{\rm NL}}{\rho_{\rm min}} \sim | \tau_{g\infty} |^2 \gg 1 ,
\label{eq:M-NL-M-min-cosine}
\ee
where we used Eq.(\ref{eq:tau-g}).
Thus, in contrast with the polynomial scenario (\ref{eq:M-NL-M-min}), the first nonlinear
structures are much greater than the smallest stable solitons and also have a greater
density.
As for the discussions below (\ref{eq:rho-NL-tau}) and (\ref{eq:M-rhoc-cosine}), this mismatch
and the appearance of the factors $ | \tau_{g\infty} |$ is due to the interplay between
the self-interactions, the kinetic terms and the quantum pressure, which cannot be neglected
in this scenario.
Moreover, we find that these initial structures $\{ M_{\rm NL}, \rho_{\rm NL} \}$ deviate
from the soliton scaling law (\ref{eq:M-rhoc-cosine}), since we obtain
\be
\rho_{\rm NL} \sim \rho_b \left( \frac{M_{\rm NL}}{M_{\rm min}} \right)^{8/11}
\gg \rho_b \left( \frac{M_{\rm NL}}{M_{\rm min}} \right)^{2/5} .
\label{eq:mismatch}
\ee
In other words, these initial structures are too dense as compared with the soliton
equilibrium profiles.
Therefore, they cannot relax to stable solitons without significant changes.
In particular, if we consider an aggregation mechanism as in Sec.~\ref{sec:aggregation},
we can no longer assume that they constitute the first steps of an aggregation process
that evolves along the soliton scaling law (\ref{eq:M-rhoc-cosine}), since this starting point
itself deviates from this scaling law.
The mismatch (\ref{eq:mismatch}) is due to the fact that the static solitons are governed by
the balance between the quantum pressure and the self-interactions, whereas the instability
that gives rise to the first nonlinear structures (\ref{eq:M-NL-cosine}) also involves the kinetic
energy, associated with the time derivatives in the equations of motion
(\ref{eq:delta-eta-cos}) or (\ref{eq:Mathieu}).

The structures of masses $M_{\rm NL}$ cannot expand within one Hubble time to lower their density
so as to fall onto the scaling law (\ref{eq:M-rhoc-cosine}), because of the conservation of mass
within large comoving volumes (there is no outer space to expand into).
Therefore, it is more natural to assume that they evolve towards the scaling law
(\ref{eq:M-rhoc-cosine}) by merging while keeping a density of the order of $\rho_{\rm NL}$.
From (\ref{eq:M-NL-M-min-cosine}) we find that this target mass $M_i$ is
\be
M_i = M_{\rm NL} | \tau_{g\infty} |^{9/4} , \;\;\;
\mbox{so that} \;\; \rho_{\rm NL} \sim \rho_b \left( \frac{M_i}{M_{\rm min}} \right)^{2/5} .
\label{eq:Mi-cosine-def}
\ee
This also means that the radius of these clumps has grown to $R_i$ with
\be
R_i = r_{\rm NL} | \tau_{g\infty} |^{3/4} .
\label{eq:Ri-rNL-cosine}
\ee
We can compare this size with the initial velocity $v_{\rm NL}$ of the structures that
enter the nonlinear regime.
At the time $t_{g\infty}$, we again estimate the typical velocity from the
continuity equation (\ref{eq:continuity-linear}), $v \sim r \frac{\partial \delta}{\partial t}$.
With a growth rate $\delta \sim e^{\mu\tau}$, we obtain when $\delta \sim 1$ and with
$\mu \sim 1$,
\be
v_{\rm NL} \sim r_{\rm NL} \frac{|\tau_{g\infty}|}{t_{g\infty}} .
\label{eq:v-NL-cosine}
\ee
Comparing with (\ref{eq:Ri-rNL-cosine}), we can see that it takes less than a Hubble time
for a disturbance to travel from $r_{\rm NL}$ to $R_i$, as $R_i-r_{\rm NL} \ll v_{\rm NL} t_{g\infty}$.
This suggests that it is indeed possible for the scalar-field structures to reach the mass $M_i$
within a Hubble time and to relax on the soliton scaling law (\ref{eq:M-rhoc-cosine}).
This stage is illustrated by the third column in Fig.~\ref{fig_Plots-cos}.
It seems difficult however to obtain a more rigorous description of this process by
analytical means. A more detailed study of this nonlinear stage is left for future works
using numerical simulations.

\subsubsection{Lack of significant aggregation}
\label{sec:aggregation-cosine}

As in the polynomial scenario discussed in Sec.~\ref{sec:aggregation}, we could expect
the scalar-field clumps $M_i$ formed by this relaxation process to grow further through collisions.
From the analysis above, we start with the initial mass $M_i$ of Eq.(\ref{eq:Mi-cosine-def}),
density $\rho_i \sim \rho_{\rm NL}$, radius $R_i \sim (M_i/\rho_i)^{1/3}$ and the velocity
$v_i \sim R_i/t_{g\infty}$. This velocity estimate is somewhat uncertain.
It is somewhat lower than the initial velocity (\ref{eq:v-NL-cosine}) at the entry into the
nonlinear regime, but this is expected as the velocity of the clouds should decrease as they merge.
We estimate the aggregation of the scalar clumps following the approach presented
in Sec.~\ref{sec:aggregation}.
We still have $M(t) \propto 1/(a^3 n)$ but the radius of the clouds no longer grows as
$M^{1/3}$, because equilibrium profiles no longer remain at a constant density $\rho_\Lambda$.
Instead, their characteristic density grows as $M^{2/5}$ with their mass, which means that
their radius only grows as $R\propto M^{1/5}$ and their cross section
as $M^{2/5}$. Then, the solution of the aggregation equation (\ref{eq:collisions-1}) becomes
\be
n(t) = n_i \left( \frac{a}{a_i} \right)^{-3} \left[ 1 + \frac{3 n_i \sigma_i v_i}{10 H_i}
\left( 1 - \left( \frac{a_i}{a} \right)^2 \right) \right]^{-5/3} ,
\label{eq:nt-aggregation-cosine}
\ee
where the subscript $i$ stands for the initial condition at the time $t_{g\infty}$.
At late times the comoving number density goes to the finite value
\be
n_c = n_{c_i} \left( 1 + \frac{3 n_i \sigma_i v_i}{10 H_i} \right)^{-5/3} ,
\ee
which corresponds to a typical size and mass of the final halos of the order of
\begin{eqnarray}
R_\infty &=& R_i \left( 1 + \frac{3 n_i \sigma_i v_i}{10 H_i} \right)^{1/3} ,
\nonumber\\
M_\infty &=& M_i \left( 1 + \frac{3n_i \sigma_i v_i}{10 H_i} \right)^{5/3} .
\label{eq:R-Ri-growth-cosine}
\end{eqnarray}
As compared with Eq.(\ref{eq:R-Ri-growth}), we can see that the slower growth of the
radius and cross section as the clumps merge significantly damps the efficiency of the
aggregation process.
Moreover, with $n_i \sim 1/R_i^3$, $\sigma_i \sim R_i^2$ and $v_i \sim R_i/t_i$
we obtain
\be
\frac{n_i\sigma_i v_i}{H_i} \sim 1 , \;\;\; \mbox{hence} \;\;\;
R_\infty \sim R_i , \;\; M_\infty \sim M_i .
\label{eq:Rinfty-cosine}
\ee
Therefore, there should be no significant aggregation through collisions.
This is quite different from the strong aggregation process found for the polynomial case
in Sec.~\ref{sec:aggregation}.
This is due to the much slower velocity, which we took as $v_i \sim R_i/t_{g\infty}$.
This is related to the different formation process of the first nonlinear structures.
Whereas in the polynomial scenario we had a tachyonic instability,
leading to the fast formation of virialized scalar clouds with a typical velocity
set by the strength of the self-interactions, $v \sim \sqrt{|\Phi_{\rm I}|}$ as in
Eq.(\ref{eq:Ri-vi-sigmai-ni}), for the Bessel-type self-interaction potential
we have a parametric resonance that is not directly set by the strength of the
self-interactions, $\rho_b/\rho_a$, but by the interplay between the kinetic terms,
the quantum pressure and the self-interactions, leading to a resonance between
the oscillatory behavior of the self-interaction potential and the oscillations of the scalar field
due to its wave-like properties (the quantum pressure term combined with the kinetic term).
This leads to very different scalings, as seen by the comparison of $v_{\rm NL}$
in Eq.(\ref{eq:v-NL-cosine}), which explicitly involves the cosmic time $t_{g\infty}$,
with $v_i$ in Eq.(\ref{eq:Ri-vi-sigmai-ni}), which only involves the self-interactions
strength $\Phi_{\rm I}$.

On the other hand, if we take the larger initial value
$v_{\rm NL}$ of Eq.(\ref{eq:v-NL-cosine}) instead of $R_i/t_{g\infty}$, which is greater by a factor
$|\tau_{g\infty}|^{1/4}$, we obtain a more significant aggregation process with
$R_\infty \sim R_i |\tau_{g\infty}|^{1/12}$ and
$M_\infty \sim M_i |\tau_{g\infty}|^{5/12}$.
The relatively small exponents show that these values are not so much larger than the
previous estimates (\ref{eq:Rinfty-cosine}), unless $|\tau_{g\infty}|$ is huge.

Another difference from the polynomial scenario of Sec.~\ref{sec:aggregation} is that
energy is no longer conserved along the soliton scaling law.
Indeed, from Eq.(\ref{eq:clumps-scaling-E-cosine}) we find for the total energy per unit mass
(apart from kinetic energy)
\be
\frac{E}{M} \sim - \frac{\rho_b}{\rho_a} + \frac{\rho_b}{\rho_a}
\left( \frac{M}{M_{\rm min}} \right)^{-2/5} .
\label{eq:Etot-M-cosine}
\ee
This means that the internal specific energy decreases as the solitons merge. This favors the
mergings towards more massive halos but also suggests that some energy is radiated away
as low-mass scalar waves. These may later form smaller objects or a continuous component,
that could be accreted at later times by the solitons.

Thus, the estimate (\ref{eq:Rinfty-cosine}) is more uncertain than for the
polynomial case (\ref{eq:R-ballistic}).
We can expect a broad range of halo masses and more complex nonlinear dynamics than
for the polynomial case studied in Sec.~\ref{sec:aggregation}.
A more detailed investigation is left for future numerical simulations.

As in the case of the tachyonic instability, the solitons that are studied here evade the large-excursion instability which can happen for dense configurations,
when the amplitude of the background field probes anharmonic parts of the scalar potential. In the case of axionic potentials, this instability could have interesting consequences such as the implosion of the solitonic configurations and potential detectable effects in the form of gravitational waves \cite{Arvanitaki:2019rax}. Here, we avoid these phenomena as the field never violates harmonicity at leading order. On the other hand, as the effective potential $\Phi_{\rm I}(\rho)$ for the axion monodromy case oscillates at large-enough density, in the nonrelativistic regime that we have considered, it is plausible that some of the extrema of the energy functional that we have found are not in fact bona fide minima but local maxima of the energy. In this case, and similarly to the large-excursion case, there could be metastable or unstable solitonic configurations. We could for instance envisage that a maximum evolves towards a minimum simply by rearrangement of its field configuration or explodes under the destabilising effect of the quantum pressure. We leave these questions to future investigations.

\subsection{Solitons dominated by gravity}
\label{sec:solitons-gravity}

In contrast with the tachyonic scenario presented in Sec.~\ref{sec:tachyonic},
the relatively weaker strength of the self-interactions in this parametric-resonance scenario
implies that, for certain values of the model parameters, the solitons formed during the
nonlinear stage become dominated by gravity rather than by the self-interactions.

First, if we consider the structures of mass $M_{\rm NL}$ and density $\rho_{g\infty}$,
at the entry into the nonlinear regime, we obtain from (\ref{eq:M-NL-M-min-cosine})
and (\ref{eq:E-grav-Q-I}) that $E_{\rm grav} \ll E_{\rm Q}$ provided we have
\be
m \gg \left( \frac{\rho_b}{\rho_a} \right)^{-1} \frac{H_0^{3/8} T_{g\infty}^{5/4} }
{3^{5/16} \Omega_{\gamma0}^{9/16} M_{\rm Pl}^{5/8} } .
\label{eq:MNL-grav-cos}
\ee
We will check in Sec.~\ref{sec:constraints-cosine} that this condition is always satisfied,
for the range of parameters that we consider.

However, as the structures merge to reach the greater mass $M_i$ of Eq.(\ref{eq:Mi-cosine-def}),
their self-gravity also grows and can dominate over the self-interactions.
Then, we find that gravity remains small as compared with the quantum pressure and the
self-interactions for these masses $M_i$ provided we have
\be
\mbox{negligible gravity:} \;\;\;
m < \left( \frac{\rho_b}{\rho_a} \right)^{-1} \frac{\Omega_{\gamma0}^{9/8} T_{g\infty}^{7/2}}
{3^{7/8} M_{\rm Pl}^{7/4} H_0^{3/4} } .
\label{eq:Mi-grav-cos}
\ee
This boundary is shown by the black dotted line in Fig.~\ref{fig_T-m-cos} below
and it is not satisfied for low values of the parameter $T_{g\infty}$.
In this case, before they reach the mass $M_i$ of Eq.(\ref{eq:Mi-cosine-def}),
the nonlinear scalar-field clumps become dominated by gravity rather than by the
self-interactions.
This leads to a different scaling law from (\ref{eq:M-rhoc-cosine}) for the resulting solitons.
Thus, the balance between self-gravity and the quantum pressure yields the new scaling
law
\be
\rho_c \sim \frac{m^6 M^4}{M_{\rm Pl}^6} ,
\label{eq:rho-M-fuzzy}
\ee
for the relationship between the core density and the soliton mass.
If the nonlinear structures now merge at the characteristic density $\rho_{g\infty}$
until they reach this new scaling law (\ref{eq:rho-M-fuzzy}), the clump mass $M_i$
of Eq.(\ref{eq:Mi-cosine-def}) is replaced by
\be
M_{i, {\rm grav}} = \frac{\rho_{g\infty}^{1/4} M_{\rm Pl}^{3/2}}{m^{3/2}} .
\label{eq:Mi-grav}
\ee
Assuming that there is no significant aggregation afterwards, as the clumps are diluted by
the expansion of the Universe, this also gives the order of magnitude of the final scalar-field
clumps that play the role of the dark matter particles.

These final relaxation and dilution phases are illustrated by the fourth column
in Fig.~\ref{fig_Plots-cos}.

{

\subsection{No collapse into black holes}
\label{sec:no-BH-cos}

\subsubsection{Solitons dominated by self-interactions}
\label{sec:BH-self-interactions}

As for the tachyonic scenario, we again check that the scalar-field clumps do not collapse
into black holes.
We first consider the case where the condition (\ref{eq:Mi-grav-cos}) is satisfied:
the solitons are dominated by the balance between the quantum pressure and the self-interactions,
while gravity is negligible.
Then, from Eq.(\ref{eq:Rinfty-cosine}) the gravitational potential at the surface of these
stable solitons reads
\be
| \Phi | \sim \frac{{\cal G} M_\infty}{R_\infty} \sim
\frac{\bar\rho_{c_s}}{M_{\rm Pl}^2 H_{c_s}^2} \left( \frac{\rho_b}{\rho_a} \right)^{3/17}
\left( \frac{m}{H_{c_s}} \right)^{-18/17} \ll 1 ,
\label{eq:Phi-small-cos-self-inter}
\ee
as all factors in the last expression are much smaller than unity.
Thus, these clumps are far in the weak-gravity regime and do not form black holes.
This is again consistent with the fact that gravity is subdominant with respect to the
scalar-field self-interactions, which are already weak.

\subsubsection{Solitons dominated by gravity}
\label{sec:BH-gravity}

However, in the regime studied in Sec.~\ref{sec:solitons-gravity} when the condition
(\ref{eq:Mi-grav-cos}) is violated, gravity dominates over the self-interactions and
the scaling law of the solitons is changed to Eq.(\ref{eq:rho-M-fuzzy}).
Together with Eq.(\ref{eq:Mi-grav}), this gives for the gravitational potential
\be
\Phi \sim \frac{\rho_{g\infty}^{1/2}}{m M_{\rm Pl}} .
\label{eq:Phi-small-cos-grav}
\ee
We will check in Sec.~\ref{sec:constraints-cosine} and in Fig.~\ref{fig_M-R-cos}
below that $| \Phi | \ll 1$ over the allowed
parameter space delimited by other constraints (parametric-resonance condition,
classicality condition, ...).
Therefore, in this case again, the clumps remain far in the Newtonian-gravity regime
and do not form black holes.

}

{

\subsection{Parameter space}
\label{sec:constraints-cosine}

\begin{figure}
\begin{center}
\epsfxsize=8.8 cm \epsfysize=6.7 cm {\epsfbox{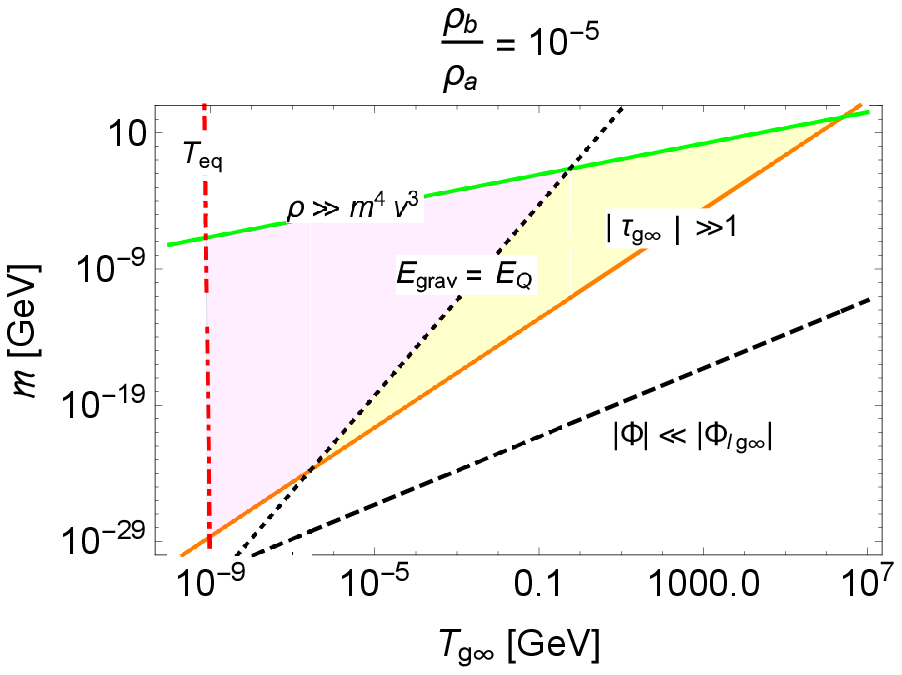}}\\
\epsfxsize=8.8 cm \epsfysize=6.7 cm {\epsfbox{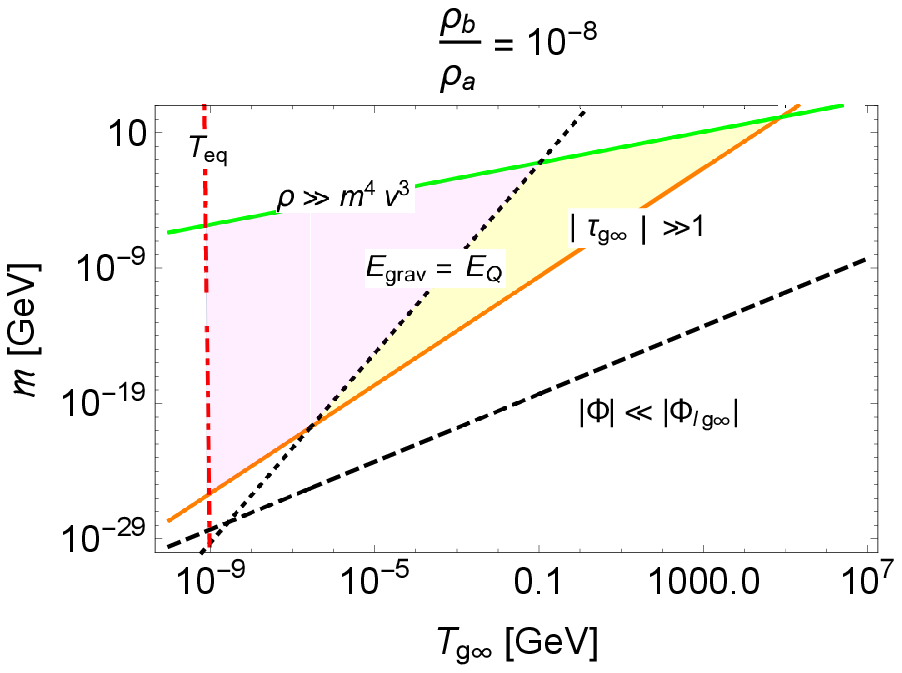}}
\end{center}
\caption{The shaded area is the domain of validity, in the plane $(T_{g\infty},m)$, of the scenario
described in this paper associated with potentials of the form (\ref{eq:V-phi-cosine}).
The upper panel shows the case $\frac{\rho_b}{\rho_a}=10^{-5}$ and the lower panel the case
$\frac{\rho_b}{\rho_a}=10^{-8}$.
From the left and turning clockwise, the constraints that delimit the allowed domain
are associated with the condition of formation before the matter-radiation equality,
the classicality condition, and the parametric-resonance condition.
The lower black dashed line is the condition for gravity to be negligible during the
parametric-resonance stage, until the density fluctuations reach the nonlinear regime.
The black dotted line labeled ``$E_{\rm grav} = E_{\rm Q}$'' is the boundary condition
(\ref{eq:Mi-grav-cos}).
Thus, the triangular parameter space is split into two allowed regions, separated by this transition line.
Nonlinear scalar-field solitons in the yellow domain to the right of this line
are governed by the balance between the quantum pressure and the self-interactions,
whereas solitons in the magenta domain to the left of this line
are governed by the balance between the quantum pressure and gravity.}
\label{fig_T-m-cos}
\end{figure}

We now study the parameter space of this parametric-resonance model for the formation
of dark matter as scalar clouds.
This is shown in Fig.~\ref{fig_T-m-cos} over the $(T_{g\infty},m)$ plane, for the choices
$\frac{\rho_b}{\rho_a} = 10^{-5}$ and $\frac{\rho_b}{\rho_a} = 10^{-8}$.
As compared with the tachyonic scenario considered in Sec.~\ref{sec:constraints},
the background temperature $T_{g\infty}$ at the formation time (when the parametric resonance
is in full swing) plays the same role as $T_{c_s}$ (when the tachyonic instability appeared).
The ratio $\frac{\rho_b}{\rho_a}$ [which sets the magnitude of the self-interactions as compared
with the quadratic term in the scalar-field potential $V(\phi)$, see Eqs.(\ref{eq:V-phi-cosine})
and (\ref{eq:rhoa-rhob})], plays the same role as $\Phi_{\rm I_{c_s}}$ (which also set the relative
magnitude of the self-interactions in the tachyonic model).

As in the tachyonic scenario, we require that the scalar clouds form before the time of
matter-radiation equality, in order to recover a standard CDM scenario at low redshifts.
Therefore, we impose the lower bound
\be
T_{g\infty} > T_{\rm eq} , \;\;\; \mbox{with} \;\;\; T_{\rm eq} \simeq 1 \, {\rm eV} ,
\label{eq:T-eq-cos}
\ee
which is shown by the red dot-dashed line labeled ``$T_{\rm eq}$'' on the left in
Fig.~\ref{fig_T-m-cos}.

We also have further theoretical self-consistency conditions.
Again, we must satisfy the condition $m \gg H$, so that the slow-roll stage ends much before
the formation of the scalar clouds and our nonrelativistic analysis is valid, far inside the
oscillatory stage of the scalar field $\phi$ at the bottom of its mainly quadratic potential.
Using Eqs.(\ref{eq:tg-infty-def}), (\ref{eq:rho-cs-rhob}), and (\ref{eq:tau-g}), we obtain the
useful relations
\be
\frac{H_{c_s}}{m} \sim \left( \frac{H_{g\infty}}{m} \right)^{17/57}
\left( \frac{\rho_b}{\rho_a} \right)^{16/57}  ,
\label{eq:H-cs-H-ginfty}
\ee
and
\ba
| \tau_{g\infty} | & \sim & \left( \frac{\rho_b}{\rho_a} \right)^{2/3}
\left( \frac{m}{H_{g\infty}} \right)^{2/3} \nonumber \\
& \sim & 3^{1/3}  \left( \frac{\rho_b}{\rho_a} \right)^{2/3}
\frac{m^{2/3} M_{\rm Pl}^{2/3}}{T_{g\infty}^{4/3}} .
\label{eq:tau-ginfty-Hginfty}
\ea
The relations (\ref{eq:H-cs-H-ginfty}) and (\ref{eq:tau-ginfty-Hginfty}) show that the conditions
$|\tau_{g\infty}| \gg 1$ and $\frac{\rho_b}{\rho_a} \ll 1$ automatically ensure $m \gg H_{g\infty}$
and $m \gg H_{c_s}$. Therefore, the condition $m \gg H$ is automatically satisfied,
once the parametric-resonance condition (\ref{eq:tau-large}), $|\tau_{g\infty}| \gg 1$, is verified.
Using Eq.(\ref{eq:tau-ginfty-Hginfty}), this gives the condition
\be
|\tau_{g\infty}| \gg 1 : \;\;\;
m \gg \left( \frac{\rho_b}{\rho_a} \right)^{-1} \frac{T_{g\infty}^{2}}{\sqrt{3} M_{\rm Pl}} .
\label{eq:m-Tginfty-lower-cos}
\ee
This is shown by the orange solid line labeled ``$|\tau_{g\infty}| \gg 1$'' in
Fig.~\ref{fig_T-m-cos}.
Here, we take a factor $10^3$ to ensure that the left and right hand sides in
Eq.(\ref{eq:m-Tginfty-lower-cos}) are separated by at least three orders of magnitude.

Second, the classicality condition (\ref{eq:classical})  provides an upper bound
on the scalar mass $m$,
\be
m \ll \rho_{g\infty}^{1/4} \, v_{\rm NL}^{-3/4} .
\label{eq:scalar-m-upper-cosine-vNL}
\ee
From Eq.(\ref{eq:v-NL-cosine}) we obtain
\be
v_{\rm NL} \sim \left( \frac{\rho_b}{\rho_a} \right)^{1/2} | \tau_{g\infty} |^{-1/4} \ll 1 ,
\ee
which shows that velocities are indeed nonrelativistic.
Then, Eq.(\ref{eq:scalar-m-upper-cosine-vNL}) gives
\be
\frac{\rho}{m^4 v^3} \gg 1: \;\;\;
m \ll \left( \frac{\rho_b}{\rho_a} \right)^{-2/7}  \frac{ M_{\rm Pl}^{2/7} H_0^{1/7} T_{g\infty}^{4/7}}
{3^{1/7} \Omega_{\gamma 0}^{3/14}} .
\label{eq:m-Tginfty-upper-cos}
\ee
This is shown by the green solid line labeled ``$\rho \gg m^4 v^3$'' in Fig.~\ref{fig_T-m-cos}.
Here, we again take a factor $10^3$ to ensure the left and right hand sides
are separated by at least three orders of magnitude.

Third, we assumed that gravity is negligible during the initial growth of the scalar-field fluctuations.
The equation of motion (\ref{eq:delta-cosmo}) shows that this is satisfied if
$4\pi {\cal G} \bar\rho \ll c_s^2 k^2/a^2$.
From Eq.(\ref{eq:cs2-I-cosine-cos}) the self-interaction contribution to the squared
sound speed is $\left. c_s^2\right|_{\rm I} \sim \frac{\rho_b}{\rho_a}
\left( \frac{\bar\rho}{\rho_b}\right)^{-1/4}$.
At time $t_{g\infty}$, for density $\rho_{g\infty}$ and wave number $k_g$, this condition gives
\be
| \Phi | \ll | \Phi_{\rm I_{g\infty}} | : \;\;\;
m \gg \left( \frac{\rho_b}{\rho_a} \right)^{-1} \frac{ H_0^{3/8} T_{g\infty}^{5/4} }
{3^{5/16} \Omega_{\gamma 0}^{9/16} M_{\rm Pl}^{5/8} }.
\label{eq:gravity-small-cos}
\ee
This corresponds to the black dashed line labeled ``$| \Phi | \ll | \Phi_{\rm I_{g\infty}} |$''
in Fig.~\ref{fig_T-m-cos}.
We can see that it is automatically verified when the previous conditions are satisfied.
As expected, Eq.(\ref{eq:gravity-small-cos}) coincides with the condition (\ref{eq:MNL-grav-cos})
that ensures that gravity is still negligible at the entry into the nonlinear regime.

Thus, as shown in Fig.~\ref{fig_T-m-cos}, the parameter space of the model takes the form
of a triangle in the $(T_{g\infty},m)$ plane. It is delimited by the background temperature
$T_{\rm eq}$ at matter-radiation equality, (\ref{eq:T-eq-cos}), the parametric-resonance condition
(\ref{eq:m-Tginfty-lower-cos}), and the classicality condition (\ref{eq:m-Tginfty-upper-cos}).
The requirement that gravity remains small during the formation process,
(\ref{eq:gravity-small-cos}), is automatically satisfied.
Thus, we can see that the scalar-field mass spans the range
\be
10^{-28} \, {\rm GeV} \lesssim m \lesssim 10 \; {\rm GeV} ,
\ee
while the background temperature at the redshift $z_{g\infty}$ covers the range
\be
1 \, {\rm eV} \lesssim T_{g\infty} \lesssim 10^6 \, {\rm GeV} .
\ee
As for the tachyonic scenarios, this gives a wide range of temperatures and masses
in the allowed parameter space.

In contrast with the tachyonic scenarios, although gravity is always negligible
during the parametric-resonance stage, where the density fluctuations grow until they reach
the nonlinear regime, gravity can become dominant in the final solitons that form
after the nonlinear collapse and the relaxation towards the soliton scaling laws.
This is the new phenomenon studied in Sec.~\ref{sec:solitons-gravity}:
for scalar masses above the threshold (\ref{eq:Mi-grav-cos}) the clumps formed at the end of
the nonlinear stage are dominated by gravity.
The transition between the regimes where gravity is negligible
or dominant with respect to the self-interactions in the final clumps is shown by the
black dotted line labeled ``$E_{\rm grav} = E_{\rm Q}$'', given by Eq.(\ref{eq:Mi-grav-cos}).
This divides the triangle of the allowed parameter space in the $(T_{g\infty},m)$ plane
in two parts.
In the right part, shown by the yellow shaded area, the final solitons are governed by the balance
between the quantum pressure and the self-interactions.
In the left part, shown by the magenta shaded area, the final solitons are governed by the balance
between the quantum pressure and their self-gravity.

We can also check that the scalar-field clumps do not form black holes.
We have seen in Sec.~\ref{sec:BH-self-interactions} that this is guaranteed by
Eq.(\ref{eq:Phi-small-cos-self-inter}) when the solitons are governed by the balance
between the self-interactions and the quantum pressure, i.e. to the right of the
black dotted line ``$E_{\rm grav} = E_{\rm Q}$'' in Fig.~\ref{fig_T-m-cos}.
For models to the left of this transition line, the soliton self-gravity dominates over the
scalar-field self-interactions and their gravitational potential is given by
Eq.(\ref {eq:Phi-small-cos-grav}).
The latter remains small provided we have:
\be
\Phi \ll 1 \;\;\; \mbox{for} \;\;\; m \gg \frac{H_0^{1/4} T_{g\infty}^{3/2}}
{ \Omega_{\gamma 0}^{3/8} M_{\rm Pl}^{3/4} } .
\label{eq:Phi-small-BH-cos}
\ee
We again checked that this boundary line is much below the shaded area in Fig.~\ref{fig_T-m-cos}.
Therefore, over all the allowed parameter space the scalar-field clumps do not collapse
into black holes.

}

{

\subsection{Mass and size of the scalar clumps}
\label{sec:mass-size-clumps-cos}

\begin{figure}
\begin{center}
\epsfxsize=8.8 cm \epsfysize=6.2 cm {\epsfbox{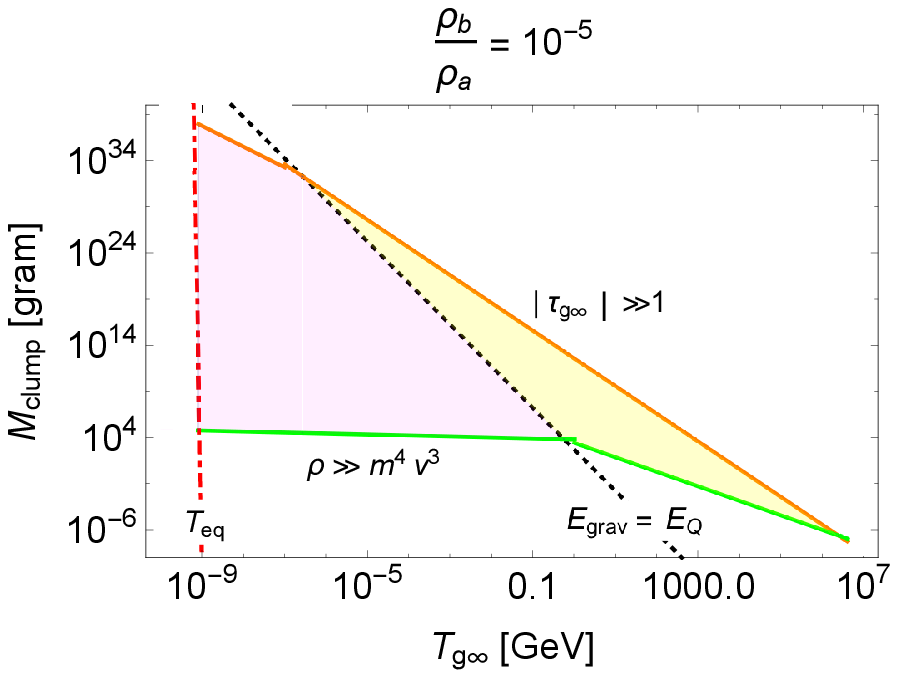}}\\
\epsfxsize=8.8 cm \epsfysize=6.2 cm {\epsfbox{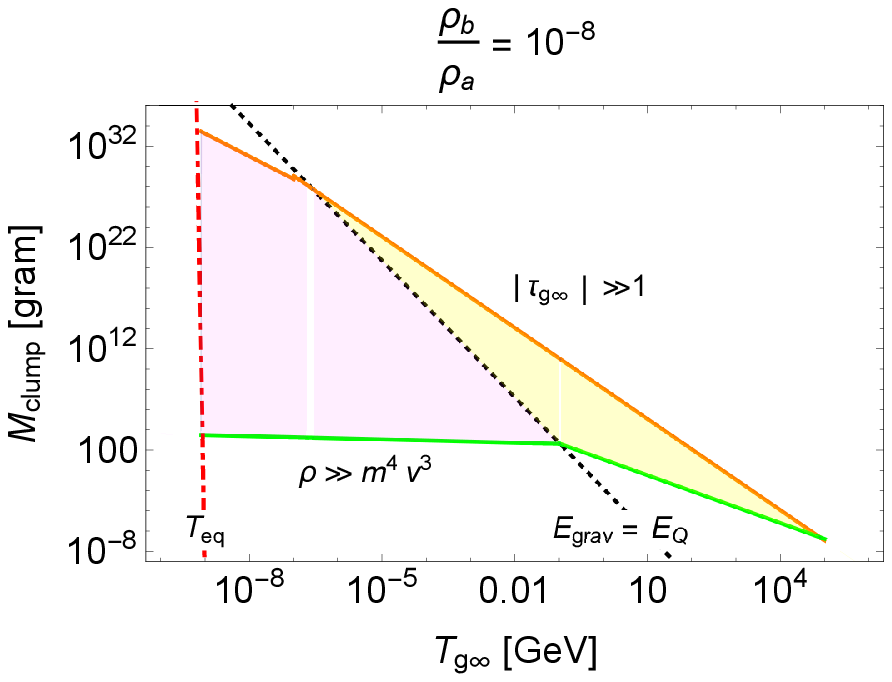}}
\end{center}
\caption{
Mass of the clumps as a function of the background temperature $T_{g\infty}$
at the peak of the parametric resonance, for $\frac{\rho_b}{\rho_a}=10^{-5}$ (upper panel)
and $\frac{\rho_b}{\rho_a}=10^{-8}$ (lower panel).
For a given $T_{g\infty}$ there is a wide range of possible clump masses $M_{\rm clump}$.
The yellow and magenta domains, on either side of the black dotted line
``$E_{\rm grav} = E_{\rm Q}$'', correspond to the yellow and magenta domains shown in
Fig.~\ref{fig_T-m-cos}.
}
\label{fig_Tg-M-cos}
\end{figure}

\begin{figure}
\begin{center}
\epsfxsize=8.8 cm \epsfysize=6.2 cm {\epsfbox{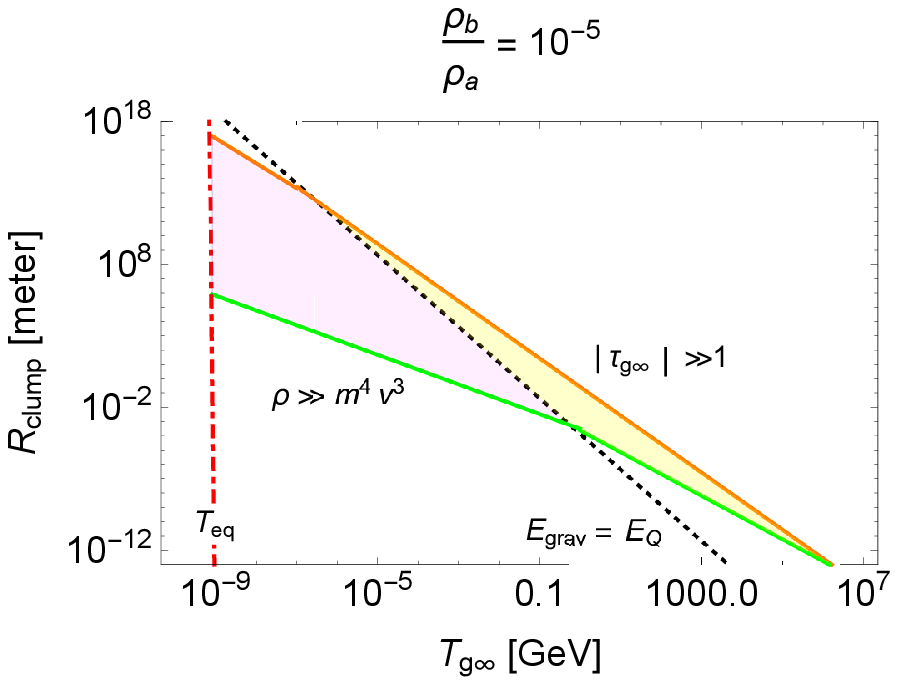}}\\
\epsfxsize=8.8 cm \epsfysize=6.2 cm {\epsfbox{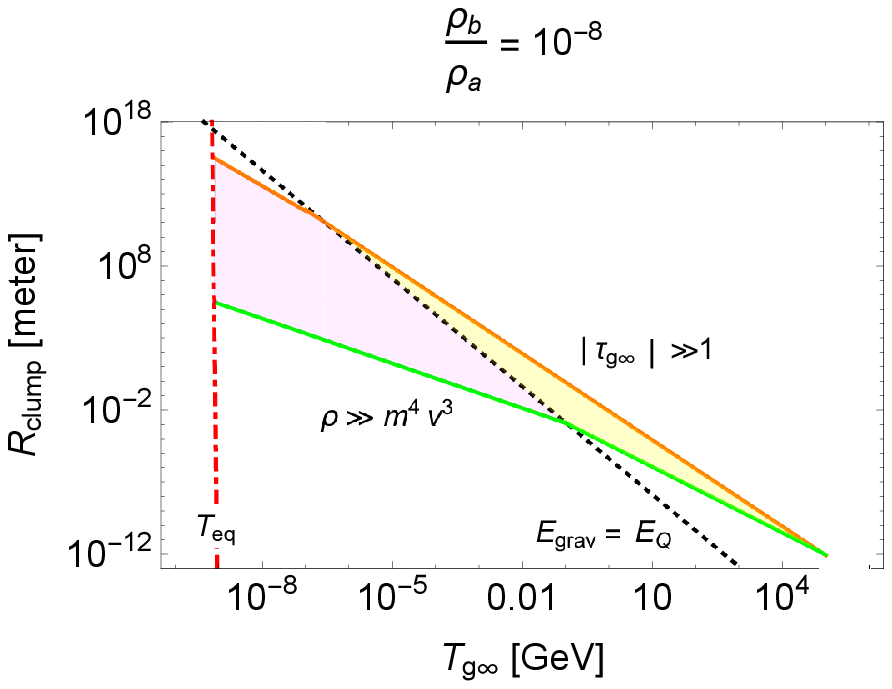}}
\end{center}
\caption{
Radius of the clumps as a function of the background temperature $T_{g\infty}$
at the peak of the parametric resonance, for $\frac{\rho_b}{\rho_a}=10^{-5}$ (upper panel)
and $\frac{\rho_b}{\rho_a}=10^{-8}$ (lower panel).
For a given $T_{g\infty}$ there is a wide range of possible clump radii $R_{\rm clump}$.
The yellow and magenta domains, on either side of the black dotted line
``$E_{\rm grav} = E_{\rm Q}$'', correspond to the yellow and magenta domains shown in
Figs.~\ref{fig_T-m-cos} and \ref{fig_Tg-M-cos}.}
\label{fig_Tg-R-cos}
\end{figure}

In the regime dominated by the self-interactions, the mass and the size of the
solitons formed at the end of the nonlinear stage are given by (\ref{eq:Rinfty-cosine}).
This yields}
\ba
&& \mbox{negligible gravity:} \;\;\;
M_{\rm clump} \sim \left( \frac{\rho_b}{\rho_a} \right)^{1/2}
\frac{3^{1/4} M_{\rm Pl}^{5/2} H_0^{1/2}}{\Omega_{\gamma 0}^{3/4} m T_{g\infty}} ,
\nonumber \\
&& R_{\rm clump} \sim \left( \frac{\rho_b}{\rho_a} \right)^{1/6}
\frac{3^{1/3} M_{\rm Pl}^{2/3}}{m^{1/3} T_{g\infty}^{4/3}} .
\label{eq:Mclump-cos-1}
\ea
{
In the regime dominated by the self-gravity, the mass and the size of the
solitons formed at the end of the nonlinear stage are given by
Eqs.(\ref{eq:rho-M-fuzzy})-(\ref{eq:Mi-grav}).
This yields}
\ba
&& \mbox{negligible self-interactions:} \;\;\;
M_{\rm clump} \sim \frac{M_{\rm Pl}^{13/8} H_0^{1/8} T_{g\infty}^{3/4}}
{(3\Omega_{\gamma 0})^{3/16} m^{3/2}} ,\nonumber \\
&& R_{\rm clump} \sim \frac{(3\Omega_{\gamma 0})^{3/16} M_{\rm Pl}^{3/8}}
{m^{1/2} H_0^{1/8} T_{g\infty}^{3/4}} .
\label{eq:Mclump-cos-2}
\ea

{
In contrast with the tachyonic case studied in Sec.~\ref{sec:mass-size-clumps},
the mass and size of the clumps depend on the scalar mass $m$, in addition to the
background temperature $T_{g\infty}$.
Therefore, there is a finite range of clump mass and radius for a given $T_{g\infty}$,
as displayed in Figs.~\ref{fig_Tg-M-cos} and \ref{fig_Tg-R-cos}.
We obtain a deformed triangular domain, which corresponds to the domain of parameter space
shown in Fig.~\ref{fig_T-m-cos}.
Its boundaries are again set by the background temperature
$T_{\rm eq}$ at matter-radiation equality, (\ref{eq:T-eq-cos}), the parametric-resonance condition
(\ref{eq:m-Tginfty-lower-cos}), and the classicality condition (\ref{eq:m-Tginfty-upper-cos}),
as labeled in the figure.
The black dotted line labeled ``$E_{\rm grav} = E_{\rm Q}$'' again divides the allowed domain
into a region where self-gravity is negligible (to the right of this transition line) and a region
where it is dominant (to the left).
The slope of the upper and lower boundaries differs on either side of the transition
because the clump mass and radius are either given by Eq.(\ref{eq:Mclump-cos-1}) or by
Eq.(\ref{eq:Mclump-cos-2}).

As for the tachyonic case shown in Fig.~\ref{fig_M-Tcs-poly}, we find that the clumps cover
a huge range of masses and radii, from microscopic to subgalactic scales.
Thus, their mass goes from $10^{-6} \, {\rm gram}$ up to
$10^{36} \, {\rm gram} \sim 10^3 \, M_\odot$, and their radius from $0.01 \, {\rm angstrom}$
to $0.1 \, {\rm parsec}$.
Again, the largest clumps are similar to galactic molecular clouds and do not correspond to the
standard stellar-mass MACHOs (massive compact halo objects).

Because the largest radius obtained in Fig.~\ref{fig_Tg-R-cos} is slightly below $1 \, {\rm pc}$,
the condition $R_{\rm clump} < R_{\max}$ is automatically satisfied for
$R_{\max} = 1 \, {\rm pc}$. This is why we did not plot this condition in Fig.~\ref{fig_T-m-cos}.

}

{

\subsection{Evading microlensing constraints}
\label{sec:micro-lensing-cos}

\begin{figure}
\begin{center}
\epsfxsize=8.8 cm \epsfysize=6. cm {\epsfbox{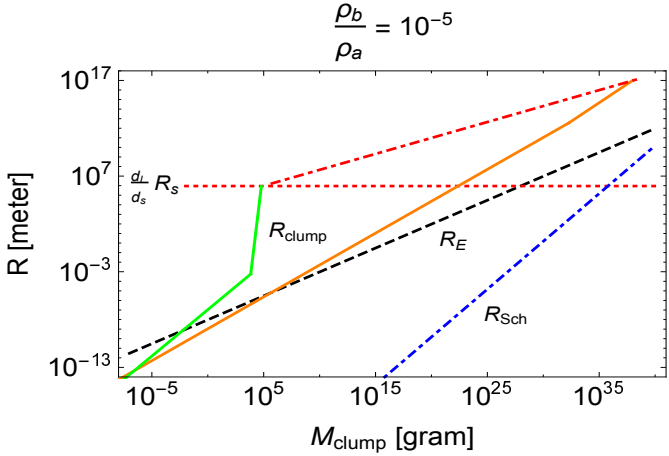}}
\end{center}
\caption{Characteristic radii in the lens plane. The range of allowed clump radii for a given
mass $M_{\rm clump}$ is the domain labeled ``$R_{\rm clump}$''
delimited by the orange, green, and red dot-dashed curves.
We also show the Schwarzschild radius $R_{\rm Sch}$ (blue dot-dashed line),
the Einstein radius $R_E$ (black dashed line), and the outer
impact parameter $\frac{d_L}{d_s} R_s$ of a source of one solar radius aligned with the lens
(red dotted line). We take $d_L = 1 \, {\rm kpc}$ and $d_s = d_{\rm M31} \simeq 770 \, {\rm kpc}$.}
\label{fig_M-R-cos}
\end{figure}

\begin{figure}
\begin{center}
\epsfxsize=8.8 cm \epsfysize=6. cm {\epsfbox{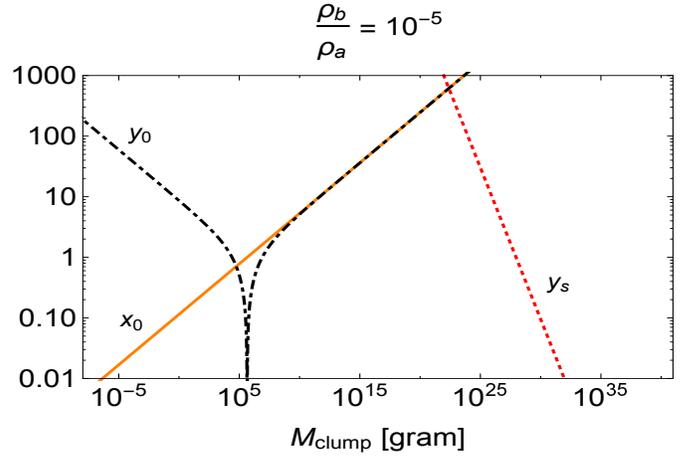}}
\end{center}
\caption{
Lensing radii normalized to the Einstein radius, in the lens plane.
We show the normalized scalar-clump radius $x_0$ associated with
the smallest clumps of a given mass in the allowed domain (orange solid line),
and the associated quantity $y_0=| x_0-1/x_0|$ (black dot-dashed line).
The red dotted line is the source radius $y_s$ for a star of one solar radius.
We take $d_L = 1 \, {\rm kpc}$ and $d_s = d_{\rm M31} \simeq 770 \, {\rm kpc}$.}
\label{fig_lens-cos}
\end{figure}

As for the tachyonic scenario, we now check whether these scalar clumps can be detected
through microlensing observations.
Considering again the lensing of a distant star of one solar radius in M31,
at $d_s = M_{31} \simeq 770 \, {\rm kpc}$, by a clump located in the Milky Way at
$d_L=1 \, {\rm kpc}$, we show in Fig.~\ref{fig_M-R-cos} the characteristic radii in the lens plane
as a function of the clump mass $M_{\rm clump}$.
The source projected radius, $\frac{d_L}{d_s} R_s \sim 10^6 \, {\rm meter}$ and
the Einstein radius $R_E$ of Eq.(\ref {eq:R_E-def}) are the same
as in Fig.~\ref{fig_M-R-poly}.
We also show the Schwarzschild radius $R_{\rm Sch}$ of Eq.(\ref{eq:R-Sch-def}).
As in Figs.~\ref{fig_Tg-M-cos} and \ref{fig_Tg-R-cos}, there is now a finite range of
clump radii for a given clump mass. The finite allowed parameter space shown in
Fig.~\ref{fig_T-m-cos} translates into the orange, green and red dot-dashed line that
enclose the label ``$R_{\rm clump}$''.
The breaks in the orange and green boundary lines, associated with the parametric-resonance
and classicality conditions (\ref{eq:m-Tginfty-lower-cos}) and (\ref{eq:m-Tginfty-upper-cos}),
are due to the transition from the self-interaction regime (\ref{eq:Mclump-cos-1}) to the
self-gravity regime (\ref{eq:Mclump-cos-2}).
The red dot-dashed curve, associated with the constraint (\ref{eq:T-eq-cos}) on the background
temperature, does not show any break because it is fully in the self-gravity regime,
see Fig.~\ref{fig_T-m-cos}.

As compared with the tachyonic case displayed in Fig.~\ref{fig_M-R-poly},
we obtain similar clump masses and radii, but with the line $R_{\rm clump}$ of
Fig.~\ref{fig_M-R-poly} being thickened towards higher radii into a finite-size band.

First, we can see that the clump radii are always much greater than the Schwarzschild radius
$R_{\rm Sch}$. This confirms that the clumps do not form black holes, in agreement
with the analysis of Sec.~\ref{sec:no-BH-cos} and Eq.(\ref{eq:Phi-small-BH-cos}).

Second, the comparison with Fig.~\ref{fig_M-R-poly} shows that we have the same lensing
properties as in the tachyonic case.
The Einstein radius $R_E$ is always much smaller than either the projected source radius,
$\frac{d_L}{d_s} R_s \sim 10^6 \, {\rm meter}$, or the lens radius, $R_{\rm clump}$.
This implies that gravitational lensing effects are very small.
The strongest lensing effects are obtained for clump radii along the lower boundary of the
allowed domain, the orange curve associated with the parametric-resonance condition
$|\tau_{g\infty}| \gg 1$ in Fig.~\ref{fig_T-m-cos}. Indeed, this minimizes the decrease
of the lensing magnification due to finite-lens effects.
We show in Fig.~\ref{fig_lens-cos} the normalized lensing radii $x_0$ and $y_0$
obtained along this lower boundary of the clump-radius domain.
We also plot the normalized source radius $y_s$.
We can see that we have the same configuration as in Fig.~\ref{fig_lens-poly}.
At low clump masses, where $x_0 < 1$, we have $y_s \gg y_0$ and $y_s \gg 1$;
using Eq.(\ref{eq:x0-m1-ys-large}) this gives again $\bar\mu_0 \simeq 1$.
At intermediate clump masses, we have $x_0>1$, $y_s > y_0$ and $y_s \gg 1$;
using Eq.(\ref{eq:x0-p1-ys-large}) this also gives $\bar\mu_0 \simeq 1$.
At large clump masses, we have $x_0 \gg 1$ and $y_s < y_0$;
using Eq.(\ref{eq:x0-p1-ys-small}) this gives again $\bar\mu_0 \simeq 1$.
We found by a numerical computation that $\bar\mu_0 - 1 \ll 10^{-6}$ over all clump masses.

Thus, as for the clumps formed in the tachyonic scenario, the clumps formed in the
parametric-resonance scenario cannot be detected by microlensing.
Again, at low clump masses this is due to the finite size of the source,
which also prevents the detection of low-mass black holes, while at large masses this
is due to the large size of the lens. In this regime, they are much bigger than both the
Schwarzschild and the Einstein radii; these large clumps are similar to galactic molecular
clouds, rather than compact objects, with shallow gravitational potential wells.

}

\subsection{Discussion}

The model we have described in this section is formally equivalent to the one discussed in \cite{Berges:2019dgr} for a different range of parameters. It is relevant to define
\be
\kappa= \frac{M_I^4}{m_0^2 f^2} \simeq 16 \frac{\rho_b}{\rho_a} \ll 1 .
\ee
The regime described in \cite{Berges:2019dgr} corresponds to $\kappa \gtrsim 1$, where a rapid growth of the perturbations and the nonlinear evolution of the scalar field have been studied using numerical simulations. The formation of clumps has been observed and the consequences for structure formation analysed. In this paper, we conduct a similar analysis in the $\kappa\ll 1$ regime. If the argument of the cosine interaction term were small, the model would reduce to the tachyonic instability case that we treated in the first part of the paper. On the contrary, as the argument of the cosine term $\tau_{g\infty} \gg 1$ is large, this regime is never attained and a parametric-resonance phase sets in first. In this case, the instability is slow initially and a long period of acoustic oscillations takes place before the onset of the parametric-resonance instability. Subsequently, we find that the result of this instability can only be the formation of solitons maintained in an equilibrium state by either the self-interactions or gravity. As our treatment is only analytical, we have no description of the intermediate steps, which we plan to investigate numerically in the future. Technically, we have obtained our description of the instabilities using the nonrelativistic approximation of the scalar-field dynamics. This should give an accurate picture as the velocities of the matter perturbations in the linear regime, up to its limit, are small. Similarly, the solitons are stable configurations where the fluid is at rest. In the intermediate regime, relativistic effects might be at play and a full numerical investigation needs to be performed. This is left for future work.

\section{Conclusion}
\label{sec:Conclusion}

We have studied the formation of clumps in scalar-field models of dark matter.
These small clumps form at very high redshift, in the radiation era, and could
be a candidate for the dark matter (in a manner similar to primordial black holes or
small compact objects).
We have explicitly worked in the nonrelativistic regime, where the homogeneous background density of dark matter is realised in the form of rapid oscillations of the scalar field around the origin and self-interactions appear as small corrections to this background behavior. In the nonrelativistic regime, the dark-matter field can be described by a fluid with non-trivial pressure. The pressure comprises two terms. The first one originates from the kinetic terms of the scalar field and appears in the nonrelativistic description as a so-called quantum pressure. The second is due to the self-interactions and leads to a pressure term that is a function of the scalar-field energy density. We have shown that the fluid equations, in particular the Euler equation, develop unstable behaviors when the effective speed of sound squared becomes negative.

We have envisaged two scenarios. In the first one, the speed of sound squared becomes negative below a certain energy density, resulting in a tachyonic instability. At the field-theory level, this instability appears when the quartic term of the field potential is negative. This is similar to the case of the axions where the cosine potential changes convexity at large enough values of the field. For axions, this implies that perturbations of the scalar field have a tachyon instability at large values of the field along its background oscillations, leading to the formation of axitons. Here, we show that a tachyonic instability due to the negative quartic interaction term in the potential is present in the nonrelativistic regime, where the oscillations of the scalar field are still almost harmonic. The resulting growth of the density contrast for the scalar energy density shows a fast exponential growth, which leads to a nonlinear regime where clumps with a non-trivial spherical profile emerge. These solitons have a well-defined density, which depends on the scalar potential of the scalar field. As a result, the clumps have a mass-radius relationship of the $M\sim R^3$ type.

A second scenario appears for axion monodromy models, where a dominant quadratic term for the scalar field is perturbed by cosine interactions. In this case, the density contrast shows a parametric-resonance instability and grows after a period of acoustic oscillations governed by the quantum pressure. In the nonlinear regime, the corresponding solitons have spherical profiles with a mass and a radius that are continuously distributed above a minimum mass threshold and obey a scaling law
that follows from the balance between the self-interactions and the quantum pressure. This results in a mass-radius relationship $M\sim R^5$.  This is reminiscent of the formation of oscillons in the relativistic regime, where a delayed formation occurs before parametric resonance takes place. Interestingly, despite gravity being always negligible during the formation mechanism,
for small scalar mass and low formation redshift, gravity can eventually dominate the final relaxation
towards the highly nonlinear solitons, and hence the properties of the scalar clumps after the aggregation phase. In this case, the mass-radius relationship is in $M\sim 1/R$.  As a result, the final scalar clumps in the axion monodromy case can be governed by the balance of the quantum
pressure with either the scalar self-interactions or the clump self-gravity.

We have been able to give an analytic description of the formation of scalar clumps using both linear and nonlinear arguments. The linear analysis shows that the instability due to the negative values of the speed of sound squared is always at the origin of the clumps considered here. We have also solved numerically for the nonlinear profiles of the final collapsed objects, which must satisfy the equations of hydrostatic equilibrium.
In the case of the tachyonic instability, we also present in appendix~\ref{sec:thermo}
a thermodynamical analysis that confirms the fragmentation of the system towards
highly inhomogeneous configurations, with clumps at the characteristic density $\rho_\Lambda$.
However, we have not followed the detailed relaxation from the entry into the nonlinear regime towards these stable spherical configurations. This would require numerical simulations which go beyond the present work and are left for future studies.

{ We have computed the allowed parameter space of these models and found that
the formation redshift and the scalar-field mass span many orders of magnitude,
$10^{-26} \, {\rm GeV} \lesssim m \lesssim 10 \, {\rm GeV}$.
The dark-matter clumps formed by the scalar-field solitons also cover a huge range of scales,
much beyond the usual MACHOs, as we find
$10^{-3} \, {\rm gram} \lesssim M_{\rm clump} \lesssim 10^3 M_\odot$
and $0.01 \, {\rm angstrom} \lesssim R_{\rm clump} \lesssim 1 \, {\rm parsec}$.
Thus, they run from the size of atoms to that of galactic molecular clouds.
Because of finite-source and finite-lens effects, we found that these dark-matter clumps
are far below the detection thresholds of microlensing observations.}

Scalar clumps are particularly interesting as they would be amenable to new tests of dark matter \cite{Arvanitaki:2019rax}.
For instance, the creation of the clumps in the nonlinear regime could lead to the emission of gravitational waves \cite{Chatrchyan:2020pzh}. Their existence could even be detected by the ultra-sensitive detectors of gravitational wave experiments\cite{Jaeckel:2020mqa}. In the future, we intend to perform a more thorough investigation of the dynamics of nonrelativistic clump formation using numerical methods \cite{Amin:2019ums}.

\appendix

\section{Thermodynamics}
\label{sec:thermo}

In Secs.~\ref{sec:perturbations} and \ref{sec:scalar-field-clumps} in the main text,
we have described the dynamics leading to the formation of scalar-field clumps using
a three-pronged approach. We have first studied the tachyonic linear instabilities leading to the
nonlinear regime, where the system can develop strong inhomogeneities.
Second, we have obtained stable static equilibrium configurations. Third, we have described
the aggregation process which yields the final masses and radii of the clumps.
In this appendix, we describe in this polynomial scenario a thermodynamic approach, where the
transition from a smooth background to a strongly inhomogeneous system, associated with the formation of clumps, can be seen as resulting
from the thermodynamics of the dark-matter fluid and its interaction potential $\Phi_{\rm I}$.
A similar analysis may be envisaged for the case of the axion monodromy models.
This is left for future work.

\subsection{Phase diagram}

\begin{figure}
\begin{center}
\epsfxsize=8.8 cm \epsfysize=5.6 cm {\epsfbox{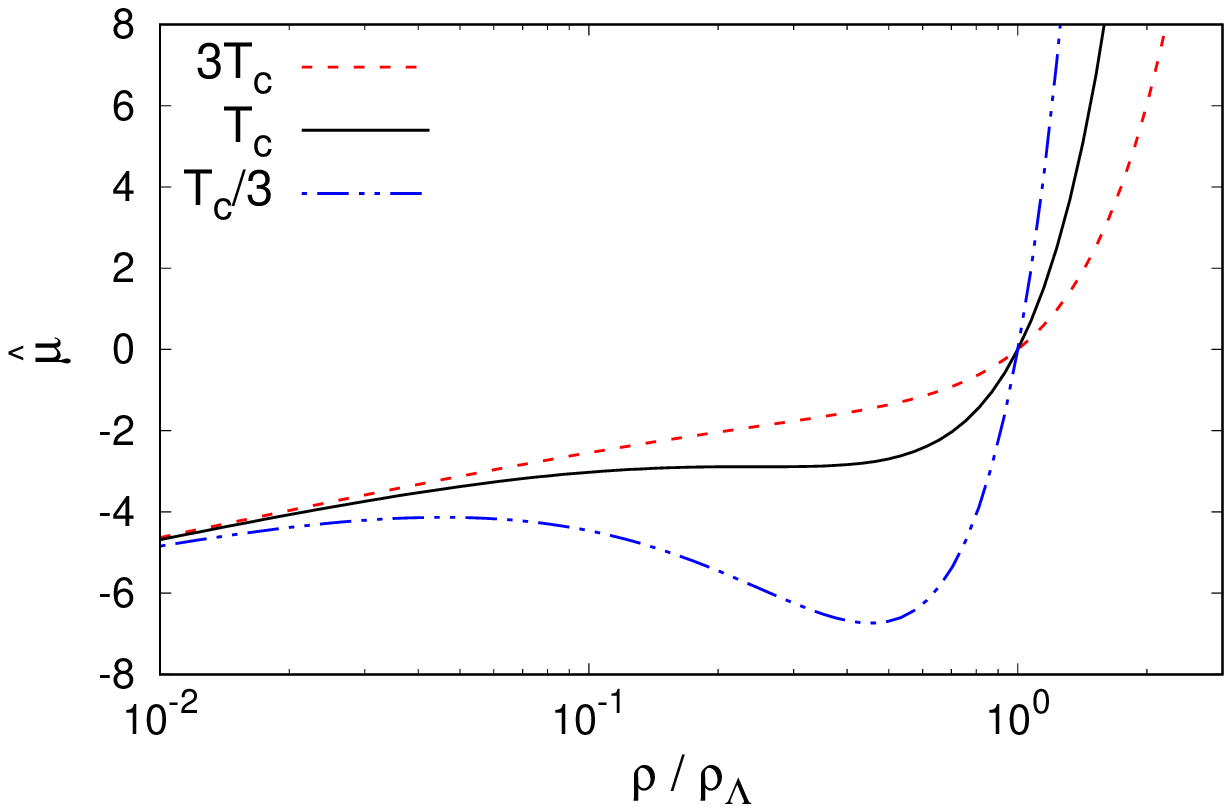}}
\epsfxsize=8.8 cm \epsfysize=5.6 cm {\epsfbox{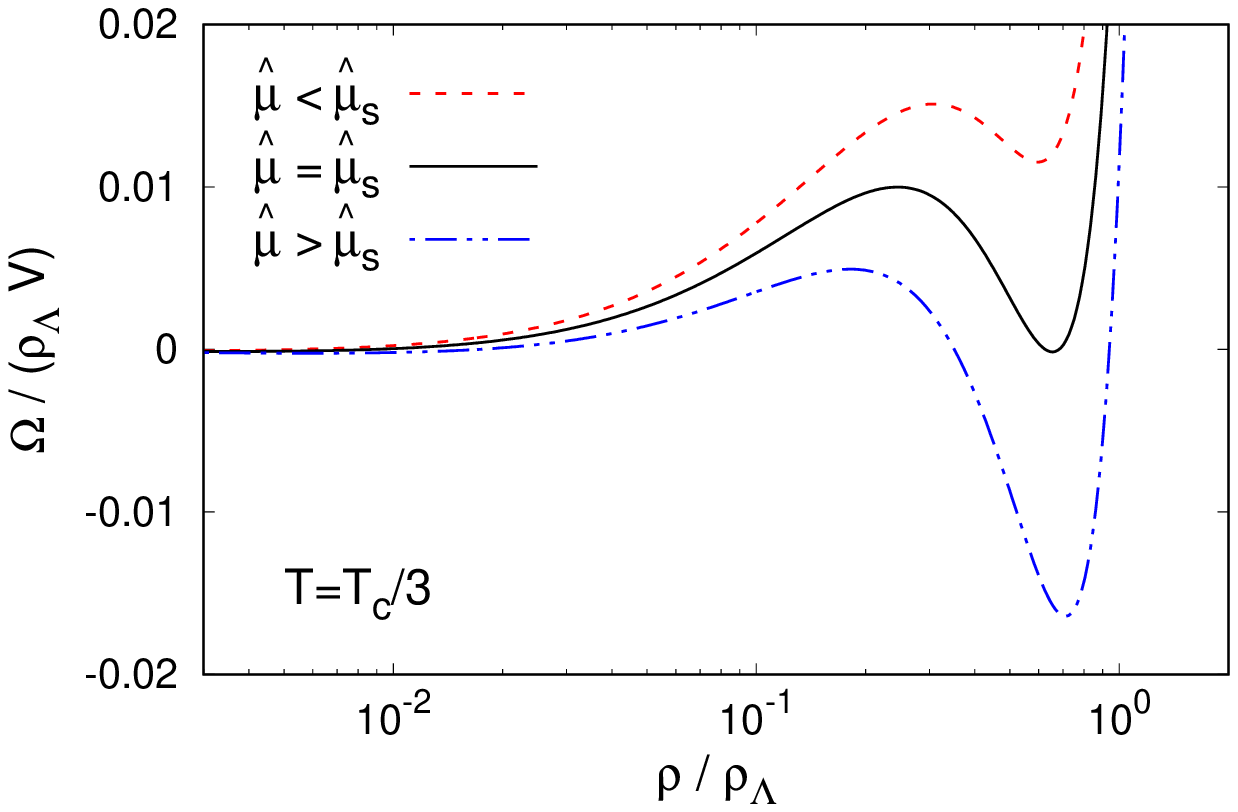}}
\epsfxsize=8.8 cm \epsfysize=5.6 cm {\epsfbox{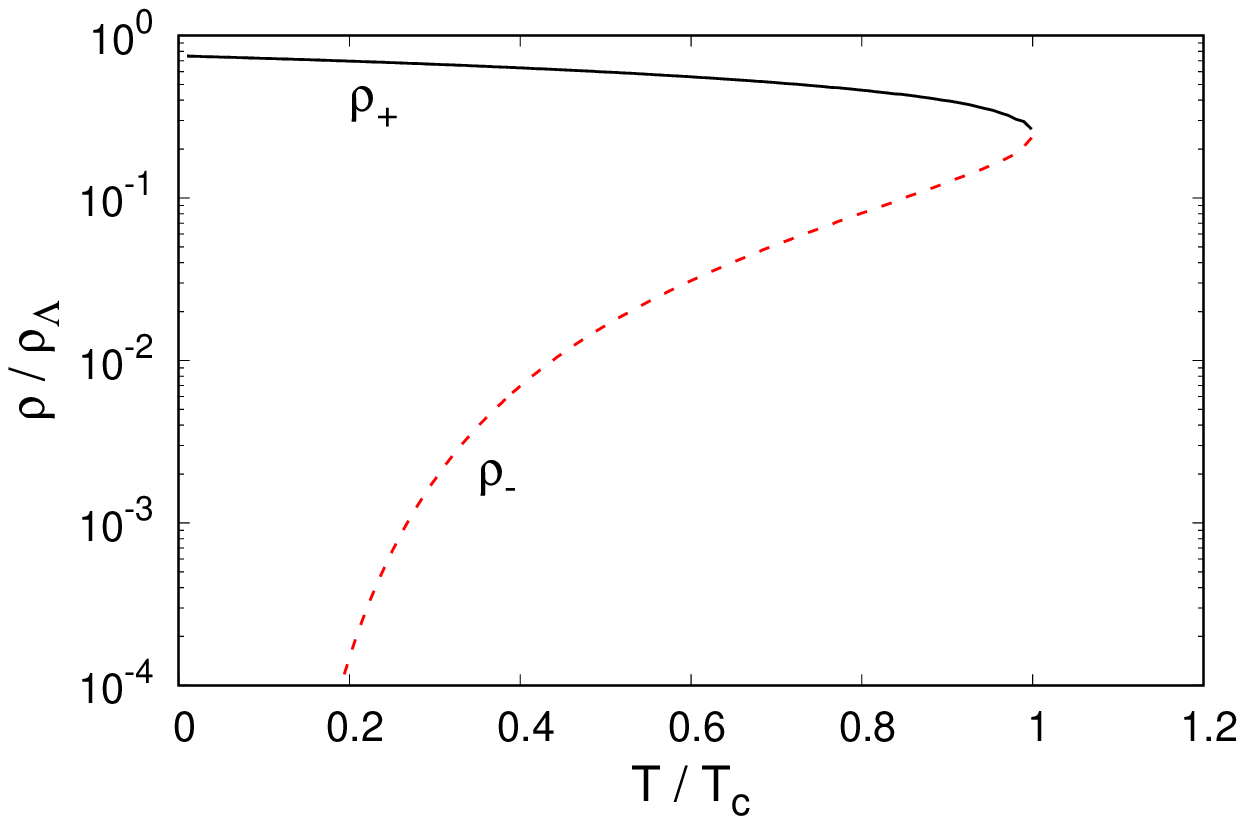}}
\end{center}
\caption{Thermodynamical diagrams for the polynomial case
(\ref{eq:Phi-I-polynomial}) with $c_1=c_2$.
{\it Upper panel:} curve $\hat\mu(\rho)$ at fixed temperature $T$ from Eq.(\ref{eq:mu-hat}),
for $T=3 T_c$, $T_c$ and $T_c/3$.
{\it Middle panel:} grand potential, normalized by $\rho_\Lambda V$, from
Eq.(\ref{eq:Omega-def}), at the low temperature $T=T_c/3$.
{\it Lower panel:} phase diagram in the plane $(T,\rho)$. At low temperature,
$T<T_c$, the system splits in two phases of densities $\rho_+$ and $\rho_-$
if $\rho_-<\bar\rho<\rho_+$.}
\label{fig_thermo}
\end{figure}

We describe here how a thermodynamical analysis shows that structures at the characteristic
density $\rho_{\Lambda}$ should form.
We discard the expansion of the Universe and use the physical coordinate ${\vec r}=a{\vec x}$
as the spatial coordinate.
The total energy $E$, conserved by the continuity and Euler equations, is given by
Eq.(\ref{eq:Etot-def}), which reads
\be
E = \int d{\vec r} \left[ \rho \frac{\vec v^{\,2}}{2} + \frac{1}{2} \rho \Phi + {\cal V}_{\rm I}
+ \frac{(\nabla\rho)^2}{8 m^2\rho} \right] .
\label{eq:E-rho-v-1}
\ee
To analyze the effect of the change of sign of the self-interactions with the density,
we neglect the quantum pressure and gravity, which only come into play at very small
and very large scales. In particular, we have seen in Sec.~\ref{sec:scalar-field-clumps} that
they are negligible for the static equilibrium configurations of interest in the case of isolated
scalar clouds (solitons).
We consider static equilibrium configurations,
with vanishing velocity field, within a given volume $V$.
Then, each state is described by the density field $\rho({\vec r})$.
It is characterized by its mass $M$, energy $E$ and entropy $S$,
\be
M = \int d{\vec r} \, \rho({\vec r}) , \;\;
E = \int d{\vec r} \; {\cal V}_{\rm I}[\rho({\vec r})] ,
\label{eq:E-rho}
\ee
\be
S = - \int d{\vec r} \, \rho({\vec r}) \ln\frac{\rho({\vec r})}{\rho_\Lambda} .
\label{eq:S-rho}
\ee
In the micro-canonical ensemble, the most likely configurations are those that maximise
the entropy at fixed values of the mass and the energy.
This means that we look for the maximum of $S-E/T+(\mu/T) M$, where $1/T$ and $(\mu/T)$
are Lagrange multipliers associated with the energy and mass constraints.
This is equivalent to the minimization of the grand potential $\Omega$ defined by
\be
\Omega = E - T S - \mu M
= \int d{\vec r} \, \left[ {\cal V}_{\rm I} + T \rho \ln\frac{\rho}{\rho_\Lambda}
- \mu\rho \right] .
\label{eq:Omega-def}
\ee
Formally, this also corresponds to the macro-canonical ensemble, where we also look for
the minimum of the grand potential $\Omega$ where $T$ and $\mu$ are the temperature
and the chemical potential.
In our case, because there is no external thermal bath or reservoir of particles,
we work in the micro-canonical ensemble and consider a fixed volume with a given mass
and energy. Then, $T$ and $\mu$ are only Lagrange multipliers.
However, we will refer to $T$ and $\mu$ as the effective temperature and chemical potential
in the following, to simplify the terminology and facilitate the intuition of the behaviors that
we obtain, which follow the standard properties of phase transitions.

The characteristic density $\rho_\Lambda$ is  introduced in Eq.(\ref{eq:S-rho})
to make the argument of the logarithm dimensionless.
Any other density could be used, as its value is irrelevant because it is degenerate
with the Lagrange multiplier $\mu$ in Eq.(\ref{eq:Omega-def}).
The thermodynamical equilibrium is given by the minimum of the grand potential.
This gives
\be
\frac{\delta\Omega}{\delta\rho} = 0 : \;\;\;
\Phi_{\rm I} + T [ \ln(\rho/\rho_\Lambda) +1 ] - \mu =  0 ,
\label{eq:thermo-eq}
\ee
where we neglect gravity and the quantum pressure. In this approximation,
we obtain a local equation in ${\vec r}$, except for the global constraints
associated with the total mass and energy.
It is convenient to introduce a reduced effective  chemical potential $\hat\mu$ by
\be
\hat\mu = \mu/T - 1 ,
\ee
so that the thermodynamical equilibrium reads
\be
\hat\mu = \ln(\rho/\rho_\Lambda) + \Phi_{\rm I} / T .
\label{eq:mu-hat}
\ee
At fixed temperature, this implicit equation determines the density
$\rho({\hat\mu})$ as a function of $\hat\mu$.
We show the curve $\hat\mu(\rho)$ for several values of $T$ in
the upper panel in Fig.~\ref{fig_thermo}, for the polynomial case
(\ref{eq:Phi-I-polynomial}) with $c_1=c_2$.

At large temperature, $T\to \infty$, the self-interactions $\Phi_{\rm I}$ become
negligible; the  grand potential is governed by the entropy. Then,
the function $\hat\mu(\rho)$ is monotonically increasing and
Eq.(\ref{eq:mu-hat}) gives the unique solution $\rho \simeq \rho_\Lambda e^{\hat\mu}$.
This implies a homogeneous system at this density.
At low temperature, $T \to 0$, the self-interactions come into play.
There is still a single solution to Eq.(\ref{eq:mu-hat}), hence a homogeneous equilibrium,
as long as $\hat\mu(\rho)$ remains a monotonic increasing function of $\rho$,
\be
{\rm homogeneous:} \;\;\; 1 + \frac{1}{T} \rho \frac{d\Phi_{\rm I}}{d\rho} > 0 .
\ee
Thus, we recover the condition (\ref{eq:q-up-def}), $\frac{d\Phi_{\rm I}}{d\rho} < 0$,
for instabilities and inhomogeneities to appear.
More precisely, let us consider self-interaction potentials $\Phi_{\rm I}(\rho)$
such that the derivative with respect to $\ln\rho$ has a finite minimum.
Then, the curve $\hat\mu(\rho)$ becomes non-monotonic below the critical temperature
$T_c$ with
\be
T_c = \max\left( - \frac{d\Phi_{\rm I}}{d\ln\rho} \right)  .
\label{eq:betac-def}
\ee
For the polynomial case (\ref{eq:Phi-I-polynomial}) this gives the critical temperature
\be
T_c = \frac{c_1^2}{8 c_2} .
\label{eq:Tc-polynomial}
\ee
As seen in the upper panel in Fig.~\ref{fig_thermo},
for $T < T_c$, there are three solutions $\rho(\hat\mu)$ to the equation
(\ref{eq:mu-hat}), $\rho_- < \rho_m < \rho_+$, when $\hat\mu$ is in the range
$\hat\mu_2<\hat\mu<\hat\mu_1$, where $\hat\mu_1$ and $\hat\mu_2$ are the local
maximum and minimum of the curve $\hat\mu(\rho)$, at densities $\rho_1<\rho_2$.
From Eq.(\ref{eq:mu-hat}), we obtain for the self-interaction potential
(\ref{eq:Phi-I-polynomial})
\be
\frac{\rho_1}{\rho_\Lambda} = \frac{c_1 - \sqrt{c_1^2-8c_2 T}}{4 c_2} , \;\;
\frac{\rho_2}{\rho_\Lambda} = \frac{c_1 + \sqrt{c_1^2-8c_2 T}}{4 c_2} .
\label{eq:rho1-rho2}
\ee
Both $\rho_- < \rho_1$ and $\rho_+>\rho_2$ are local minima of the grand potential
$\Omega$ whereas $\rho_m$ is a local maximum, as seen in the middle panel in
Fig.~\ref{fig_thermo}. The low-density minimum $\rho_-$ is not easily seen in the figure
because the potential $\Omega$ is only slightly below zero.
However, its presence is easily seen from the fact that the grand potential
(\ref{eq:Omega-def}) behaves as
$\Omega \simeq V T \rho \ln\frac{\rho}{\rho_\Lambda}$
at low densities, because ${\cal V}_{\rm I}(\rho)$ goes to zero as $\rho^2$.
This means that $\Omega(\rho)$ is a decreasing function of $\rho$ at low densities,
which implies that there is a local minimum $\rho_-$ in the middle panel in
Fig.~\ref{fig_thermo} at a density below $\rho_\Lambda/10$.
The physical solution is the deepest minimum among $\{\rho_-,\rho_+\}$.
For $\hat\mu \simeq \hat\mu_2$ (upper dashed line in the middle panel in
Fig.~\ref{fig_thermo}), close to the low-density monotonic branch,
this is $\rho_-$, whereas for $\hat\mu \simeq \hat\mu_1$ (lower dot-dashed line),
close to the high-density monotonic branch, this is $\rho_+$.
In-between these two regimes, there is a critical value
$\hat\mu_s$ (solid line), $\hat\mu_2<\hat\mu_s<\hat\mu_1$, where we make the transition from
$\rho_-$ to $\rho_+$ as the values $\Omega_-$ and $\Omega_+$ of the grand potential
cross each other. This gives a first-order phase transition, with a finite density jump
at the critical chemical potential $\hat\mu_s$.
In the limit of low temperature, we can obtain the analytic behavior of $\rho_-$ and $\rho_+$
at the critical chemical potential $\hat\mu_s$ as follows.
From Eqs.(\ref{eq:Omega-def}) and (\ref{eq:thermo-eq}), we find that the grand potential
at equilibrium reads
\be
\Omega_{\rm eq} = V [ {\cal V}_{\rm I} - \rho \Phi_{\rm I} - \rho T ] .
\ee
For the polynomial case (\ref{eq:Phi-I-polynomial}), this gives
\be
\Omega_{\rm eq} = V \left[ - \rho T + \frac{c_1}{2} \frac{\rho^2}{\rho_\Lambda}
- \frac{2 c_2}{3} \frac{\rho^3}{\rho_\Lambda^2} \right] .
\label{eq:Omega-eq-a1a2}
\ee
On the other hand, from Eq.(\ref{eq:rho1-rho2}) we obtain the asymptotic behaviors
\be
T \to 0 : \;\;\; \frac{\rho_1}{\rho_\Lambda} \simeq \frac{T}{c_1} \to 0 , \;\;
\frac{\rho_2}{\rho_\Lambda} \to \frac{c_1}{2 c_2} .
\label{eq:rho1-rho2-low-T}
\ee
Then, from $\rho_-<\rho_1$ we obtain $\rho_-\to 0$ and $\Omega_- \to 0$.
Therefore, at the critical chemical potential $\hat\mu_s$, where $\Omega_-=\Omega_+$,
we also have  $\Omega_+ \to 0$ at low temperature.
From Eq.(\ref{eq:Omega-eq-a1a2}), together with the finite lower bound $\rho_+>\rho_2$,
this implies
$\frac{c_1}{2} \frac{\rho_+}{\rho_\Lambda} - \frac{2 c_2}{3} \frac{\rho_+^2}{\rho_\Lambda^2} \to 0$.
Thus, we obtain the low-temperature asymptotes
\be
T \to 0 : \;\;\; \rho_- \to 0 \;\; \mbox{and} \;\; \rho_+ \to \rho_\infty
\;\; \mbox{at} \;\; \hat\mu_s ,
\label{eq:rho_-_rho_+}
\ee
with
\be
\rho_\infty = \frac{3 c_1}{4 c_2} \rho_\Lambda .
\label{eq:rho-infty}
\ee
We show the densities $\rho_-(T)$ and $\rho_+(T)$ of these two phases
in the lower panel in Fig.~\ref{fig_thermo}. The curves agree with the
asymptotic limits (\ref{eq:rho_-_rho_+}).

Therefore, at high temperature, $T>T_c$, the system is homogeneous with the
density $\rho=M/V$. At low temperature, $T<T_c$, the system shows a phase transition
with a coexistence of two phases at densities $\rho_-<\rho_+$, with the chemical potential
given by the critical value $\hat\mu_s(T)$ (to coexist the two phases must have the
same value of $\Omega$).
Thus, if the mean density $\bar\rho$ in the volume $V$ is below $\rho_-$ or above $\rho_+$,
the system is homogeneous at the density $\bar\rho$. If we have
$\rho_-<\bar\rho<\rho_+$, the system is inhomogeneous, with a coexistence of the two phases
at densities $\rho_-$ and $\rho_+$.
Their relative abundance is then given by the constraint on the total mass,
\be
M = \bar\rho V = \rho_- V_- + \rho_+ V_+  .
\ee
Since $V_- < V$ is bounded and $\rho_- \to 0$, we find at low temperature
\be
T \to 0 : \;\;\; \rho_+ V_+ \simeq M , \;\;\; V_+ \simeq \frac{\bar\rho}{\rho_+} V \ll V .
\label{eq:rho+_rho-_rhob}
\ee
In other words, at low temperature and density below $\rho_\Lambda$,
the system goes to a configuration where most of the volume is empty and a small fraction
of the volume is at the characteristic density $\rho_\infty$.
This characteristic density is slightly above the density $\rho_{c_s}$ of
Eq.(\ref{eq:rho-cs}) where $d\Phi_{\rm I}/d\rho$ vanishes.
However, this thermodynamical analysis does not predict the size of the high-density
clumps.

\subsection{Evolution in the phase diagram}

We now go back to the minimization problem (\ref{eq:Omega-def}) within the context
of the micro-canonical ensemble and of the cosmological scalar-field dynamics
studied in the main text, in Sec.~\ref{sec:perturbations}.
Let us consider a constant scalar-field mass $M$ within a constant large comoving volume $V$,
as is appropriate for cosmological dynamics.
The system is homogeneous until the redshift $z_{c_s}$ where the tachyonic instability sets in
and quickly leads to nonlinear density contrasts.
The thermodynamical analysis above is then meant as a shortcut to predict the final state
of the relaxation associated with the highly nonlinear dynamics that follow the entry into the
nonlinear regime.
To do so, we must find where the initial configurations and their subsequent evolution lies in the  phase diagram shown by the lower panel
in Fig.~\ref{fig_thermo}. As we are interested in  times after $z_{c_s}$,
and the expansion of the Universe dilutes the mean density $\bar\rho$ below the
initial value $\rho_{c_s} \sim \rho_\Lambda$, see Eq.(\ref{eq:rho-cs}), we have that
$\bar\rho$ becomes increasingly small as compared with $\rho_\Lambda$ and with the upper branch
$\rho_+\sim\rho_\Lambda$ of the phase diagram.
To find out whether the system is in the strongly inhomogeneous region to the left or in the homogeneous
region to the right of the boundary curve $\rho_-$ we need the energy of the system
(indeed, the effective temperature $T$ is the Lagrange multiplier associated with the energy).
From Eq.(\ref{eq:E-rho}), the energy that corresponds to the homogeneous configuration is
\be
\bar\rho \ll \rho_\Lambda: \;\;\;
E(\bar\rho) = V \, {\cal V}_{\rm I}(\bar\rho) \simeq - \frac{c_1}{2} \frac{\bar\rho}{\rho_\Lambda} M .
\label{eq:E-bar-rho}
\ee
On the other hand, the energy that corresponds to inhomogeneous configurations, with
domains at $\rho_+ \simeq \rho_\infty$ from Eq.(\ref{eq:rho-infty}) and at $\rho_- \ll \rho_\Lambda$,
is
\be
E(\rho_+,\rho_-) = V_+ \, {\cal V}_{\rm I}(\bar\rho_+) + V_- \, {\cal V}_{\rm I}(\bar\rho_-)
\simeq - \frac{3}{16} \frac{c_1^2}{c_2} M ,
\label{eq:E_rho+_rho-}
\ee
where we used Eq.(\ref{eq:rho+_rho-_rhob}).
Thus, we obtain $E(\rho_+,\rho_-) \ll E(\bar\rho) < 0$ and as expected the inhomogeneous
configuration associated with low $T$ is also associated with  a low energy, in our case
a large negative energy.
On the other hand, at the entry into the nonlinear regime at the redshift $z_{c_s}$ we have
$\rho \simeq \rho_{c_s} \sim \rho_\Lambda$ and the initial energy is
$E_{c_s} \sim - c_1 M$.
As long as gravity is negligible, that is, until gravitational clustering develops
at redshifts $z \lesssim 10$, the local self-interactions associated with the potential
${\cal V}_{\rm I}$ conserve the energy within large comoving volumes, which are essentially
independent of each other (it is simply the sum of the internal energies of the scalar-field
solitons contained within each comoving volume).
Therefore, we keep $E \sim - c_1 M$, which selects the inhomogeneous configuration
(\ref{eq:E_rho+_rho-}), whereas the homogeneous configuration (\ref{eq:E-bar-rho})
corresponds to an increasingly far high-energy configuration, with
$E(\bar\rho) \to 0^-$.

Thus, we can conclude that at the entry in the nonlinear regime, at $z_{c_s}$,
the system is close to the upper-right point in the phase diagram shown by the lower panel
in Fig.~\ref{fig_thermo}, where the curves $\rho_+$ and $\rho_-$ meet with
$\rho_+ \sim \rho_- \sim \bar\rho \sim \rho_\Lambda$, and that at later times the system moves
to the lower-left part of the diagram, increasingly far into the inhomogeneous region to the left
of the boundary curve $\rho_-$.
Hence  this simple thermodynamical analysis suggests that after the tachyonic instabililty
studied in Sec.~\ref{sec:tachyonic-exponential} has reached the nonlinear regime the complex dynamics
that follow will lead to a fragmentation of the system over domains of density of the order
of $\rho_+ \sim \rho_\Lambda$, which contain most of the mass, and domains of density
$\rho_- \ll \rho_\Lambda$, which make most of the volume.
This agrees with a simple halo model where the scalar field is clustered into the stable solitons
obtained in Sec.~\ref{sec:scalar-field-clumps} amidst empty space.

\section{Time-dependent Mathieu equation}
\label{sec:Mathieu-1}

The evolution equation (\ref{eq:delta-eta-cos}) reads
\be
\frac{d^2\delta}{d\eta^2} + \omega^2 \delta + \epsilon \, e^{11(\eta-\eta_{c_s})/6}
\cos\left( 2 e^{-\eta} \right) \delta = 0 ,
\label{eq:d-delta2-d-eta2-epsilon}
\ee
with
\be
\omega = \frac{H_{c_s} k^2}{3 m k_{c_s}^2} , \;\;\;
\epsilon = \frac{H_{c_s}^2 k^2}{9 m^2 k_{c_s}^2} .
\ee
By assumption, for the asymptotic behavior (\ref{eq:cs2-I-cosine-cos}) of the Bessel
function to be valid, we restrict ourselves to the range
\be
e^{-\eta} \gg 1 .
\ee
We typically have $\omega \ll 1$ and $\epsilon \ll 1$ as $H_{c_s} \ll m$,
except for very large wave numbers.
We can look for a perturbative expansion in $\epsilon$ of the form
\be
\delta(\eta) = \sum_{n=0}^{\infty} \epsilon^n \delta^{(n)}(\eta) .
\label{eq:delta-eta-perturbative}
\ee
The zeroth-order solution is
\be
\delta^{(0)}(\eta) = \delta_i \cos[\omega(\eta-\eta_i)] ,
\label{eq:delta-0-cos}
\ee
with the initial conditions $\{ \delta= \delta_i, \delta' = 0\}$ at the initial time $\eta_i$.
Thus, when the self-interactions are negligible the density contrast shows acoustic
oscillations of constant amplitude because of the quantum pressure term.
In this regime, the density perturbations do not grow.
Using for instance the method of variation of parameters or Green's function
\cite{Bender-Orszag}, we obtain the solution of Eq.(\ref{eq:d-delta2-d-eta2-epsilon})
up to order $n$ as
\ba
&&
\delta^{(n)}(\eta) = \cos[(\omega(\eta-\eta_i)] \int_{\eta_i}^{\eta} d\eta'
\frac{ \sin[\omega(\eta'-\eta_i)] } {\omega} \nonumber \\
&& \times e^{11(\eta'-\eta_{c_s})/6}
\cos\left( 2 e^{-\eta'} \right) \delta^{(n-1)}(\eta')
- \sin[(\omega(\eta-\eta_i)] \nonumber \\
&& \times \int_{\eta_i}^{\eta} d\eta' \frac{ \cos[\omega(\eta'-\eta_i)] } {\omega}
e^{11(\eta'-\eta_{c_s})/6} \cos\left( 2 e^{-\eta'} \right) \nonumber \\
&& \times \delta^{(n-1)}(\eta') .
\label{eq:delta-n-delta-n-1}
\ea
From this recursion it is easy to obtain the upper bound
\be
\left| \epsilon^n \delta^{(n)}(\eta) \right| \leq \frac{\left| \delta_i \right|}{n!}
\left( \frac{12 \epsilon}{11\omega} e^{11(\eta-\eta_{c_s})/6} \right)^n .
\label{eq:delta-n-upper}
\ee
Therefore, the perturbative expansion (\ref{eq:delta-eta-perturbative}) converges for
all values of $\omega$, $\epsilon$ and $\eta$.
Moreover, we have
\be
| \delta | < 3 | \delta_i |  \;\;\; \mbox{for} \;\;\;
\frac{12 \epsilon}{11\omega} e^{11(\eta-\eta_{c_s})/6} < 1 .
\label{eq:bound-delta}
\ee
Thus, for any wave number the solution is well described by the zeroth-order
acoustic oscillations (\ref {eq:delta-0-cos}) at sufficiently early times.

From Eq.(\ref{eq:delta-n-delta-n-1}) we obtain the first-order correction $\delta^{(1)}$
in terms of incomplete Gamma functions. For moderate values of $\omega$,
and large values of $e^{-\eta}$, this gives
\ba
\omega \ll e^{-\eta} : \;\;\; \delta^{(1)}(\eta) & \simeq & \frac{\delta_i}{4}
e^{11(\eta-\eta_{c_s})/6}  e^{2\eta} \cos[(\omega(\eta-\eta_i)]
\nonumber \\
&& \times \cos\left( 2 e^{-\eta} \right) ,
\label{eq:delta-1-Gamma}
\ea
whereas for large values of $\omega$ we obtain
\ba
\omega \gg e^{-\eta} : \;\;\; \delta^{(1)}(\eta) & \simeq & \frac{\delta_i}{4 \omega}
e^{11(\eta-\eta_{c_s})/6}  e^{\eta} \sin[(\omega(\eta-\eta_i)]
\nonumber \\
&& \times \sin\left( 2 e^{-\eta} \right) .
\label{eq:delta-2-Gamma}
\ea
We can directly check on the equation of motion (\ref{eq:d-delta2-d-eta2-epsilon})
that these are the first-order perturbative corrections associated with the zeroth-order term
(\ref{eq:delta-0-cos}) in these two regimes.
The amplitudes (\ref{eq:delta-1-Gamma}) and (\ref{eq:delta-2-Gamma})
are smaller than the conservative upper bound (\ref{eq:delta-n-upper}) by factors
$e^{\eta} \ll 1$.
This is due to the fast oscillating factor $\cos(2e^{-\eta})$ in the perturbative term of the
equation of motion (\ref{eq:d-delta2-d-eta2-epsilon}), which damps its impact on the dynamics.
Then, the density contrast is well described by the zeroth-order solution (\ref{eq:delta-0-cos})
until the first order correction, given by either (\ref{eq:delta-1-Gamma}) or
(\ref{eq:delta-2-Gamma}), becomes of the same order.

\section{Soliton radial profile as a damped trajectory in a potential}
\label{app:damped-trajectory}

\subsection{Polynomial case}
\label{app:trajectory-polynomial}

\begin{figure}
\begin{center}
\epsfxsize=8.8 cm \epsfysize=6 cm {\epsfbox{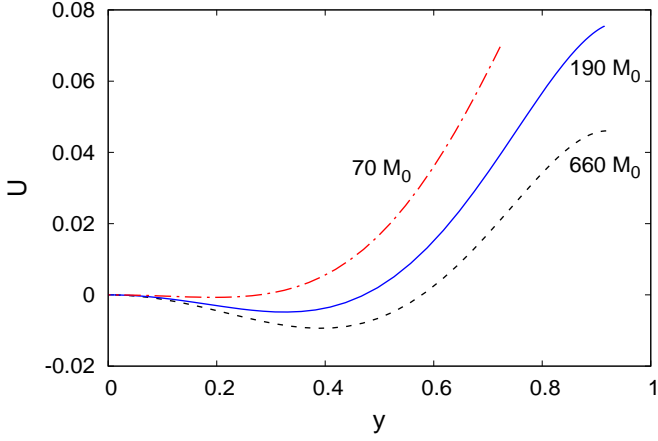}}
\end{center}
\caption{Trajectories in the potentials $U(y)$ for different soliton masses, corresponding
to the soliton radial profiles of Fig.~\ref{fig_rho_x_poly}.}
\label{fig_U_y_poly}
\end{figure}

The differential equation (\ref{eq:y-x-profile}) that determines the radial profile of the
scalar-field solitons, defined as a minimum of the energy at fixed mass, can also be interpreted
as the trajectory $y(x)$ of particle that rolls down a potential $U(y)$, with a time-dependent
friction,
\be
\frac{d^2y}{dx^2} + \frac{2}{x} \frac{dy}{dx} = - \frac{dU}{dy} ,
\label{eq:trajectory-y-x-U}
\ee
where $x$ is interpreted as a time coordinate, running from $0$ to $+\infty$.
In particular, we have
\be
\frac{d}{dx} \left[ \frac{1}{2} \left( \frac{dy}{dx} \right)^2 + U(y) \right] = - \frac{2}{x}
\left( \frac{dy}{dx}\right)^2 \leq 0 ,
\ee
which shows that the total energy of the particle, defined as the sum of its kinetic and potential
energies, decreases with the time $x$. The boundary conditions of the trajectory are
$\frac{dy}{dx}=0$ at $x=0$, because we require a regular profile at the origin, and
$y(x) = 0$ at $x\to\infty$, as the density must decrease at large radii to obtain a finite mass.
With $U(0)=0$, this means that at late times the particle must settle to the point $y=0$
and that it starts at $x=0$ from a value $y_0>0$ with $U(y_0)>0$ and a vanishing velocity.

For the polynomial scalar-field potential (\ref{eq:Phi-I-polynomial}), associated with the
differential equation (\ref{eq:y-x-profile}), the effective particle potential $U(y)$ reads
\be
U(y) = \frac{1}{2} y^4 - \frac{c_2}{3 c_1} y^6 - \tilde\alpha y^2 .
\label{eq:U-y-poly}
\ee
It depends on the unknown parameter $\tilde\alpha$, which is a function of the soliton mass $M$.
This parameter $\tilde\alpha$ is strictly positive so that the density shows an exponential
tail at large radii with $y \sim e^{-\sqrt{2\tilde\alpha}x}$.

We show in Fig.~\ref{fig_U_y_poly} the potentials $U(y)$ for the soliton profiles displayed
in Fig.~\ref{fig_rho_x_poly}, over the range $0 \leq y \leq y_0$ covered by the particle as it rolls
down its potential from the starting point $y_0$.
As $\tilde\alpha > 0$, we can see from (\ref{eq:U-y-poly}) and Fig.~\ref{fig_U_y_poly}
that the potential $U(y)$ first decreases as $- \tilde\alpha y^2$ at low $y$.
This corresponds to the fact that the particle coming from the right must take an infinite time
($x\to\infty$) to reach the zero-density point $y=0$, by slowly climbing upward the potential
$U(y)$. Note that the friction becomes negligible at late times because of the factor $2/x$.
Thanks to the attractive self-interaction term $y^4/2$, the potential $U(y)$ turns upward
to positive values at larger $y$, and next turns downward because of the large-density
repulsive self-interaction term $- \frac{c_2}{3 c_1} y^6$.
In particular, there is only one local minimum $y_-$ and a global maximum $y_+>y_-$ over the
range $0 \leq y < +\infty$. The particle must start slightly to the left of the maximum $y_+$
to roll down to $y=0$, which is reached at infinite time.
Therefore, we can see that the density profile can only reach large masses by having the
particle start very close to the maximum $y_+$, so that it stays there for a very long time
of the order of $x_+$, until it rolls down the potential $U(y)$ to finally settle at $y=0$.
This convergence of the starting point to the maximum $y_+$ is clearly seen in Fig.~\ref{fig_U_y_poly}
as we increase the soliton mass. This in turns means that we have a constant density core
at $\rho = \rho_\Lambda y_0^2$ up to an increasingly large core radius $x_+$, beyond which
the density falls off to converge to its exponential tail.
This agrees with the profiles found in Fig.~\ref{fig_rho_x_poly}.
The position of the maximum $y_+$ is set by the balance between the attractive and repulsive
self-interactions, at $y^2 \sim \frac{c_1}{c_2} \sim 1$, and it does not significantly depend
on the parameter $\tilde\alpha$, and hence nor on the soliton mass $M$.
This means that the core density remains of the order of $\rho_\Lambda$ and stabilizes
to a finite value for large masses, in agreement with Fig.~\ref{fig_rho_x_poly}.

\subsection{Cosine model}
\label{app:trajectory-cosine}

\begin{figure*}
\begin{center}
\epsfxsize=5.9 cm \epsfysize=5 cm {\epsfbox{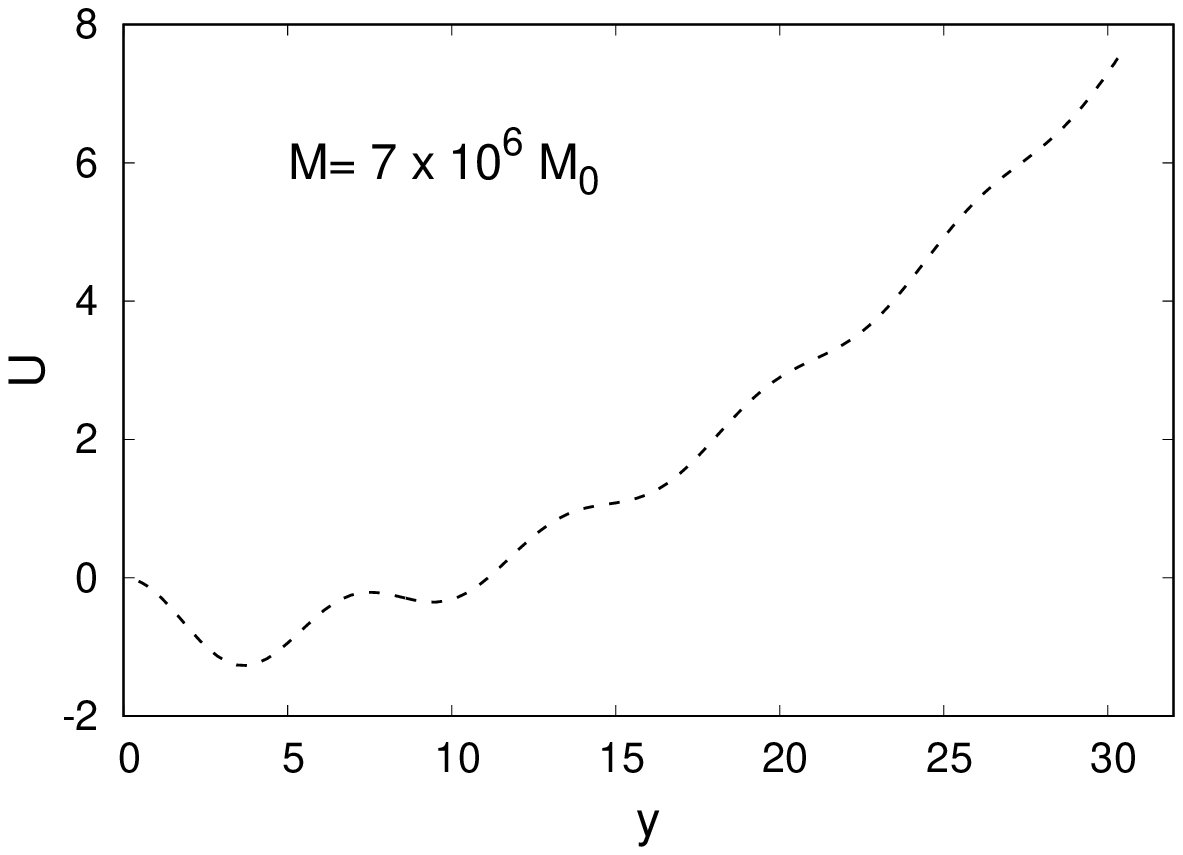}}
\epsfxsize=5.9 cm \epsfysize=5 cm {\epsfbox{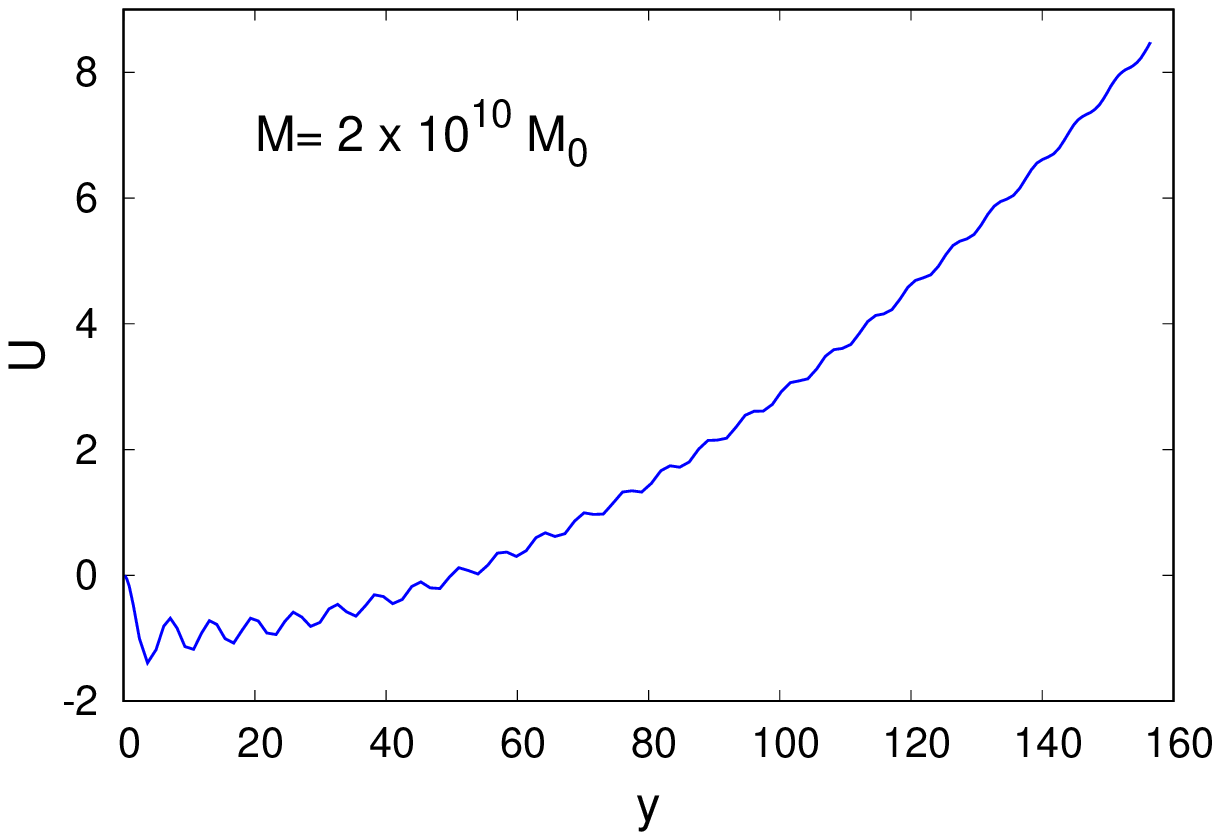}}
\epsfxsize=5.9 cm \epsfysize=5 cm {\epsfbox{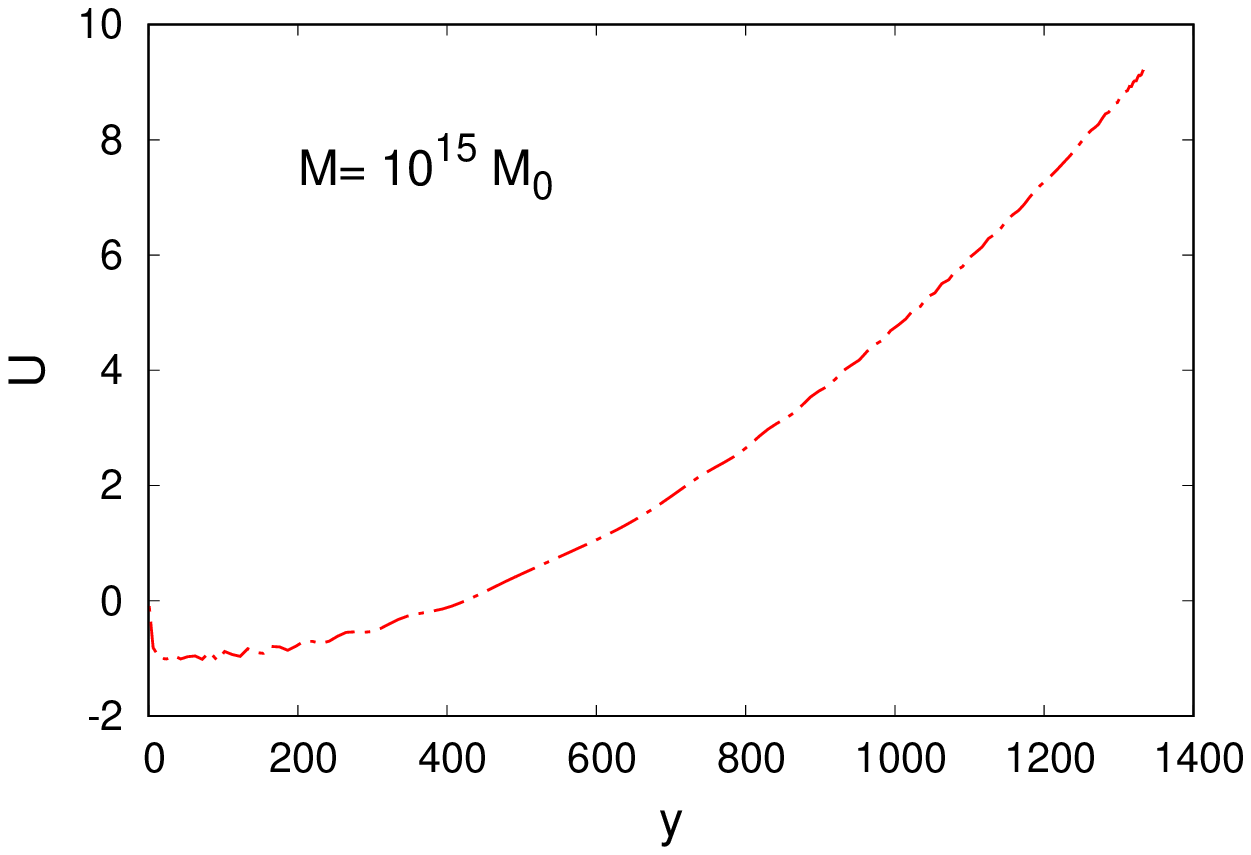}}
\end{center}
\caption{Trajectories in the potentials $U(y)$ for different soliton masses, corresponding
to the soliton radial profiles of Fig.~\ref{fig_rho_x_cosine}.}
\label{fig_U_y_cosine}
\end{figure*}

For the cosine scalar-field potential (\ref{eq:V-phi-cosine}), the soliton radial density profiles
are given by Eq.(\ref{eq:y-x-profile-cosine}). As in Eq.(\ref{eq:trajectory-y-x-U}), this can be
interpreted as the trajectory of a particle $y(x)$ over time $x$, rolling down a potential
$U(y)$ with a time-dependent friction term $\frac{2}{x} \frac{dy}{dx}$.
The potential $U(y)$ is now given by
\be
U(y) = J_0(y) -1 - \frac{\tilde\alpha}{2} y^2 .
\label{eq:U-y-cosine}
\ee
We show in Fig.~\ref{fig_U_y_cosine} the trajectories in the potentials $U(y)$ associated
with the density profiles of Fig.~\ref{fig_rho_x_cosine}.
The parameter $\tilde\alpha$ is now negative, in the range $-\frac{1}{2} < \tilde\alpha < 0$.
The density shows an exponential tail at large radii, with $y \sim e^{-\sqrt{\tilde\alpha+1/2}\, x}$.
This again gives a potential $U(y)$ that decreases with $y$ near the origin,
as $U(y) \simeq - (1+2\tilde\alpha) y^2/4$, so that the particle climbs the potential upward
at late time to reach the origin at infinite time.
In contrast with the polynomial case studied in Fig.~\ref{fig_U_y_poly}, the soliton no longer
gains mass by starting increasingly close to a maximum $y_+$ but by starting at increasingly
large values $y_0$. There, the Bessel function $J_0(y)$ and its oscillations are negligible and
the potential is dominated by the quadratic term $-\tilde\alpha y^2/2$.
Thus, for large masses, the particle slowly rolls down the quadratic potential
$-\tilde\alpha y^2/2$
from increasingly large initial values $y_0$, until $y$ becomes of order unity with $U \simeq -1$.
It next takes an infinite time to climb up to $(y=0,U=0)$.
This agrees with the radial density profiles of Fig.~\ref{fig_rho_x_cosine}, with a core density
and a core radius that grow with the soliton mass.

We can also see that the oscillations of the Bessel function $J_0(y)$ in $U(y)$,
which arise from the Bessel function $J_1(\sqrt{\rho/\rho_b})$ in the self-interaction potential
$\Phi_{\rm I}(\rho)$ in Eq.(\ref{eq:Phi-I-J1}), have a negligible impact at large mass.
This explains why the soliton mass-density relation shown in Fig.~\ref{fig_rho_M_cosine}
takes the power-law form (\ref{eq:M-rhoc-cosine}) predicted by the simple analytic ansatz
(\ref{eq:EI-Gaussian-cosine})-(\ref{eq:EI-top-hat-cosine}).
The Bessel function is not scale free but it only brings negligible deviations from the
power-law behaviors arising from the main quadratic component
$U(y) \simeq -\tilde\alpha y^2/2$.

We can infer that a similar behavior will be found for other models where the self-interaction
component $U_{\rm I}(y)$ goes to a constant or grows more slowly than $y^2$ at large $y$,
in contrast with the polynomial case of Sec.~\ref{app:trajectory-polynomial} where we had
$U_{\rm I}(y) \sim - \frac{c_2}{3c_1} y^6$ at large $y$.
In the general case, $U(y)$ is related to the self-interaction potential ${\cal V}_{\rm I}$ by
\be
U(y) = \frac{\alpha}{2} y^2 - \frac{1}{2\rho_\Lambda} {\cal V}_{\rm I}(\rho_\Lambda y^2) ,
\ee
where we defined $y$ and the dimensionless radius $x$ by
\be
y = \sqrt{\frac{\rho}{\rho_\Lambda}} , \;\;\; x = \sqrt{2} m r ,
\ee
and $\rho_\Lambda$ is a characteristic density.
Thus, models where ${\cal V}_{\rm I}(\rho)$ is bounded or grows more slowly than $\rho$,
will have solitonic density profiles with a core density that grows at high masses,
as in Figs.~\ref{fig_U_y_cosine} and \ref{fig_rho_x_cosine}.
In contrast, models where  ${\cal V}_{\rm I}(\rho)$ has a minimum at a characteristic density
$\rho_\Lambda$ and grows faster than $\rho$ at larger densities will have solitonic density profiles
with a core density that converges to a finite value of order $\rho_\Lambda$ and a radius
that grows as $M^{1/3}$ at high masses, as in Figs.~\ref{fig_U_y_poly} and \ref{fig_rho_x_poly}.

\vspace{-.3cm}

\bibliography{ref3}

\begin{thebibliography}{71}
\expandafter\ifx\csname natexlab\endcsname\relax\def\natexlab#1{#1}\fi
\expandafter\ifx\csname bibnamefont\endcsname\relax
  \def\bibnamefont#1{#1}\fi
\expandafter\ifx\csname bibfnamefont\endcsname\relax
  \def\bibfnamefont#1{#1}\fi
\expandafter\ifx\csname citenamefont\endcsname\relax
  \def\citenamefont#1{#1}\fi
\expandafter\ifx\csname url\endcsname\relax
  \def\url#1{\texttt{#1}}\fi
\expandafter\ifx\csname urlprefix\endcsname\relax\def\urlprefix{URL }\fi
\providecommand{\bibinfo}[2]{#2}
\providecommand{\eprint}[2][]{\url{#2}}

\bibitem[{\citenamefont{Bergstr{\"o}m}(2013)}]{Bergstrom:2012np}
\bibinfo{author}{\bibfnamefont{L.}~\bibnamefont{Bergstr{\"o}m}}, in
  \emph{\bibinfo{booktitle}{{40th Saas-Fee Course}: {Astrophysics at Very-High
  Energies}}} (\bibinfo{year}{2013}), Saas-Fee Advanced Course, pp.
  \bibinfo{pages}{123--222}, \eprint{1202.1170}.

\bibitem[{\citenamefont{Kowalska and Sessolo}(2018)}]{Kowalska:2018toh}
\bibinfo{author}{\bibfnamefont{K.}~\bibnamefont{Kowalska}} \bibnamefont{and}
  \bibinfo{author}{\bibfnamefont{E.~M.} \bibnamefont{Sessolo}},
  \bibinfo{journal}{Adv. High Energy Phys.} \textbf{\bibinfo{volume}{2018}},
  \bibinfo{pages}{6828560} (\bibinfo{year}{2018}), \eprint{1802.04097}.

\bibitem[{\citenamefont{Dine et~al.}(1981)\citenamefont{Dine, Fischler, and
  Srednicki}}]{Dine:1981rt}
\bibinfo{author}{\bibfnamefont{M.}~\bibnamefont{Dine}},
  \bibinfo{author}{\bibfnamefont{W.}~\bibnamefont{Fischler}}, \bibnamefont{and}
  \bibinfo{author}{\bibfnamefont{M.}~\bibnamefont{Srednicki}},
  \bibinfo{journal}{Phys. Lett. B} \textbf{\bibinfo{volume}{104}},
  \bibinfo{pages}{199} (\bibinfo{year}{1981}).

\bibitem[{\citenamefont{Abbott and Sikivie}(1983)}]{Abbott:1982af}
\bibinfo{author}{\bibfnamefont{L.}~\bibnamefont{Abbott}} \bibnamefont{and}
  \bibinfo{author}{\bibfnamefont{P.}~\bibnamefont{Sikivie}},
  \bibinfo{journal}{Phys. Lett. B} \textbf{\bibinfo{volume}{120}},
  \bibinfo{pages}{133} (\bibinfo{year}{1983}).

\bibitem[{\citenamefont{Preskill et~al.}(1983)\citenamefont{Preskill, Wise, and
  Wilczek}}]{Preskill:1982cy}
\bibinfo{author}{\bibfnamefont{J.}~\bibnamefont{Preskill}},
  \bibinfo{author}{\bibfnamefont{M.~B.} \bibnamefont{Wise}}, \bibnamefont{and}
  \bibinfo{author}{\bibfnamefont{F.}~\bibnamefont{Wilczek}},
  \bibinfo{journal}{Phys. Lett. B} \textbf{\bibinfo{volume}{120}},
  \bibinfo{pages}{127} (\bibinfo{year}{1983}).

\bibitem[{\citenamefont{Peccei and Quinn}(1977)}]{Peccei:1977hh}
\bibinfo{author}{\bibfnamefont{R.~D.} \bibnamefont{Peccei}} \bibnamefont{and}
  \bibinfo{author}{\bibfnamefont{H.~R.} \bibnamefont{Quinn}},
  \bibinfo{journal}{Phys. Rev. Lett.} \textbf{\bibinfo{volume}{38}},
  \bibinfo{pages}{1440} (\bibinfo{year}{1977}).

\bibitem[{\citenamefont{Wilczek}(1978)}]{Wilczek:1977pj}
\bibinfo{author}{\bibfnamefont{F.}~\bibnamefont{Wilczek}},
  \bibinfo{journal}{Phys. Rev. Lett.} \textbf{\bibinfo{volume}{40}},
  \bibinfo{pages}{279} (\bibinfo{year}{1978}).

\bibitem[{\citenamefont{Weinberg}(1978)}]{Weinberg:1977ma}
\bibinfo{author}{\bibfnamefont{S.}~\bibnamefont{Weinberg}},
  \bibinfo{journal}{Phys. Rev. Lett.} \textbf{\bibinfo{volume}{40}},
  \bibinfo{pages}{223} (\bibinfo{year}{1978}).

\bibitem[{\citenamefont{Vysotsky et~al.}(1978)\citenamefont{Vysotsky,
  Zeldovich, Khlopov, and Chechetkin}}]{Vysotsky:1978dc}
\bibinfo{author}{\bibfnamefont{M.}~\bibnamefont{Vysotsky}},
  \bibinfo{author}{\bibfnamefont{Y.}~\bibnamefont{Zeldovich}},
  \bibinfo{author}{\bibfnamefont{M.}~\bibnamefont{Khlopov}}, \bibnamefont{and}
  \bibinfo{author}{\bibfnamefont{V.}~\bibnamefont{Chechetkin}},
  \bibinfo{journal}{Pisma Zh. Eksp. Teor. Fiz.} \textbf{\bibinfo{volume}{27}},
  \bibinfo{pages}{533} (\bibinfo{year}{1978}).

\bibitem[{\citenamefont{Marsh}(2016)}]{Marsh:2015xka}
\bibinfo{author}{\bibfnamefont{D.~J.~E.} \bibnamefont{Marsh}},
  \bibinfo{journal}{Phys. Rept.} \textbf{\bibinfo{volume}{643}},
  \bibinfo{pages}{1} (\bibinfo{year}{2016}), \eprint{1510.07633}.

\bibitem[{\citenamefont{Ure{\~n}a-L{\'o}pez}(2019)}]{Urena-Lopez:2019kud}
\bibinfo{author}{\bibfnamefont{L.~A.} \bibnamefont{Ure{\~n}a-L{\'o}pez}},
  \bibinfo{journal}{Front. Astron. Space Sci.} \textbf{\bibinfo{volume}{6}},
  \bibinfo{pages}{47} (\bibinfo{year}{2019}).

\bibitem[{\citenamefont{Hu et~al.}(2000)\citenamefont{Hu, Barkana, and
  Gruzinov}}]{Hu:2000ke}
\bibinfo{author}{\bibfnamefont{W.}~\bibnamefont{Hu}},
  \bibinfo{author}{\bibfnamefont{R.}~\bibnamefont{Barkana}}, \bibnamefont{and}
  \bibinfo{author}{\bibfnamefont{A.}~\bibnamefont{Gruzinov}},
  \bibinfo{journal}{Phys. Rev. Lett.} \textbf{\bibinfo{volume}{85}},
  \bibinfo{pages}{1158} (\bibinfo{year}{2000}), \eprint{astro-ph/0003365}.

\bibitem[{\citenamefont{Hui et~al.}(2017)\citenamefont{Hui, Ostriker, Tremaine,
  and Witten}}]{Hui:2016ltb}
\bibinfo{author}{\bibfnamefont{L.}~\bibnamefont{Hui}},
  \bibinfo{author}{\bibfnamefont{J.~P.} \bibnamefont{Ostriker}},
  \bibinfo{author}{\bibfnamefont{S.}~\bibnamefont{Tremaine}}, \bibnamefont{and}
  \bibinfo{author}{\bibfnamefont{E.}~\bibnamefont{Witten}},
  \bibinfo{journal}{Phys. Rev.} \textbf{\bibinfo{volume}{D95}},
  \bibinfo{pages}{043541} (\bibinfo{year}{2017}), \eprint{1610.08297}.

\bibitem[{\citenamefont{Sakharov and Khlopov}(1994)}]{Sakharov:1994id}
\bibinfo{author}{\bibfnamefont{A.}~\bibnamefont{Sakharov}} \bibnamefont{and}
  \bibinfo{author}{\bibfnamefont{M.}~\bibnamefont{Khlopov}},
  \bibinfo{journal}{Phys. Atom. Nucl.} \textbf{\bibinfo{volume}{57}},
  \bibinfo{pages}{485} (\bibinfo{year}{1994}).

\bibitem[{\citenamefont{Sakharov et~al.}(1996)\citenamefont{Sakharov, Sokoloff,
  and Khlopov}}]{Sakharov:1996xg}
\bibinfo{author}{\bibfnamefont{A.}~\bibnamefont{Sakharov}},
  \bibinfo{author}{\bibfnamefont{D.}~\bibnamefont{Sokoloff}}, \bibnamefont{and}
  \bibinfo{author}{\bibfnamefont{M.}~\bibnamefont{Khlopov}},
  \bibinfo{journal}{Phys. Atom. Nucl.} \textbf{\bibinfo{volume}{59}},
  \bibinfo{pages}{1005} (\bibinfo{year}{1996}).

\bibitem[{\citenamefont{Johnson and Kamionkowski}(2008)}]{Johnson:2008se}
\bibinfo{author}{\bibfnamefont{M.~C.} \bibnamefont{Johnson}} \bibnamefont{and}
  \bibinfo{author}{\bibfnamefont{M.}~\bibnamefont{Kamionkowski}},
  \bibinfo{journal}{Phys. Rev.} \textbf{\bibinfo{volume}{D78}},
  \bibinfo{pages}{063010} (\bibinfo{year}{2008}), \eprint{0805.1748}.

\bibitem[{\citenamefont{Hwang and Noh}(2009)}]{Hwang:2009js}
\bibinfo{author}{\bibfnamefont{J.-c.} \bibnamefont{Hwang}} \bibnamefont{and}
  \bibinfo{author}{\bibfnamefont{H.}~\bibnamefont{Noh}},
  \bibinfo{journal}{Phys. Lett.} \textbf{\bibinfo{volume}{B680}},
  \bibinfo{pages}{1} (\bibinfo{year}{2009}), \eprint{0902.4738}.

\bibitem[{\citenamefont{Park et~al.}(2012)\citenamefont{Park, Hwang, and
  Noh}}]{Park:2012ru}
\bibinfo{author}{\bibfnamefont{C.-G.} \bibnamefont{Park}},
  \bibinfo{author}{\bibfnamefont{J.-c.} \bibnamefont{Hwang}}, \bibnamefont{and}
  \bibinfo{author}{\bibfnamefont{H.}~\bibnamefont{Noh}},
  \bibinfo{journal}{Phys. Rev.} \textbf{\bibinfo{volume}{D86}},
  \bibinfo{pages}{083535} (\bibinfo{year}{2012}), \eprint{1207.3124}.

\bibitem[{\citenamefont{Hlozek et~al.}(2015)\citenamefont{Hlozek, Grin, Marsh,
  and Ferreira}}]{Hlozek:2014lca}
\bibinfo{author}{\bibfnamefont{R.}~\bibnamefont{Hlozek}},
  \bibinfo{author}{\bibfnamefont{D.}~\bibnamefont{Grin}},
  \bibinfo{author}{\bibfnamefont{D.~J.~E.} \bibnamefont{Marsh}},
  \bibnamefont{and} \bibinfo{author}{\bibfnamefont{P.~G.}
  \bibnamefont{Ferreira}}, \bibinfo{journal}{Phys. Rev.}
  \textbf{\bibinfo{volume}{D91}}, \bibinfo{pages}{103512}
  (\bibinfo{year}{2015}), \eprint{1410.2896}.

\bibitem[{\citenamefont{Cembranos et~al.}(2016)\citenamefont{Cembranos, Maroto,
  and N{\'u}{\~n}ez~Jare{\~n}o}}]{Cembranos:2015oya}
\bibinfo{author}{\bibfnamefont{J.~A.~R.} \bibnamefont{Cembranos}},
  \bibinfo{author}{\bibfnamefont{A.~L.} \bibnamefont{Maroto}},
  \bibnamefont{and} \bibinfo{author}{\bibfnamefont{S.~J.}
  \bibnamefont{N{\'u}{\~n}ez~Jare{\~n}o}}, \bibinfo{journal}{JHEP}
  \textbf{\bibinfo{volume}{03}}, \bibinfo{pages}{013} (\bibinfo{year}{2016}),
  \eprint{1509.08819}.

\bibitem[{\citenamefont{Cembranos et~al.}(2017)\citenamefont{Cembranos, Maroto,
  and N{\'u}{\~n}ez~Jare{\~n}o}}]{Cembranos:2016ugq}
\bibinfo{author}{\bibfnamefont{J.~A.~R.} \bibnamefont{Cembranos}},
  \bibinfo{author}{\bibfnamefont{A.~L.} \bibnamefont{Maroto}},
  \bibnamefont{and} \bibinfo{author}{\bibfnamefont{S.~J.}
  \bibnamefont{N{\'u}{\~n}ez~Jare{\~n}o}}, \bibinfo{journal}{JHEP}
  \textbf{\bibinfo{volume}{02}}, \bibinfo{pages}{064} (\bibinfo{year}{2017}),
  \eprint{1611.03793}.

\bibitem[{\citenamefont{Schive et~al.}(2014)\citenamefont{Schive, Chiueh, and
  Broadhurst}}]{Schive:2014dra}
\bibinfo{author}{\bibfnamefont{H.-Y.} \bibnamefont{Schive}},
  \bibinfo{author}{\bibfnamefont{T.}~\bibnamefont{Chiueh}}, \bibnamefont{and}
  \bibinfo{author}{\bibfnamefont{T.}~\bibnamefont{Broadhurst}},
  \bibinfo{journal}{Nature Phys.} \textbf{\bibinfo{volume}{10}},
  \bibinfo{pages}{496} (\bibinfo{year}{2014}), \eprint{1406.6586}.

\bibitem[{\citenamefont{Broadhurst et~al.}(2018)\citenamefont{Broadhurst, Luu,
  and Tye}}]{Broadhurst:2018fei}
\bibinfo{author}{\bibfnamefont{T.}~\bibnamefont{Broadhurst}},
  \bibinfo{author}{\bibfnamefont{H.~N.} \bibnamefont{Luu}}, \bibnamefont{and}
  \bibinfo{author}{\bibfnamefont{S.~H.~H.} \bibnamefont{Tye}}
  (\bibinfo{year}{2018}), \eprint{1811.03771}.

\bibitem[{\citenamefont{Ostriker and Steinhardt}(2003)}]{Ostriker:2003qj}
\bibinfo{author}{\bibfnamefont{J.~P.} \bibnamefont{Ostriker}} \bibnamefont{and}
  \bibinfo{author}{\bibfnamefont{P.~J.} \bibnamefont{Steinhardt}},
  \bibinfo{journal}{Science} \textbf{\bibinfo{volume}{300}},
  \bibinfo{pages}{1909} (\bibinfo{year}{2003}), \eprint{astro-ph/0306402}.

\bibitem[{\citenamefont{Cembranos et~al.}(2005)\citenamefont{Cembranos, Feng,
  Rajaraman, and Takayama}}]{Cembranos:2005us}
\bibinfo{author}{\bibfnamefont{J.~A.~R.} \bibnamefont{Cembranos}},
  \bibinfo{author}{\bibfnamefont{J.~L.} \bibnamefont{Feng}},
  \bibinfo{author}{\bibfnamefont{A.}~\bibnamefont{Rajaraman}},
  \bibnamefont{and} \bibinfo{author}{\bibfnamefont{F.}~\bibnamefont{Takayama}},
  \bibinfo{journal}{Phys. Rev. Lett.} \textbf{\bibinfo{volume}{95}},
  \bibinfo{pages}{181301} (\bibinfo{year}{2005}), \eprint{hep-ph/0507150}.

\bibitem[{\citenamefont{Weinberg et~al.}(2015)\citenamefont{Weinberg, Bullock,
  Governato, Kuzio~de Naray, and Peter}}]{Weinberg:2013aya}
\bibinfo{author}{\bibfnamefont{D.~H.} \bibnamefont{Weinberg}},
  \bibinfo{author}{\bibfnamefont{J.~S.} \bibnamefont{Bullock}},
  \bibinfo{author}{\bibfnamefont{F.}~\bibnamefont{Governato}},
  \bibinfo{author}{\bibfnamefont{R.}~\bibnamefont{Kuzio~de Naray}},
  \bibnamefont{and} \bibinfo{author}{\bibfnamefont{A.~H.~G.}
  \bibnamefont{Peter}}, \bibinfo{journal}{Proc. Nat. Acad. Sci.}
  \textbf{\bibinfo{volume}{112}}, \bibinfo{pages}{12249}
  (\bibinfo{year}{2015}), \bibinfo{note}{[Proc. Nat. Acad.
  Sci.112,2249(2015)]}, \eprint{1306.0913}.

\bibitem[{\citenamefont{Pontzen and Governato}(2014)}]{Pontzen:2014lma}
\bibinfo{author}{\bibfnamefont{A.}~\bibnamefont{Pontzen}} \bibnamefont{and}
  \bibinfo{author}{\bibfnamefont{F.}~\bibnamefont{Governato}},
  \bibinfo{journal}{Nature} \textbf{\bibinfo{volume}{506}},
  \bibinfo{pages}{171} (\bibinfo{year}{2014}), \eprint{1402.1764}.

\bibitem[{\citenamefont{Boylan-Kolchin
  et~al.}(2011)\citenamefont{Boylan-Kolchin, Bullock, and
  Kaplinghat}}]{BoylanKolchin:2011de}
\bibinfo{author}{\bibfnamefont{M.}~\bibnamefont{Boylan-Kolchin}},
  \bibinfo{author}{\bibfnamefont{J.~S.} \bibnamefont{Bullock}},
  \bibnamefont{and}
  \bibinfo{author}{\bibfnamefont{M.}~\bibnamefont{Kaplinghat}},
  \bibinfo{journal}{Mon. Not. Roy. Astron. Soc.}
  \textbf{\bibinfo{volume}{415}}, \bibinfo{pages}{L40} (\bibinfo{year}{2011}),
  \eprint{1103.0007}.

\bibitem[{\citenamefont{Moore et~al.}(1999)\citenamefont{Moore, Ghigna,
  Governato, Lake, Quinn, Stadel, and Tozzi}}]{Moore:1999nt}
\bibinfo{author}{\bibfnamefont{B.}~\bibnamefont{Moore}},
  \bibinfo{author}{\bibfnamefont{S.}~\bibnamefont{Ghigna}},
  \bibinfo{author}{\bibfnamefont{F.}~\bibnamefont{Governato}},
  \bibinfo{author}{\bibfnamefont{G.}~\bibnamefont{Lake}},
  \bibinfo{author}{\bibfnamefont{T.~R.} \bibnamefont{Quinn}},
  \bibinfo{author}{\bibfnamefont{J.}~\bibnamefont{Stadel}}, \bibnamefont{and}
  \bibinfo{author}{\bibfnamefont{P.}~\bibnamefont{Tozzi}},
  \bibinfo{journal}{Astrophys. J.} \textbf{\bibinfo{volume}{524}},
  \bibinfo{pages}{L19} (\bibinfo{year}{1999}), \eprint{astro-ph/9907411}.

\bibitem[{\citenamefont{de~Blok}(2010)}]{deBlok:2009sp}
\bibinfo{author}{\bibfnamefont{W.~J.~G.} \bibnamefont{de~Blok}},
  \bibinfo{journal}{Adv. Astron.} \textbf{\bibinfo{volume}{2010}},
  \bibinfo{pages}{789293} (\bibinfo{year}{2010}), \eprint{0910.3538}.

\bibitem[{\citenamefont{Cembranos et~al.}(2018)\citenamefont{Cembranos, Maroto,
  N{\'u}{\~n}ez~Jare{\~n}o, and Villarrubia-Rojo}}]{Cembranos:2018ulm}
\bibinfo{author}{\bibfnamefont{J.~A.~R.} \bibnamefont{Cembranos}},
  \bibinfo{author}{\bibfnamefont{A.~L.} \bibnamefont{Maroto}},
  \bibinfo{author}{\bibfnamefont{S.~J.}
  \bibnamefont{N{\'u}{\~n}ez~Jare{\~n}o}}, \bibnamefont{and}
  \bibinfo{author}{\bibfnamefont{H.}~\bibnamefont{Villarrubia-Rojo}},
  \bibinfo{journal}{JHEP} \textbf{\bibinfo{volume}{08}}, \bibinfo{pages}{073}
  (\bibinfo{year}{2018}), \eprint{1805.08112}.

\bibitem[{\citenamefont{Armengaud et~al.}(2017)\citenamefont{Armengaud,
  Palanque-Delabrouille, Y{\`e}che, Marsh, and Baur}}]{Armengaud:2017nkf}
\bibinfo{author}{\bibfnamefont{E.}~\bibnamefont{Armengaud}},
  \bibinfo{author}{\bibfnamefont{N.}~\bibnamefont{Palanque-Delabrouille}},
  \bibinfo{author}{\bibfnamefont{C.}~\bibnamefont{Y{\`e}che}},
  \bibinfo{author}{\bibfnamefont{D.~J.~E.} \bibnamefont{Marsh}},
  \bibnamefont{and} \bibinfo{author}{\bibfnamefont{J.}~\bibnamefont{Baur}},
  \bibinfo{journal}{Mon. Not. Roy. Astron. Soc.}
  \textbf{\bibinfo{volume}{471}}, \bibinfo{pages}{4606} (\bibinfo{year}{2017}),
  \eprint{1703.09126}.

\bibitem[{\citenamefont{Brax et~al.}(2019)\citenamefont{Brax, Cembranos, and
  Valageas}}]{Brax:2019fzb}
\bibinfo{author}{\bibfnamefont{P.}~\bibnamefont{Brax}},
  \bibinfo{author}{\bibfnamefont{J.~A.~R.} \bibnamefont{Cembranos}},
  \bibnamefont{and} \bibinfo{author}{\bibfnamefont{P.}~\bibnamefont{Valageas}},
  \bibinfo{journal}{Phys. Rev.} \textbf{\bibinfo{volume}{D100}},
  \bibinfo{pages}{023526} (\bibinfo{year}{2019}), \eprint{1906.00730}.

\bibitem[{\citenamefont{Brax et~al.}(2020{\natexlab{a}})\citenamefont{Brax,
  Cembranos, and Valageas}}]{Brax:2019npi}
\bibinfo{author}{\bibfnamefont{P.}~\bibnamefont{Brax}},
  \bibinfo{author}{\bibfnamefont{J.~A.} \bibnamefont{Cembranos}},
  \bibnamefont{and} \bibinfo{author}{\bibfnamefont{P.}~\bibnamefont{Valageas}},
  \bibinfo{journal}{Phys. Rev. D} \textbf{\bibinfo{volume}{101}},
  \bibinfo{pages}{023521} (\bibinfo{year}{2020}{\natexlab{a}}),
  \eprint{1909.02614}.

\bibitem[{\citenamefont{Brax et~al.}(2020{\natexlab{b}})\citenamefont{Brax,
  Cembranos, and Valageas}}]{Brax:2020tuk}
\bibinfo{author}{\bibfnamefont{P.}~\bibnamefont{Brax}},
  \bibinfo{author}{\bibfnamefont{J.~A.} \bibnamefont{Cembranos}},
  \bibnamefont{and} \bibinfo{author}{\bibfnamefont{P.}~\bibnamefont{Valageas}},
  \bibinfo{journal}{Phys. Rev. D} \textbf{\bibinfo{volume}{101}},
  \bibinfo{pages}{063510} (\bibinfo{year}{2020}{\natexlab{b}}),
  \eprint{2001.06873}.

\bibitem[{\citenamefont{Chavanis}(2012)}]{Chavanis:2011uv}
\bibinfo{author}{\bibfnamefont{P.-H.} \bibnamefont{Chavanis}},
  \bibinfo{journal}{Astron. Astrophys.} \textbf{\bibinfo{volume}{537}},
  \bibinfo{pages}{A127} (\bibinfo{year}{2012}), \eprint{1103.2698}.

\bibitem[{\citenamefont{Carr and Kuhnel}(2020)}]{Carr:2020xqk}
\bibinfo{author}{\bibfnamefont{B.}~\bibnamefont{Carr}} \bibnamefont{and}
  \bibinfo{author}{\bibfnamefont{F.}~\bibnamefont{Kuhnel}}
  (\bibinfo{year}{2020}), \eprint{2006.02838}.

\bibitem[{\citenamefont{{Tisserand} et~al.}(2007)\citenamefont{{Tisserand}, {Le
  Guillou}, {Afonso}, {Albert}, {Andersen}, {Ansari}, {Aubourg}, {Bareyre},
  {Beaulieu}, {Charlot} et~al.}}]{2007A&A...469..387T}
\bibinfo{author}{\bibfnamefont{P.}~\bibnamefont{{Tisserand}}},
  \bibinfo{author}{\bibfnamefont{L.}~\bibnamefont{{Le Guillou}}},
  \bibinfo{author}{\bibfnamefont{C.}~\bibnamefont{{Afonso}}},
  \bibinfo{author}{\bibfnamefont{J.~N.} \bibnamefont{{Albert}}},
  \bibinfo{author}{\bibfnamefont{J.}~\bibnamefont{{Andersen}}},
  \bibinfo{author}{\bibfnamefont{R.}~\bibnamefont{{Ansari}}},
  \bibinfo{author}{\bibfnamefont{{\'E}.}~\bibnamefont{{Aubourg}}},
  \bibinfo{author}{\bibfnamefont{P.}~\bibnamefont{{Bareyre}}},
  \bibinfo{author}{\bibfnamefont{J.~P.} \bibnamefont{{Beaulieu}}},
  \bibinfo{author}{\bibfnamefont{X.}~\bibnamefont{{Charlot}}},
  \bibnamefont{et~al.}, \bibinfo{journal}{\aap} \textbf{\bibinfo{volume}{469}},
  \bibinfo{pages}{387} (\bibinfo{year}{2007}), \eprint{astro-ph/0607207}.

\bibitem[{\citenamefont{Chavanis}(2018)}]{Chavanis:2017loo}
\bibinfo{author}{\bibfnamefont{P.-H.} \bibnamefont{Chavanis}},
  \bibinfo{journal}{Phys. Rev.} \textbf{\bibinfo{volume}{D98}},
  \bibinfo{pages}{023009} (\bibinfo{year}{2018}), \eprint{1710.06268}.

\bibitem[{\citenamefont{Arvanitaki et~al.}(2020)\citenamefont{Arvanitaki,
  Dimopoulos, Galanis, Lehner, Thompson, and Van~Tilburg}}]{Arvanitaki:2019rax}
\bibinfo{author}{\bibfnamefont{A.}~\bibnamefont{Arvanitaki}},
  \bibinfo{author}{\bibfnamefont{S.}~\bibnamefont{Dimopoulos}},
  \bibinfo{author}{\bibfnamefont{M.}~\bibnamefont{Galanis}},
  \bibinfo{author}{\bibfnamefont{L.}~\bibnamefont{Lehner}},
  \bibinfo{author}{\bibfnamefont{J.~O.} \bibnamefont{Thompson}},
  \bibnamefont{and}
  \bibinfo{author}{\bibfnamefont{K.}~\bibnamefont{Van~Tilburg}},
  \bibinfo{journal}{Phys. Rev. D} \textbf{\bibinfo{volume}{101}},
  \bibinfo{pages}{083014} (\bibinfo{year}{2020}), \eprint{1909.11665}.

\bibitem[{\citenamefont{Kolb and Tkachev}(1993)}]{Kolb:1993zz}
\bibinfo{author}{\bibfnamefont{E.~W.} \bibnamefont{Kolb}} \bibnamefont{and}
  \bibinfo{author}{\bibfnamefont{I.~I.} \bibnamefont{Tkachev}},
  \bibinfo{journal}{Phys. Rev. Lett.} \textbf{\bibinfo{volume}{71}},
  \bibinfo{pages}{3051} (\bibinfo{year}{1993}), \eprint{hep-ph/9303313}.

\bibitem[{\citenamefont{Schiappacasse and
  Hertzberg}(2018)}]{Schiappacasse:2017ham}
\bibinfo{author}{\bibfnamefont{E.~D.} \bibnamefont{Schiappacasse}}
  \bibnamefont{and} \bibinfo{author}{\bibfnamefont{M.~P.}
  \bibnamefont{Hertzberg}}, \bibinfo{journal}{JCAP}
  \textbf{\bibinfo{volume}{01}}, \bibinfo{pages}{037} (\bibinfo{year}{2018}),
  \bibinfo{note}{[Erratum: JCAP 03, E01 (2018)]}, \eprint{1710.04729}.

\bibitem[{\citenamefont{Amin et~al.}(2010)\citenamefont{Amin, Easther, and
  Finkel}}]{Amin:2010dc}
\bibinfo{author}{\bibfnamefont{M.~A.} \bibnamefont{Amin}},
  \bibinfo{author}{\bibfnamefont{R.}~\bibnamefont{Easther}}, \bibnamefont{and}
  \bibinfo{author}{\bibfnamefont{H.}~\bibnamefont{Finkel}},
  \bibinfo{journal}{JCAP} \textbf{\bibinfo{volume}{12}}, \bibinfo{pages}{001}
  (\bibinfo{year}{2010}), \eprint{1009.2505}.

\bibitem[{\citenamefont{Amin et~al.}(2012)\citenamefont{Amin, Easther, Finkel,
  Flauger, and Hertzberg}}]{Amin:2011hj}
\bibinfo{author}{\bibfnamefont{M.~A.} \bibnamefont{Amin}},
  \bibinfo{author}{\bibfnamefont{R.}~\bibnamefont{Easther}},
  \bibinfo{author}{\bibfnamefont{H.}~\bibnamefont{Finkel}},
  \bibinfo{author}{\bibfnamefont{R.}~\bibnamefont{Flauger}}, \bibnamefont{and}
  \bibinfo{author}{\bibfnamefont{M.~P.} \bibnamefont{Hertzberg}},
  \bibinfo{journal}{Phys. Rev. Lett.} \textbf{\bibinfo{volume}{108}},
  \bibinfo{pages}{241302} (\bibinfo{year}{2012}), \eprint{1106.3335}.

\bibitem[{\citenamefont{Oll{\'e} et~al.}(2020)\citenamefont{Oll{\'e},
  Pujol{\`a}s, and Rompineve}}]{Olle:2019kbo}
\bibinfo{author}{\bibfnamefont{J.}~\bibnamefont{Oll{\'e}}},
  \bibinfo{author}{\bibfnamefont{O.}~\bibnamefont{Pujol{\`a}s}},
  \bibnamefont{and}
  \bibinfo{author}{\bibfnamefont{F.}~\bibnamefont{Rompineve}},
  \bibinfo{journal}{JCAP} \textbf{\bibinfo{volume}{02}}, \bibinfo{pages}{006}
  (\bibinfo{year}{2020}), \eprint{1906.06352}.

\bibitem[{\citenamefont{Zhang et~al.}(2020)\citenamefont{Zhang, Amin, Copeland,
  Saffin, and Lozanov}}]{Zhang:2020bec}
\bibinfo{author}{\bibfnamefont{H.-Y.} \bibnamefont{Zhang}},
  \bibinfo{author}{\bibfnamefont{M.~A.} \bibnamefont{Amin}},
  \bibinfo{author}{\bibfnamefont{E.~J.} \bibnamefont{Copeland}},
  \bibinfo{author}{\bibfnamefont{P.~M.} \bibnamefont{Saffin}},
  \bibnamefont{and} \bibinfo{author}{\bibfnamefont{K.~D.}
  \bibnamefont{Lozanov}} (\bibinfo{year}{2020}), \eprint{2004.01202}.

\bibitem[{\citenamefont{Amin and Mocz}(2019)}]{Amin:2019ums}
\bibinfo{author}{\bibfnamefont{M.~A.} \bibnamefont{Amin}} \bibnamefont{and}
  \bibinfo{author}{\bibfnamefont{P.}~\bibnamefont{Mocz}},
  \bibinfo{journal}{Phys. Rev. D} \textbf{\bibinfo{volume}{100}},
  \bibinfo{pages}{063507} (\bibinfo{year}{2019}), \eprint{1902.07261}.

\bibitem[{\citenamefont{Guth}(2015)}]{Guth2015}
\bibinfo{author}{\bibfnamefont{A.~H.} \bibnamefont{Guth}},
  \bibinfo{journal}{Physical Review D} \textbf{\bibinfo{volume}{92}}
  (\bibinfo{year}{2015}).

\bibitem[{\citenamefont{Madelung}(1927)}]{Madelung_1927}
\bibinfo{author}{\bibfnamefont{E.}~\bibnamefont{Madelung}},
  \bibinfo{journal}{Zeitschrift fur Physik} \textbf{\bibinfo{volume}{40}},
  \bibinfo{pages}{322} (\bibinfo{year}{1927}), ISSN \bibinfo{issn}{1434-601X},
  \urlprefix\url{http://dx.doi.org/10.1007/BF01400372}.

\bibitem[{\citenamefont{Chavanis}(2011)}]{Chavanis:2011zi}
\bibinfo{author}{\bibfnamefont{P.-H.} \bibnamefont{Chavanis}},
  \bibinfo{journal}{Phys. Rev.} \textbf{\bibinfo{volume}{D84}},
  \bibinfo{pages}{043531} (\bibinfo{year}{2011}), \eprint{1103.2050}.

\bibitem[{\citenamefont{{Gorbunov} and {Rubakov}}(2011)}]{2011itec.book.....G}
\bibinfo{author}{\bibfnamefont{D.~S.} \bibnamefont{{Gorbunov}}}
  \bibnamefont{and} \bibinfo{author}{\bibfnamefont{V.~A.}
  \bibnamefont{{Rubakov}}}, \emph{\bibinfo{title}{{Introduction to the Theory
  of the Early Universe: Cosmological Perturbations and Inflationary Theory}}}
  (\bibinfo{publisher}{WSPC}, \bibinfo{year}{2011}).

\bibitem[{\citenamefont{Visinelli et~al.}(2018)\citenamefont{Visinelli, Baum,
  Redondo, Freese, and Wilczek}}]{Visinelli:2017ooc}
\bibinfo{author}{\bibfnamefont{L.}~\bibnamefont{Visinelli}},
  \bibinfo{author}{\bibfnamefont{S.}~\bibnamefont{Baum}},
  \bibinfo{author}{\bibfnamefont{J.}~\bibnamefont{Redondo}},
  \bibinfo{author}{\bibfnamefont{K.}~\bibnamefont{Freese}}, \bibnamefont{and}
  \bibinfo{author}{\bibfnamefont{F.}~\bibnamefont{Wilczek}},
  \bibinfo{journal}{Phys. Lett. B} \textbf{\bibinfo{volume}{777}},
  \bibinfo{pages}{64} (\bibinfo{year}{2018}), \eprint{1710.08910}.

\bibitem[{\citenamefont{Guzm{\'a}n and Avilez}(2018)}]{Guzman:2018evm}
\bibinfo{author}{\bibfnamefont{F.}~\bibnamefont{Guzm{\'a}n}} \bibnamefont{and}
  \bibinfo{author}{\bibfnamefont{A.~A.} \bibnamefont{Avilez}},
  \bibinfo{journal}{Phys. Rev. D} \textbf{\bibinfo{volume}{97}},
  \bibinfo{pages}{116003} (\bibinfo{year}{2018}), \eprint{1804.08670}.

\bibitem[{\citenamefont{Schwabe et~al.}(2016)\citenamefont{Schwabe, Niemeyer,
  and Engels}}]{Schwabe:2016rze}
\bibinfo{author}{\bibfnamefont{B.}~\bibnamefont{Schwabe}},
  \bibinfo{author}{\bibfnamefont{J.~C.} \bibnamefont{Niemeyer}},
  \bibnamefont{and} \bibinfo{author}{\bibfnamefont{J.~F.}
  \bibnamefont{Engels}}, \bibinfo{journal}{Phys. Rev. D}
  \textbf{\bibinfo{volume}{94}}, \bibinfo{pages}{043513}
  (\bibinfo{year}{2016}), \eprint{1606.05151}.

\bibitem[{\citenamefont{Cotner}(2016)}]{Cotner:2016aaq}
\bibinfo{author}{\bibfnamefont{E.}~\bibnamefont{Cotner}},
  \bibinfo{journal}{Phys. Rev. D} \textbf{\bibinfo{volume}{94}},
  \bibinfo{pages}{063503} (\bibinfo{year}{2016}), \eprint{1608.00547}.

\bibitem[{\citenamefont{Hertzberg et~al.}(2020)\citenamefont{Hertzberg, Li, and
  Schiappacasse}}]{Hertzberg:2020dbk}
\bibinfo{author}{\bibfnamefont{M.~P.} \bibnamefont{Hertzberg}},
  \bibinfo{author}{\bibfnamefont{Y.}~\bibnamefont{Li}}, \bibnamefont{and}
  \bibinfo{author}{\bibfnamefont{E.~D.} \bibnamefont{Schiappacasse}},
  \bibinfo{journal}{JCAP} \textbf{\bibinfo{volume}{07}}, \bibinfo{pages}{067}
  (\bibinfo{year}{2020}), \eprint{2005.02405}.

\bibitem[{\citenamefont{Niikura et~al.}(2019)}]{Niikura:2017zjd}
\bibinfo{author}{\bibfnamefont{H.}~\bibnamefont{Niikura}} \bibnamefont{et~al.},
  \bibinfo{journal}{Nat. Astron.} \textbf{\bibinfo{volume}{3}},
  \bibinfo{pages}{524} (\bibinfo{year}{2019}), \eprint{1701.02151}.

\bibitem[{\citenamefont{Sugiyama et~al.}(2020)\citenamefont{Sugiyama, Kurita,
  and Takada}}]{Sugiyama:2019dgt}
\bibinfo{author}{\bibfnamefont{S.}~\bibnamefont{Sugiyama}},
  \bibinfo{author}{\bibfnamefont{T.}~\bibnamefont{Kurita}}, \bibnamefont{and}
  \bibinfo{author}{\bibfnamefont{M.}~\bibnamefont{Takada}},
  \bibinfo{journal}{Mon. Not. Roy. Astron. Soc.}
  \textbf{\bibinfo{volume}{493}}, \bibinfo{pages}{3632} (\bibinfo{year}{2020}),
  \eprint{1905.06066}.

\bibitem[{\citenamefont{Smyth et~al.}(2020)\citenamefont{Smyth, Profumo,
  English, Jeltema, McKinnon, and Guhathakurta}}]{Smyth:2019whb}
\bibinfo{author}{\bibfnamefont{N.}~\bibnamefont{Smyth}},
  \bibinfo{author}{\bibfnamefont{S.}~\bibnamefont{Profumo}},
  \bibinfo{author}{\bibfnamefont{S.}~\bibnamefont{English}},
  \bibinfo{author}{\bibfnamefont{T.}~\bibnamefont{Jeltema}},
  \bibinfo{author}{\bibfnamefont{K.}~\bibnamefont{McKinnon}}, \bibnamefont{and}
  \bibinfo{author}{\bibfnamefont{P.}~\bibnamefont{Guhathakurta}},
  \bibinfo{journal}{Phys. Rev. D} \textbf{\bibinfo{volume}{101}},
  \bibinfo{pages}{063005} (\bibinfo{year}{2020}), \eprint{1910.01285}.

\bibitem[{\citenamefont{Schneider et~al.}(1992)\citenamefont{Schneider, Ehlers,
  and Falco}}]{Schneider_1992}
\bibinfo{author}{\bibfnamefont{P.}~\bibnamefont{Schneider}},
  \bibinfo{author}{\bibfnamefont{J.}~\bibnamefont{Ehlers}}, \bibnamefont{and}
  \bibinfo{author}{\bibfnamefont{E.~E.} \bibnamefont{Falco}},
  \emph{\bibinfo{title}{Gravitational Lenses}} (\bibinfo{publisher}{Springer
  Berlin Heidelberg}, \bibinfo{year}{1992}),
  \urlprefix\url{https://doi.org/10.1007%2F978-3-662-03758-4}.

\bibitem[{\citenamefont{Bartelmann}(2010)}]{Bartelmann:2010fz}
\bibinfo{author}{\bibfnamefont{M.}~\bibnamefont{Bartelmann}},
  \bibinfo{journal}{Class. Quant. Grav.} \textbf{\bibinfo{volume}{27}},
  \bibinfo{pages}{233001} (\bibinfo{year}{2010}), \eprint{1010.3829}.

\bibitem[{\citenamefont{McAllister et~al.}(2014)\citenamefont{McAllister,
  Silverstein, Westphal, and Wrase}}]{McAllister:2014mpa}
\bibinfo{author}{\bibfnamefont{L.}~\bibnamefont{McAllister}},
  \bibinfo{author}{\bibfnamefont{E.}~\bibnamefont{Silverstein}},
  \bibinfo{author}{\bibfnamefont{A.}~\bibnamefont{Westphal}}, \bibnamefont{and}
  \bibinfo{author}{\bibfnamefont{T.}~\bibnamefont{Wrase}},
  \bibinfo{journal}{JHEP} \textbf{\bibinfo{volume}{09}}, \bibinfo{pages}{123}
  (\bibinfo{year}{2014}), \eprint{1405.3652}.

\bibitem[{\citenamefont{Berges et~al.}(2019)\citenamefont{Berges, Chatrchyan,
  and Jaeckel}}]{Berges:2019dgr}
\bibinfo{author}{\bibfnamefont{J.}~\bibnamefont{Berges}},
  \bibinfo{author}{\bibfnamefont{A.}~\bibnamefont{Chatrchyan}},
  \bibnamefont{and} \bibinfo{author}{\bibfnamefont{J.}~\bibnamefont{Jaeckel}},
  \bibinfo{journal}{JCAP} \textbf{\bibinfo{volume}{08}}, \bibinfo{pages}{020}
  (\bibinfo{year}{2019}), \eprint{1903.03116}.

\bibitem[{\citenamefont{{Abramowitz}}(1974)}]{Abramowitz}
\bibinfo{author}{\bibfnamefont{M.}~\bibnamefont{{Abramowitz}}},
  \emph{\bibinfo{title}{Handbook of Mathematical Functions, With Formulas,
  Graphs, and Mathematical Tables}} (\bibinfo{publisher}{Dover Publications,
  Inc.}, \bibinfo{address}{USA}, \bibinfo{year}{1974}), ISBN
  \bibinfo{isbn}{0486612724}.

\bibitem[{\citenamefont{{McLachlan}}(1947)}]{McLachlan}
\bibinfo{author}{\bibfnamefont{N.~W.} \bibnamefont{{McLachlan}}},
  \emph{\bibinfo{title}{Theory and Application of Mathieu Functions}}
  (\bibinfo{publisher}{Oxford University Press}, \bibinfo{year}{1947}).

\bibitem[{\citenamefont{{Bender} and {Orszag}}(1999)}]{Bender-Orszag}
\bibinfo{author}{\bibfnamefont{C.}~\bibnamefont{{Bender}}} \bibnamefont{and}
  \bibinfo{author}{\bibfnamefont{S.}~\bibnamefont{{Orszag}}},
  \emph{\bibinfo{title}{Advanced Mathematical Methods for Scientists and
  Engineers}} (\bibinfo{publisher}{Springer-Verlag New York},
  \bibinfo{year}{1999}).

\bibitem[{\citenamefont{Avron and Simon}(1981)}]{AVRON198176}
\bibinfo{author}{\bibfnamefont{J.}~\bibnamefont{Avron}} \bibnamefont{and}
  \bibinfo{author}{\bibfnamefont{B.}~\bibnamefont{Simon}},
  \bibinfo{journal}{Annals of Physics} \textbf{\bibinfo{volume}{134}},
  \bibinfo{pages}{76 } (\bibinfo{year}{1981}), ISSN \bibinfo{issn}{0003-4916},
  \urlprefix\url{http://www.sciencedirect.com/science/article/pii/0003491681900051}.

\bibitem[{\citenamefont{Anahtarci and Djakov}(2012)}]{ANAHTARCI2012243}
\bibinfo{author}{\bibfnamefont{B.}~\bibnamefont{Anahtarci}} \bibnamefont{and}
  \bibinfo{author}{\bibfnamefont{P.}~\bibnamefont{Djakov}},
  \bibinfo{journal}{Journal of Mathematical Analysis and Applications}
  \textbf{\bibinfo{volume}{396}}, \bibinfo{pages}{243 } (\bibinfo{year}{2012}),
  ISSN \bibinfo{issn}{0022-247X},
  \urlprefix\url{http://www.sciencedirect.com/science/article/pii/S0022247X12005082}.

\bibitem[{\citenamefont{Fukunaga et~al.}(2019)\citenamefont{Fukunaga, Kitajima,
  and Urakawa}}]{Fukunaga:2019unq}
\bibinfo{author}{\bibfnamefont{H.}~\bibnamefont{Fukunaga}},
  \bibinfo{author}{\bibfnamefont{N.}~\bibnamefont{Kitajima}}, \bibnamefont{and}
  \bibinfo{author}{\bibfnamefont{Y.}~\bibnamefont{Urakawa}},
  \bibinfo{journal}{JCAP} \textbf{\bibinfo{volume}{06}}, \bibinfo{pages}{055}
  (\bibinfo{year}{2019}), \eprint{1903.02119}.

\bibitem[{\citenamefont{Chatrchyan and Jaeckel}(2020)}]{Chatrchyan:2020pzh}
\bibinfo{author}{\bibfnamefont{A.}~\bibnamefont{Chatrchyan}} \bibnamefont{and}
  \bibinfo{author}{\bibfnamefont{J.}~\bibnamefont{Jaeckel}}
  (\bibinfo{year}{2020}), \eprint{2004.07844}.

\bibitem[{\citenamefont{Jaeckel et~al.}(2020)\citenamefont{Jaeckel, Schenk, and
  Spannowsky}}]{Jaeckel:2020mqa}
\bibinfo{author}{\bibfnamefont{J.}~\bibnamefont{Jaeckel}},
  \bibinfo{author}{\bibfnamefont{S.}~\bibnamefont{Schenk}}, \bibnamefont{and}
  \bibinfo{author}{\bibfnamefont{M.}~\bibnamefont{Spannowsky}}
  (\bibinfo{year}{2020}), \eprint{2004.13724}.

\end{thebibliography}


\end{document}